\documentclass[twocolumn,preprintnumbers,amsmath,aps]{revtex4}
\usepackage{graphicx}
\usepackage{dcolumn}
\usepackage{bm}
\usepackage[usenames,dvipsnames]{xcolor}
\usepackage{multirow}
\usepackage{amsmath}
\usepackage{amssymb}
\usepackage{tikz}
\usetikzlibrary{arrows.meta}
\tikzset{%
  >={Latex[width=2mm,length=2mm]},
            base/.style = {rectangle, rounded corners, draw=black,
                           minimum width=2cm, minimum height=1cm,
                           text centered, font=\sffamily},
  activityStarts/.style = {base, fill=blue!10},
       startstop/.style = {base},
    activityRuns/.style = {base, fill=green!10},
         process/.style = {base, minimum width=2.5cm, fill=orange!15,
                           font=\ttfamily},
}
\begin{document}
\newcommand{\kvec}{\mbox{{\scriptsize {\bf k}}}}
\newcommand{\qvec}{\mbox{{\scriptsize {\bf q}}}}
\def\eq#1{Eq.\hspace{1mm}(\ref{#1})}
\def\fig#1{Fig.\hspace{1mm}\ref{#1}}
\def\tab#1{Tab.\hspace{1mm}\ref{#1}}
\def\app#1{Appx.\hspace{1mm}\ref{#1}}
%
%
\title{Carbonaceous sulfur hydride system: the strong-coupled room-temperature superconductor with a low value of Ginzburg-Landau parameter}
\author{I. A. Wrona$^{\left(1\right)}$}
\author{M. Kostrzewa$^{\left(1\right)}$}
\author{K. A. Krok$^{\left(1\right)}$}
\author{A. P. Durajski$^{\left(2\right)}$}
\author{R. Szcz{\c{e}}{\'s}niak$^{\left(1,2\right)}$}
\affiliation{$^1$ Division of Theoretical Physics, 
                  Jan D{\l}ugosz University in Cz{\c{e}}stochowa, Ave. Armii Krajowej 13/15, 42-200 Cz{\c{e}}stochowa, Poland}
\affiliation{$^2$ Division of Physics, 
                  Cz{\c{e}}stochowa University of Technology, Ave. Armii Krajowej 19, 42-200 Cz{\c{e}}stochowa, Poland}  
%
%
\begin{abstract}
The superconducting state in Carbonaceous Sulfur Hydride (C-S-H) system is characterized by the record-high critical temperature of $288$~K 
experimentally observed at $\sim$267 GPa. Herein, we determined the properties of the \mbox{C-S-H} superconducting phase within the scope of both classical Eliashberg equations (CEE) and the Eliashberg equations with vertex corrections (VCEE). We took into account the scenarios pertinent to either the intermediate or the high value of electron-phonon coupling constant ($\lambda\sim 0.75$ or $\lambda\sim 3.3$, respectively). The scenario for the intermediate value, however, cannot be actually realized due to the anomally high value of logarithmic phonon frequency ($\omega_{\rm ln}/k_{B}=7150$~K) it would require. 
On the other hand, we found it possible to reproduce correctly the value of $T_{C}$ and other thermodynamic quantities in the case of strong coupling. 
However, the vertex corrections lower the order parameter values within the range from $\sim 50$~K to $\sim275$~K. 
For the upper critical field $H_{C2}\sim 27$~T, the Ginzburg-Landau parameter $\kappa$ is of the order of $1.7$. 
This correlates well with the sharp drop of resistance observed by Hirsch and Marsiglio at the critical temperature. 
The strong-coupling scenario for C-S-H system is also suggested by the high values of $\lambda$ estimated for  
${\rm H_{3}S}$ ($\lambda\sim 2.1$, $\kappa\sim 1.5$), ${\rm LaH_{10}}$ ($\lambda\sim 2.8$-$3.9$, $\kappa\sim 1.6$), and ${\rm YH_{6}}$ 
($\lambda\sim 1.7$, $\kappa\sim 1.3$) compounds.
\end{abstract}
\maketitle
%
%

According to the Ashcroft thesis from 1968 \textcolor{blue}{\cite{Ashcroft1968}}, hydrogen exposed to extremely high pressure should turn into metal and become the high-temperature superconductor. However, due to the metalization pressure greater than $500$~GPa, these predictions have not been verified experimentally yet \textcolor{blue}{\cite{McMahon2012A, Dias2017}}. In 2004, Ashcroft pointed to the possibility of obtaining stable structures with properties similar to the metallic hydrogen, but achieved at significantly lower pressure, by adding atoms of heavier elements to hydrogen and inducing the effect of chemical pre-compression \textcolor{blue}{\cite{Ashcroft2004}}. The first experiment confirming these theoretical predictions \textcolor{blue}{\cite{Duan2014}} was conducted for $\rm{H_{3}S}$ compound, which electrical resistance drops to zero at $203$~K under the pressure of $155$~GPa \textcolor{blue}{\cite{Drozdov2015A, Drozdov2014A}}. This discovery led to the increase of interest in hydrogen-rich compounds, resulting in many theoretical papers. The most distinctive compounds described later are: 
$\rm{YH_{10}}$ ($T_{C}=326$~K, $250$~GPa) \textcolor{blue}{\cite{Liu2017}}, 
$\rm{LaH_{10}}$ ($T_{C}=288$~K, $200$~GPa) \textcolor{blue}{\cite{Peng2017}}, 
$\rm{ThH_{10}}$ ($T_{C}=241$~K, $100$~GPa) \textcolor{blue}{\cite{Kvashnin2018}}, and 
$\rm{ScH_{9}}$ ($T_{C}=233$~K, $300$~GPa) \textcolor{blue}{\cite{Ye2018}}. 
Among the above systems, $\rm{ThH_{10}}$ and $\rm{LaH_{10}}$ have been experimentally tested with the result: 
$\left[T_{C}\right]_{\rm{ThH_{10}}}=161$~K ($175$~GPa) \textcolor{blue}{\cite{Semenok2020}} and $\left[T_{C}\right]_{\rm{LaH_{10}}}=250$-$260$~K 
($170$-$200$~GPa) \textcolor{blue}{\cite{Drozdov2019A, Somayazulu2019A}}. 

\begin{figure*}
\includegraphics[width=0.6\columnwidth]{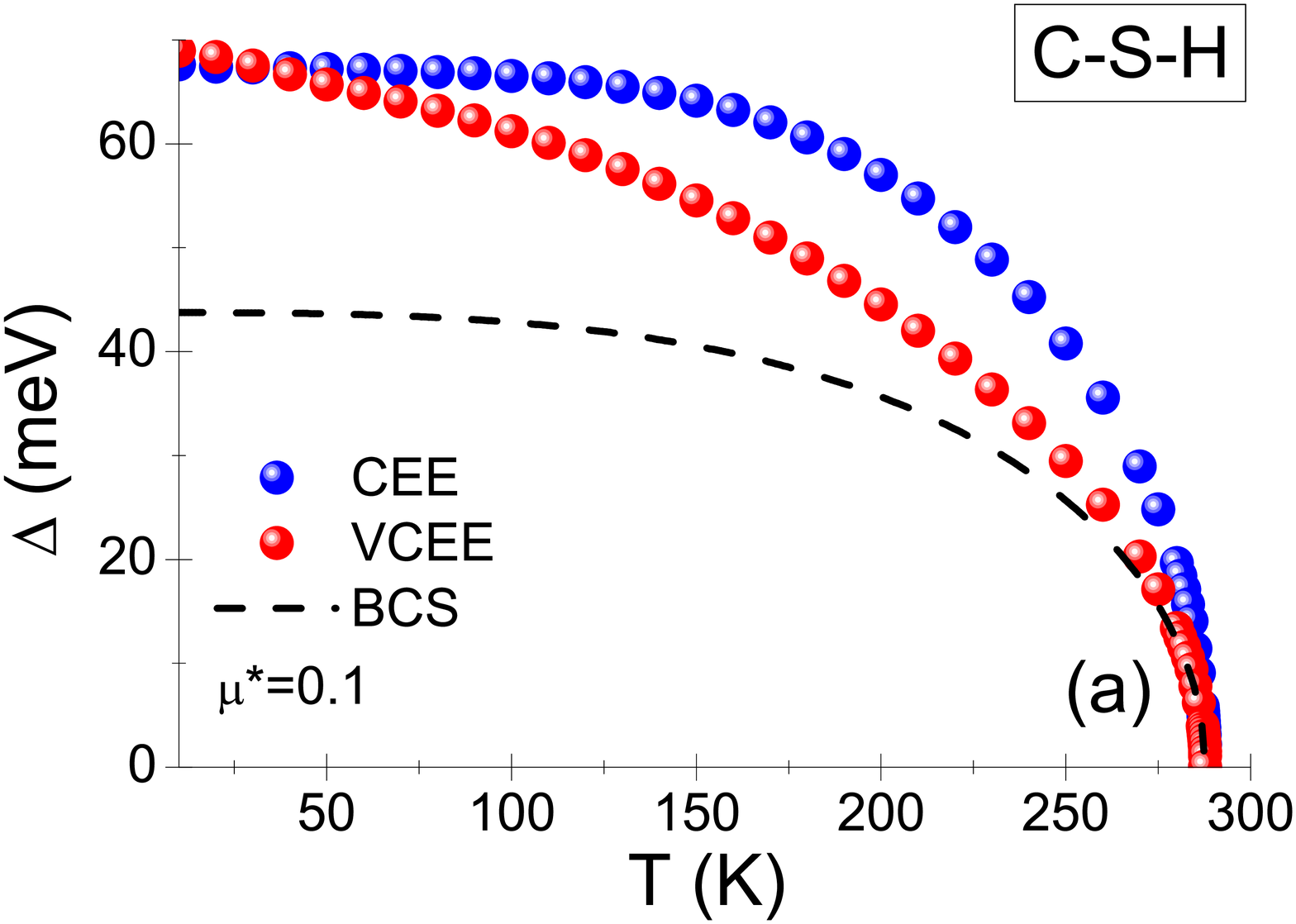}
\includegraphics[width=0.6\columnwidth]{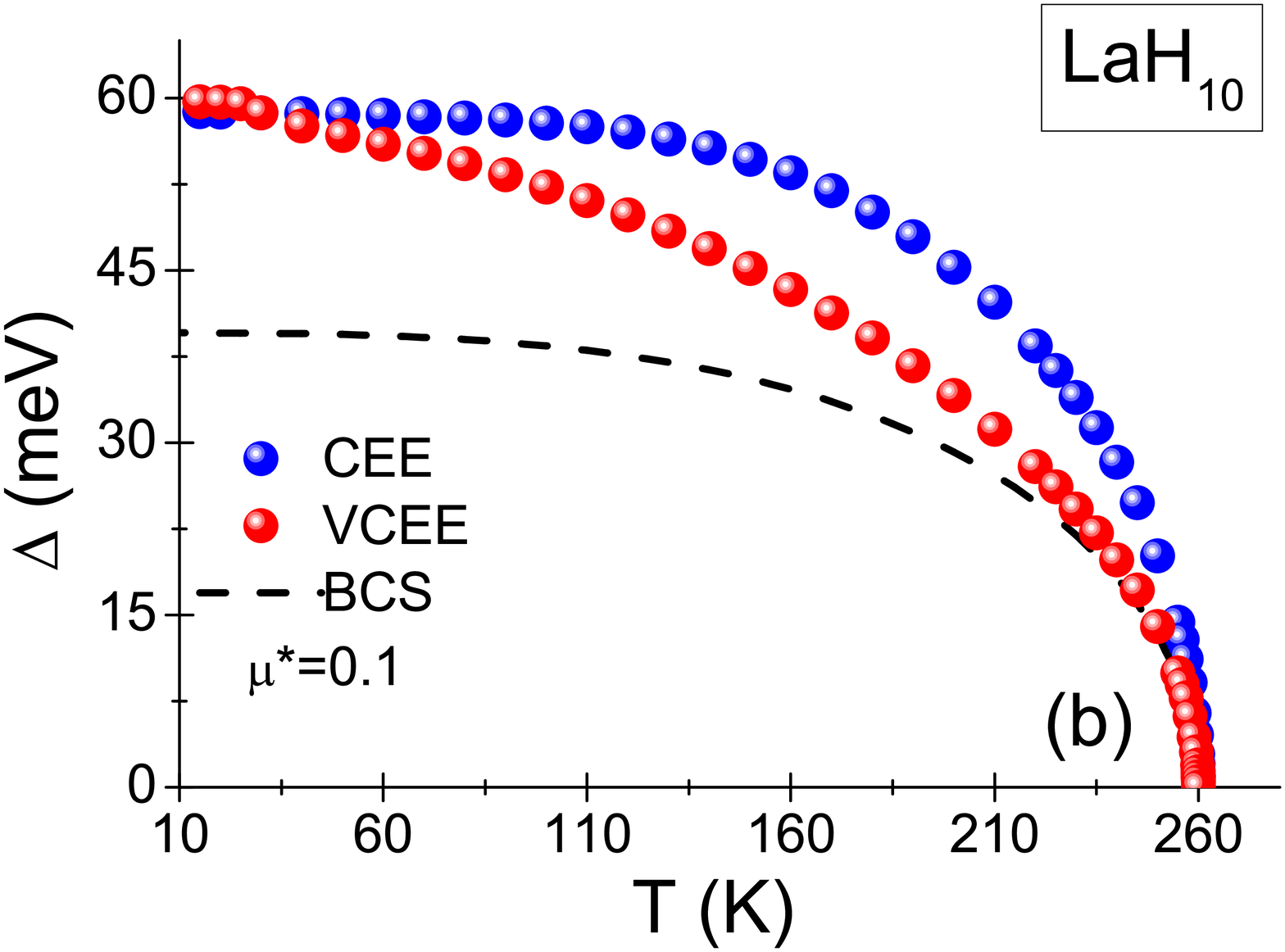}
\includegraphics[width=0.6\columnwidth]{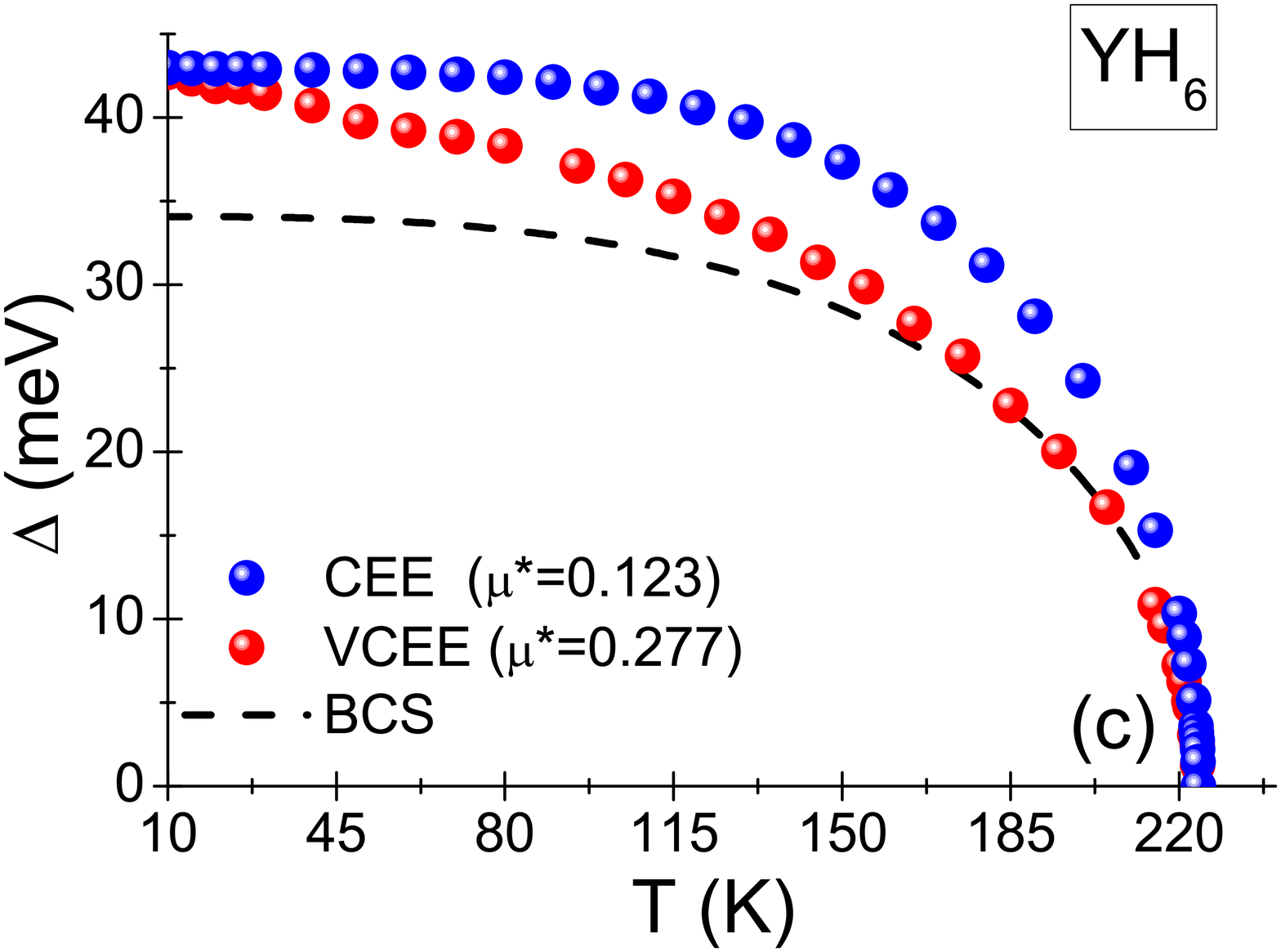}
\caption{
         The order parameter as a function of temperature for C-S-H, ${\rm LaH_{10}}$ and ${\rm YH_{6}}$ systems, respectively. 
         The results were obtained from the classical Eliashberg equations (CEE) 
         and the Eliashberg equations with vertex corrections (VCEE).  
         The colored spheres represent the numerical data.  
         The dashed lines are the BCS predictions. 
}
\label{Fig02-IIIa}
\end{figure*}

The latest experimental data obtained for the \mbox{C-S-H} system under the pressure of $267\pm 10$~GPa prove the existence of the superconducting state 
characterized by the record-high value of critical temperature (\mbox{$T_{C}=287.7$}$\pm 1.2$~K). 
This temperature noticeably exceeds the superconducting transition temperature obtained for the 
${\rm H_{3}S}$ \textcolor{blue}{\cite{Drozdov2015A, Drozdov2014A}} and the
${\rm{LaH_{10}}}$ \textcolor{blue}{\cite{Somayazulu2019A, Drozdov2019A}} compounds. 
The C-S-H system exhibits superconducting properties over a wide range of pressure, from about $140$~GPa to about $275$~GPa. 
For the lowest pressure, the critical temperature is equal to $147$~K and then increases gradually. Near the pressure of $225$~GPa, the characteristic inflection of $T_{C}\left(p\right)$ is observed, which may be related to the structural transition. Above the pressure of $225$~GPa, the increase in critical temperature is very fast. The maximum value of $T_{C}\sim 288$~K was observed at the pressure of $\sim 267$~GPa \textcolor{blue}{\cite{Snider2020A}}. 

It is suspected that the superconducting state in the \mbox{C-S-H} system is induced by the electron-phonon interaction, as in the strongly-coupled systems 
${\rm H_{3}S}$ \mbox{($\left[\lambda\right]_{\rm H_{3}S}\sim 2$)} \textcolor{blue}{\cite{Duan2014, Durajski2016A}} and
${\rm{LaH_{10}}}$ ($\left[\lambda\right]_{\rm LaH_{10}}\sim 3$) \textcolor{blue}{\cite{Liu2017, Kruglov2020A}}. 
The above hypothesis is in line with the recent results obtained with help of DFT calculations. 
In particular, Hu {\it et. al} \textcolor{blue}{\cite{Hu2020A}} showed that the enhancement of electron-phonon coupling can be induced in compounds such as ${\rm C_{1}S_{15}H_{48}}$ and ${\rm C_{1}S_{17}H_{54}}$ 
by replacing a small amount of sulfur atoms with carbon. 
At the same time, this results in the higher averaged phonon frequency, which increases with pressure. As a result, the critical temperature reaches 
the value of the room temperature at $270$~GPa. Additionally, the calculations of critical temperature of both ${\rm C_{1}S_{15}H_{48}}$ and 
${\rm C_{1}S_{17}H_{54}}$, regarded as the function of pressure, are in good agreement with the experimental data \textcolor{blue}{\cite{Snider2020A}}. 
However, according to the Hirsch and Marsiglio's observation \textcolor{blue}{\cite{Hirsch2020A}}, confirmed also by  Dogan and Cohen 
\textcolor{blue}{\cite{Dogan2020A}}, the sharp change in C-S-H resistance at superconducting transition \textcolor{blue}{\cite{Snider2020A}} may indicate that the recorded results mirror some physical mechanisms not related to the superconductivity. Worse still, the same argument 
can be given for compounds such as ${\rm H_{3}S}$ and ${\rm LaH_{10}}$. 

The aim of the presented work is the thorough analysis of Hirsch and Marsiglio's argument on the basis of correctly determined thermodynamic properties 
of the C-S-H, ${\rm H_{3}S}$, ${\rm LaH_{10}}$, and ${\rm YH_{6}}$ compounds. Let us begin with introducing the reasoning adopted by these authors. 

Snider {\it et. al} \textcolor{blue}{\cite{Snider2020A}}, apart from finding the critical temperature of the C-S-H system, estimated also the upper critical field $H_{C2}$. It was shown within the Ginzburg-Landau (GL) approach that $H_{C2}\left(0\right)=61.88$~T, 
with the Pippard coherence length of $\zeta\left(0\right)=2.31$~nm. For the conventional Werthamer-Helfand-Hohenberg (WHH) approach (the dirty limit), 
the value of $H_{C2}(0)=85.34$~T was extrapolated from the slope of the $H$-$T$ curve as: 
$H_{C2}\left(0\right)=0.693T_{C}|\frac{dH_{C2}}{dT}|_{T=T_{C}}$. The coherence length $\zeta\left(0\right)$ is equal to $1.96$~nm in this case.  
For both the GL and the WHH models, the coherence length was calculated from the formula:
$\zeta\left(0\right)=\left[\phi_{0}/2\pi H_{C2}\left(0\right)\right]^{1/2}$, where $\phi_{0}=h/2e=2.068\cdot 10^{-15}$~Wb is the flux quantum. 
Hirsch and Marsiglio noticed \textcolor{blue}{\cite{Hirsch2020A}} that the value of the London penetration depth ($\lambda_{L}\left(0\right)=3.8$~nm
\textcolor{blue}{\cite{Snider2020A}}) was calculated incorrectly. The correctly made estimation of $\lambda_{L}\left(0\right)$ within the GL and BCS model gives $113$~nm. This may suggest that C-S-H is the strongly type-II superconductor with the Ginzburg-Landau parameter $\kappa=\lambda_{L}\left(0\right)/\zeta\left(0\right)$ of $\sim 50$. Therefore it can be placed between the cuprate superconductors ($\kappa\sim 100$) and ${\rm MgB_{2}}$ ($\kappa\sim 28$). 
If so, the sharp drop in resistance observed at the critical temperature cannot be related to superconducting transition, because such a drop in resistance exhibited by the strongly type-II superconductors is evidently milder - especially in the external magnetic field - 
see for example the data reported for ${\rm MgB_{2}}$ \textcolor{blue}{\cite{Canfield2003A}}, YBCO \textcolor{blue}{\cite{Iye1988A}}, or NbN \textcolor{blue}{\cite{Hazra2016A}}.

In the folowing paragraphs, we report our results to show that Hirsch and Marsiglio's argument is incorrect due to the fact that their calculations \textcolor{blue}{\cite{Hirsch2020A}} were carried out by using the weak-coupling model. 

Firstly, we proved that the C-S-H system belongs to the group of superconductors with the high value of coupling constant $\lambda$. 
Hence it follows that its thermodynamic properties can be correctly reproduced within the Eliashberg formalism (the strong-coupling approach) 
\textcolor{blue}{\cite{Eliashberg1960A, Migdal1958A, Carbotte1990A, Freericks1997A}} (Suppl.~I and Suppl.~II). 
Secondly, taking into account the observed sharp drop in C-S-H resistance at the critical temperature, it was assumed that the superconducting phase of the discussed system is characterized by the low value of the Ginzburg-Landau parameter. It should be noted that this scenario is in line with predictions of the Eliashberg formalism. 

The thermodynamic properties of the \mbox{C-S-H} system in the superconducting state were calculated for both the intermediate-coupling ($\lambda=0.75$) and the strong-coupling ($\lambda=3.26$) approach (Suppl.~III and Suppl.~IV). The logarithmic phonon frequency $\omega_{\rm ln}$
was estimated using the formula \textcolor{blue}{\cite {Mitrovic1984A}}:
$R_{\Delta}=3.53\left[1+12.5\left[T_{C}/\omega_{\rm ln}\right]^{2}\ln\left(\omega_{\rm ln}/2T_ {C}\right)\right]$. 
The dimensionless ratio $R_{\Delta}$ was calculated within the Eliashberg approach (see Tab.~I in Suppl.~III). 
The critical temperature $T_{C}=287.7$~K was taken from the experiment \textcolor{blue}{\cite{Snider2020A}}. For the intermediate coupling, we obtained 
$\omega_{\rm ln}/k_{B}=7150$~K. This anomally high value precludes the scenario of the intermediate (or weak) electron-phonon coupling 
in the case of the C-S-H system. In the strong-coupling case, the ratio $\omega_{\rm ln}/k_{B}$ is equal to $1130$~K. 
Similar values of $\omega_{\rm ln}/k_{B}$ ($800$-$1600$~K) were obtained also for the
${\rm H_{2}S}$, ${\rm H_{3}S}$, ${\rm LaH_{10}}$, and ${\rm YH_ {6}}$ compounds (Suppl.~V). 
Please note the parameter $r=k_{B}T_{C}/\omega_{\rm ln}$, which characterizes retardation and strong-coupling effects. 
The value of $r$ equals $0.04$ for $\lambda=0.75$ (in the BSC limit $r=0$), and it reaches $0.25$ in the case of strong-coupling. 
Thus the deviation from the prediction of BCS theory is clearly visible. 

\begin{table*}
\caption{\label{Tab01-IV} The values of upper critical field for the C-S-H system in the superconducting state ($p=267$~GPa). Results obtained by using the CEE model for the intermediate-coupling and the strong-coupling cases.}
\begin{ruledtabular}
\begin{tabular}{|c|c|c|c|c|c|c|c|}
             &          &        &         &      &     &    &               \\
$\lambda$                         & $\omega_{\rm ln}/k_{B}$~(K)        & $h^{cl}_{C2}\left(0\right)$    &  $H^{cl}_{C2}\left(0\right)$~(T)  & 
$h^{di(1)}_{C2}\left(0\right)$    & $H^{di(1)}_{C2}\left(0\right)$~(T) & $h^{di(2)}_{C2}\left(0\right)$ &  $H^{di(2)}_{C2}\left(0\right)$~(T) \\
          &         &         &         &         &        &         &            \\
\hline
          &         &         &               &         &              &         &                  \\
 0.75     & 7150.5  & 0.72631 &      89.44    & 0.65603 &      80.79   & 0.66157 &      81.47       \\          
          &         &         &               &         &              &         &                  \\
\hline
          &         &         &               &         &              &         &                  \\
 3.26     & 1130.5  & 0.93388 &       115     & 0.56886 &      70.0    & 0.22370 & {\bf 27.5}       \\           
          &         &         &               &         &              &         &                  \\

\end{tabular}
\end{ruledtabular}
\end{table*}
The upper critical field at zero Kelvin was estimated by the formula \textcolor{blue}{\cite{Carbotte1990A}}:  
$H_{C2}\left(0\right)=\eta\left(T_{C}\right) h_{C2}\left(0\right)$, where the coefficient 
$\eta\left(T_{C}\right)=T_{C}|\frac{dH_{C2}(T)}{dT}|_{T_{C}}|$ was calculated on the basis of experimental data \textcolor{blue}{\cite{Snider2020A}}. 
We identified the value of $h_{C2}\left(0\right)$ in the clean ({\it cl}) and dirty ({\it di}) limit. In the {\it cl} case, 
the formula for $h_{C2}\left(0\right)$ takes the form \textcolor{blue}{\cite{Carbotte1990A}}: 
$h^{cl}_{C2}\left(0\right)=0.727\left[1-2.7\left[T_{C}/\omega_{\rm ln}\right]^{2}\ln\left(\omega_{\rm ln}/20T_{C}\right)\right]$. 
For the dirty limit, two expressions were given \textcolor{blue}{\cite{Carbotte1990A}}, that represent the limits within which the experimental data should fall: 
$h^{di(1)}_{C2}\left(0\right)=0.69\left[1-1.5T_{C}/\omega_{\rm ln}+2\left[T_{C}/\omega_{\rm ln}\right]^{2}\ln\left(\omega_{\rm ln}/0.8T_{C}\right)\right]$, and 
$h^{di(2)}_{C2}\left(0\right)=0.69\left[1-T_{C}/\omega_{\rm ln}+3.2\left[T_{C}/\omega_{\rm ln}\right]^{2}\ln\left(\omega_{\rm ln}/30T_{C}\right)\right]$. We obtained the results which are collected \mbox{in \tab{Tab01-IV}}. It is clearly visible that, for the intermediate coupling, the theoretical values of upper critical field ($H^{di(1)}_{C2}\left(0\right)=80.79$~T and 
$H^{di(1)}_{C2}\left(0\right)=81.47$~T) agree qualitatively with the Snider's data: $H_{C2}(0)=85.34$~T (the WHH approach). 
This result should not come as a~surprise, because in the case of intermediate coupling, the Eliashberg equations give results comparable to 
the BCS mean-field theory and related phenomenological models. However, as we already mentioned, the method of estimating the upper critical field based on the intermediate-coupling approach should be rejected due to the required anomally high value of logarithmic phonon frequency. 
In the strong-coupling limit, the upper critical field computed for the {\it cl} case has a very high value of $115$~T. 
It is hard to suppose, however, that this case would take place in such a complex system as C-S-H. 
In the dirty limit, the Eliashberg theory predicts a wide range of upper critical field, from $\sim 27.5$~T to $\sim 70$~T. 
\begin{table*}
\caption{\label{Tab02-IV} The values of $\zeta$, $\lambda_{L}$ and $\kappa$ for the C-S-H system ($p=267$~GPa). Results obtained by using the CEE model for the intermediate-coupling and the strong-coupling cases.}
\begin{ruledtabular}
\begin{tabular}{|c|c|c|c|c|c|c|}
          &                  &               &         &              &         &                  \\
$\lambda$ & $\zeta^{di(1)}\left(0\right)$~(nm) & $\zeta^{di(2)}\left(0\right)$~(nm) & $\lambda_{L}^{di(1)}\left(0\right)$~(nm) & 
$\lambda_{L}^{di(2)}\left(0\right)$~(nm) & $\kappa^{di(1)}$ & $\kappa^{di(2)}$                     \\
          &                  &               &         &              &         &                  \\
\hline
          &                  &               &         &              &         &                  \\
 0.75     &          2.02    & 2.01          & 124.01  & 124.79       & 61.41   & 62.06            \\          
          &                  &               &         &              &         &                  \\
\hline
          &                  &               &         &              &         &                  \\
 3.26     &          2.17    & {\bf 3.46}    & 30.57   & {\bf 15.17}  & 14.09   & {\bf 4.38}       \\           
          &                  &               &         &              &         &                  \\

\end{tabular}
\end{ruledtabular}
\end{table*}
The computed values of the Pippard coherence length $\zeta\left(0\right)$, the London penetration depth
$\lambda_{L}\left(0\right)=1935\left(\Delta\left(0\right)\zeta\left(0\right)\right)^{-3/2}\left(m_{e}/m_{e}^{\star}\right)$ 
\textcolor{blue}{\cite{Hirsch2020A}}, and the Ginzburg-Landau parameter $\kappa$ are gathered in \tab{Tab02-IV}. 
The sharp change in resistance observed experimentally for C-S-H at the transition temperature strongly suggests that, for the system in question, one should take into account the low value of Ginzburg-Landau parameter of the order of $4$ (\tab{Tab02-IV}). 
Finally, the Ginzburg-Landau parameter for the Eliashberg approach can also be calculated directly from the formula 
\textcolor{blue}{\cite{Carbotte1990A}}: $\kappa=1.2\left[1+2.3\left[T_{C}/\omega_{\rm ln}\right]^{2}\ln\left(\omega_{\rm ln}/0.2T_{C}\right)\right]$. 
In this case, for the value of $\lambda$ befitting the strong-coupling, we get $\kappa=1.73$. The above result correlates well with $\kappa\sim 4$ obtained within the more qualitative approach. 

The presented analysis of C-S-H properties is consistent with theoretical results based on the DFT method applied to the ${\rm C_{1}S_{15}H_{48}}$ and ${\rm C_{1}S_{17}H_{54}}$ systems \textcolor{blue}{\cite{Hu2020A}}. Additionaly, it should be kept in mind that 
in all hydrogen-rich systems with high $T_{C}$, only the scenario of strong-coupling has been realized so far (Suppl.~V). 
For example, the electron-phonon coupling constant $\lambda$ is of the order of $2$-$3$ in the case of ${\rm LaH_{10}}$ \textcolor{blue}{\cite{Kostrzewa2020A, Kruglov2020A}} (Suppl.~VI), and it was found to be $\sim 2$ for ${\rm H_{3}S}$ \textcolor{blue}{\cite{Duan2014, Durajski2016A}}. 
Let us also mention that another experimental detection of high-temperature superconducting state was recently reported \textcolor{blue}{\cite{Troyan2020A}}, this time in the $\rm YH_{6}$ compound ($T_{C}=224$~K at $166$~GPa). We provided the detailed description of ${\rm YH_{6}}$ properties by using  
the classical Eliashberg equations (CEE) and the Eliashberg equations with vertex corrections (VCEE) (Suppl.~VII). It turns out that 
the estimated value of electron-phonon coupling constant for $\rm YH_{6}$ is also relatively high and amounts to $1.71$ 
\textcolor{blue}{\cite{Troyan2020A}}. 

In the considered cases, the correctly calculated values of Ginzburg-Landau parameter are: 
$\left[\kappa\right]_{{\rm LaH_{10}}}=1.60$ ($T_{C}=260$~K; $p=190$~GPa), 
$\left[\kappa\right]_{{\rm H_{3}S}}=1.53$ ($T_{C}=203$~K; $p=155$~GPa) 
and 
$\left[\kappa\right]_{{\rm YH_{6}}}=1.34$ ($T_{C}=224$~K; \mbox{$p=166$~GPa}), respectively.

The temperature dependence of order parameter determined within the CEE formalism distinctly differs from the predictions of BCS theory for all hydrogen-rich high-temperature superconductors (see \fig{Fig02-IIIa}). In addition, the vertex corrections are important for the region of intermediate temperature and lower the values of $\Delta\left(T\right)$. The low-temperature value of C-S-H order parameter 
$\Delta\left(0\right)$ is between $67.64$~meV and $68.94$~meV (Tab.~I in Suppl.~III). This means that the dimensionless ratio 
$R_{\Delta}=2\Delta\left(0\right)/k_{B}T_{C}$ is from $5.46$ to $5.56$. 
The electron effective mass at $T_{C}$ ranges from $2.69$~$m_{e}$ to $4.26$~$m_{e}$. 
The dimensionless parameters $R_{H}=T_{C}C^{N}\left(T_{C}\right)/H_{C}^{2}\left(0\right)$ and $R_{C}=\Delta C\left(T_{C}\right)/C^{N}\left(T_{C}\right)$  are equal to $0.177$ and $2.37$, respectively. Taking the above into account, it can be clearly shown by using experimental methods that the C-S-H system 
belongs to the family of superconductors with high value of electron-phonon coupling constant. 

Hirsch and Marsiglio noted in their paper \textcolor{blue}{\cite{Hirsch2020A}} that for high values of electron-phonon coupling constant 
($\lambda>2$) the Eliashberg formalism may not apply due to the formation of polarons. Please note, however, that the standard Eliashberg equations do not 
take into account all mechanisms that may contribute to the reduction of coupling constant. In particular, attention should be paid to anharmonic effects \textcolor{blue}{\cite{Errea2015A}}, which should be taken into account both in the Eliashberg function and in the form of equations themselves. 
The many-body effects also play the significant role, lowering the value of $\lambda$, while the form of function $\Delta\left(T\right)$ does not change (Suppl.~VIII). This means that the Eliashberg approach is, with high probability, sufficient to determine correctly the properties of hydrogen-rich high-temperature superconductors. 
 
Summarizing, the conducted analysis showed that the superconducting state in the C-S-H system can be induced by strong electron-phonon interaction, as it is for the ${\rm LaH_{10}}$, ${\rm H_{3}S}$, and ${\rm YH_{6}}$ superconductors. For the C-S-H system, the Eliashberg formalism predicts that the value of 
Ginzburg-Landau parameter is low \mbox{($\kappa=1.73$)}. We got equally low values also in other cases, i.e. $\left[\kappa\right]_{{\rm LaH_{10}}}=1.60$, 
$\left[\kappa\right]_{{\rm H_{3}S}}=1.53$, 
and 
$\left[\kappa\right]_{{\rm YH_{6}}}=1.34$. 
This means that the experimental results obtained for hydrogen-rich compounds do not contradict the theory of superconducting state. 
Referring to the DFT-Eliashberg formalism, it should be strongly emphasized that this is the approach that allows for the correct 
prediction of properties of superconducting state before performing experimental measurements. 
   
%
\bibliography{Bibliography2}
\clearpage
 \onecolumngrid
\section*{Supporting Information for:
       {\it Carbonaceous sulfur hydride system: the strong-coupled room-temperature superconductor with a low value of Ginzburg-Landau parameter}}

\section{\label{Dod0A} Classical Eliashberg formalism}

Basic equations used to analyse the thermodynamic properties of superconducting state in high-pressure hydrogen-containing systems 
are the classical Eliashberg equations (CEE) \textcolor{blue}{\cite{Eliashberg1960A, Migdal1958A, Carbotte1990A}}. 
The mentioned formalism allows to take into account the retardation and strong-coupling effects related to linear electron-phonon 
interaction. Equally important is the fact that input parameters to the pairing kernel in Eliashberg equations can be 
calculated with high accuracy using the DFT method \textcolor{blue}{\cite{Parr1989}}. 

On imaginary axis ($i=\sqrt{-1}$), the classical Eliashberg equations take following form: 
\begin{equation}
\varphi_{n}=\pi k_BT\sum_{m=-M}^{M}
\frac{\lambda_{n,m}-\mu^{\star}\left(\omega_m\right)}
{\sqrt{\omega_m^2Z_m^2+\varphi_{m}^2}}\varphi_{m},
\label{r01-A}
\end{equation}
\begin{equation}
Z_{n}=1+\pi k_BT\sum_{m=-M}^{M}
\frac{\lambda_{n,m}}{\sqrt{\omega_m^2Z_m^2+\varphi_{m}^2}}\frac{\omega_m}{\omega_n}Z_{m},
\label{r02-A}
\end{equation}
where: $\varphi_{n}=\varphi_{n}(i\omega_n)$ and $Z_{n}=Z(i\omega_n)$ denote the order parameter function and the wave
 function renormalization factor, respectively. The order parameter is defined by: $\Delta_{n}=\varphi_{n}/Z_{n}$. 
The other symbols have the following meanings: $\omega_{n}$ is the $n$-th fermionic Matsubara frequency expressed by formula: 
$\omega_n=\pi k_BT(2n-1)$, where $k_{B}$ is the Boltzmann constant. The function $\mu^{\star}(\omega_m)$ models depairing 
interaction between electrons: 
$\mu^{\star}(\omega_m)=\mu^{\star}\theta(\Omega_{C}-|\omega_m|)$, $\mu^{\star}$ is the Coulomb pseudopotential 
\textcolor{blue}{\cite{Morel1962A,Bauer2012A}}, $\theta$ denotes the Heaviside function and $\Omega_{C}$ is so-called cut-off frequency. 
In numerical calculations, we have assumed $\Omega_{C}=1$~eV. The electron-phonon pairing kernel can be defined as follows:
\begin{eqnarray}
\label{r03-A}
\lambda_{n,m}&=&2\int^{\omega_{D}}_{0}d\omega\frac{\alpha^{2}F\left(\omega\right)\omega}{\left(\omega_{n}-\omega_{m}\right)^{2}+\omega^{2}}
=2\int^{\omega_{D}}_{0}d\omega\frac{\omega^{2}}{\left(\omega_{n}-\omega_{m}\right)^{2}+\omega^{2}}\frac{\alpha^{2}F\left(\omega\right)}{\omega}
\\ \nonumber
&\simeq&\frac{\omega_{0}^{2}}{\left(\omega_{n}-\omega_{m}\right)^{2}+\omega_{0}^{2}} 
2\int^{\omega_{D}}_{0}d\omega\frac{\alpha^{2}F\left(\omega\right)}{\omega}
=\lambda\frac{\omega_{0}^{2}}{\left(\omega_{n}-\omega_{m}\right)^{2}+\omega_{0}^{2}}, 
\end{eqnarray}
where: $\lambda=2\int^{\omega_{0}}_{0}d\Omega\frac{\alpha^{2}F\left(\Omega\right)}{\Omega}$, denotes the electron-phonon coupling constant, 
$\alpha^{2}F\left(\Omega\right)$ is the Eliashberg function and $\omega_{0}$ denotes the characteristic phonon frequency.   

After determining the values of order parameter $\Delta_{n}$ and wave function renormalization factor $Z_{n}$, 
the following thermodynamic parameters of superconducting state can be calculated: 

\begin{itemize}

\item The half-width of energy gap:  
\begin{equation}
\Delta\left(T\right)={\rm Re}\left[\Delta\left(\omega=\Delta\left(T\right)\right)\right],
\label{r04-A}
\end{equation}
where $\Delta\left(\omega\right)$ is calculated by analytical continuation of the solutions of Eliashberg equations on real axis \textcolor{blue}{\cite{Beach2000A}}.
On this basis, the dimensionless ratio $R_{\Delta}=2\Delta\left(0\right)/k_{B}T_{C}$ is determined,
where $\Delta(0)=\Delta\left(T_{0}\right)$. In the \fig{Fig01-A} for C-S-H, we presented the example form of order parameter and 
wave function renormalization factor on real axis. We have taken into account the lowest temperature: $T=T_{0}$. 
We can notice that order parameter is the complex function and in the range of lower frequencies,
corresponding to physical values of energy gap, the non-zero is only Re$\left[\Delta\left(\omega\right)\right]$.
From the physical point of view this result indicates the absence of damping effects, which are modeled by 
Im$\left[\Delta\left(\omega\right)\right]$. Based on presented data, it is also possible to calculate the quasiparticle density 
of states: $\frac{N^{S}(\omega)}{N^{N}(\omega)}={\rm Re}\left[\frac{\omega-i\Gamma}{\sqrt{\left(\omega-i\Gamma\right)^2-\Delta^2(\omega)}}
\right]$, where the pair-breaking parameter $\Gamma$ equals $0.1$~meV. In \fig{Fig02-A}, we plotted $N^{S}(\omega)/N^{N}(\omega)$ 
for the cases $\omega_{0}=\omega_{D}$ and $\omega_{0}=100$~meV. The characteristic maxima of $N^{S}\left(\omega\right)/N^{N}\left(\omega\right)$ 
are formed at the points $\omega=\pm\Delta$.
\begin{figure}
\includegraphics[width=0.48\columnwidth]{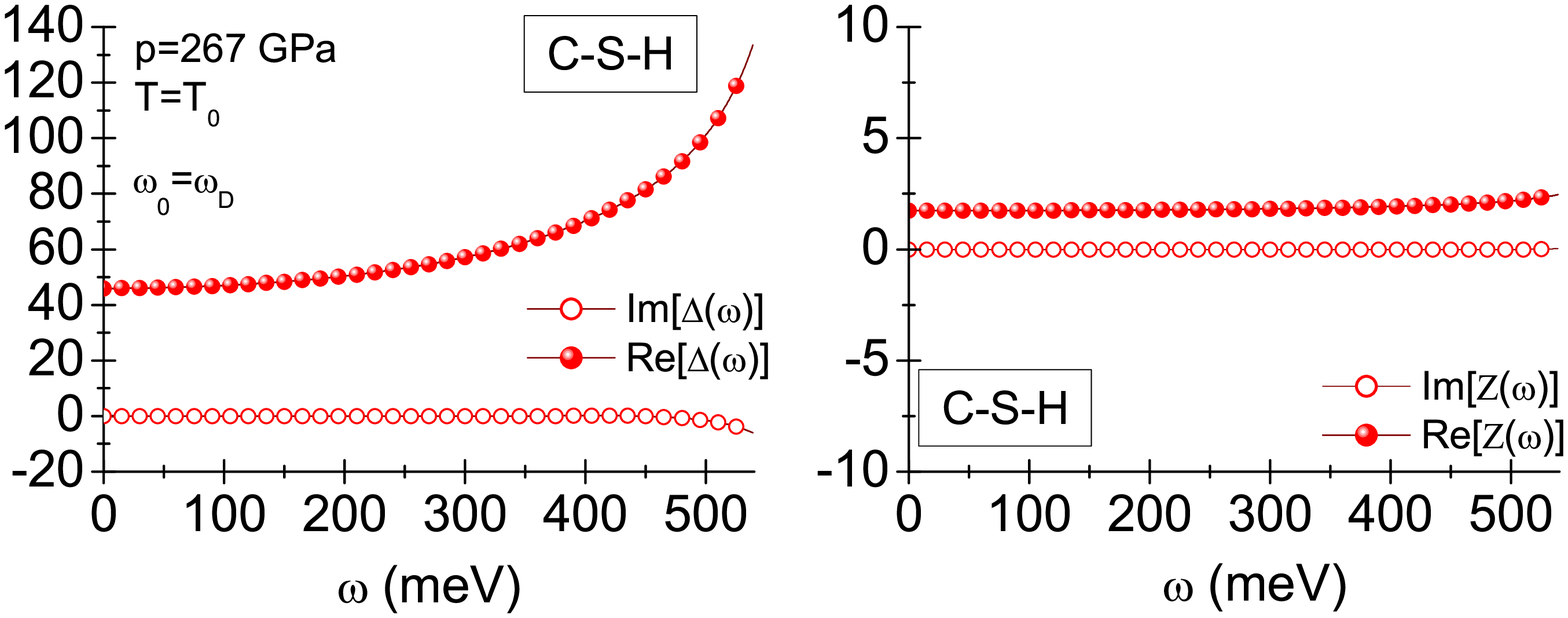}
\includegraphics[width=0.48\columnwidth]{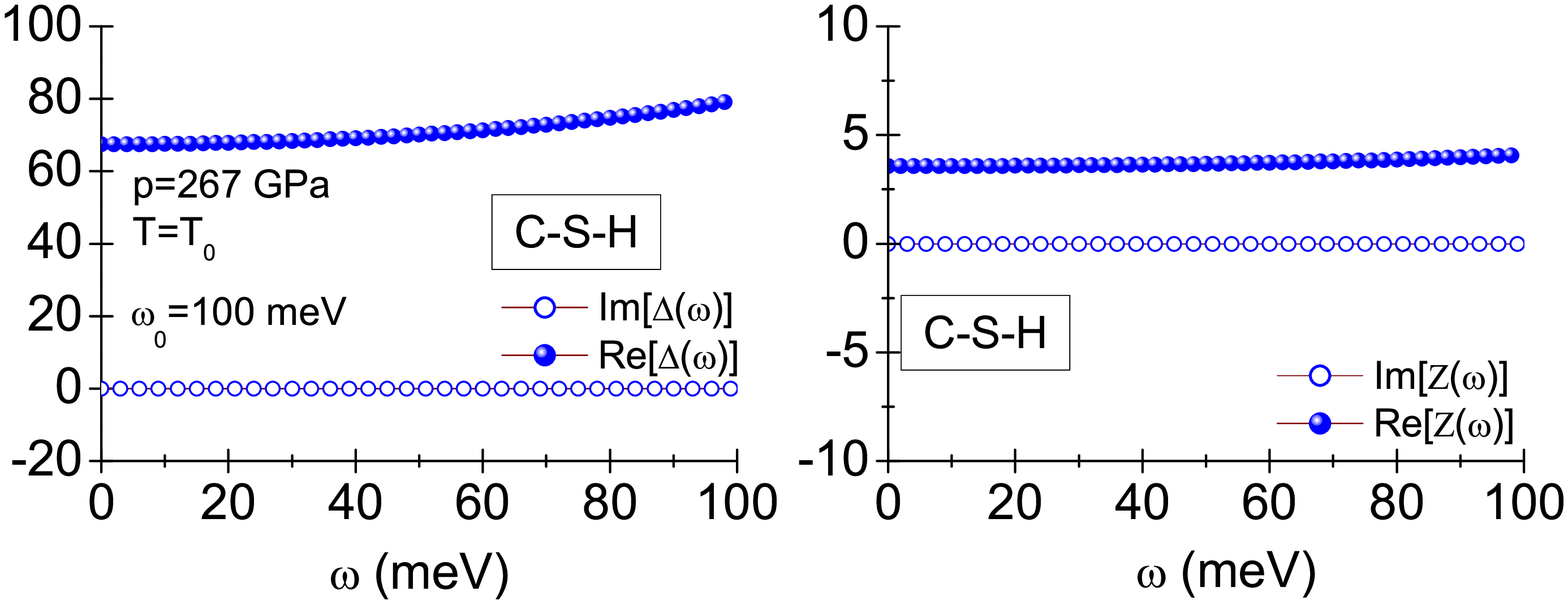}
\caption
{        The order parameter and the wave function renormalization factor on real axis for C-S-H ($p=267$~GPa and $T=T_{0}$).
         Results were obtained for $\omega_{0}=\omega_{D}$ and $\omega_{0}=100$~meV.
}
\label{Fig01-A}
\end{figure}
\begin{figure}
\includegraphics[width=0.48\columnwidth]{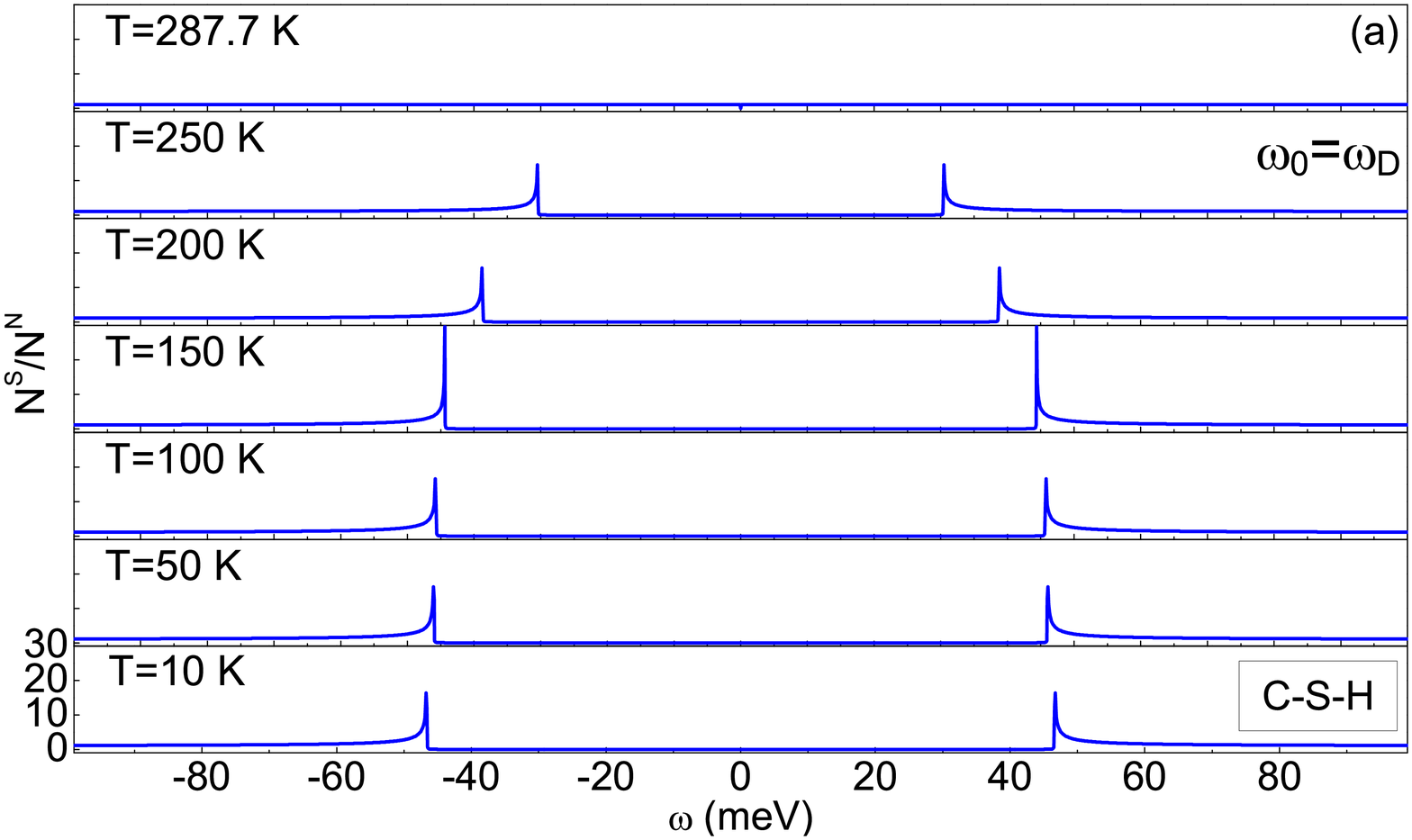}
\includegraphics[width=0.48\columnwidth]{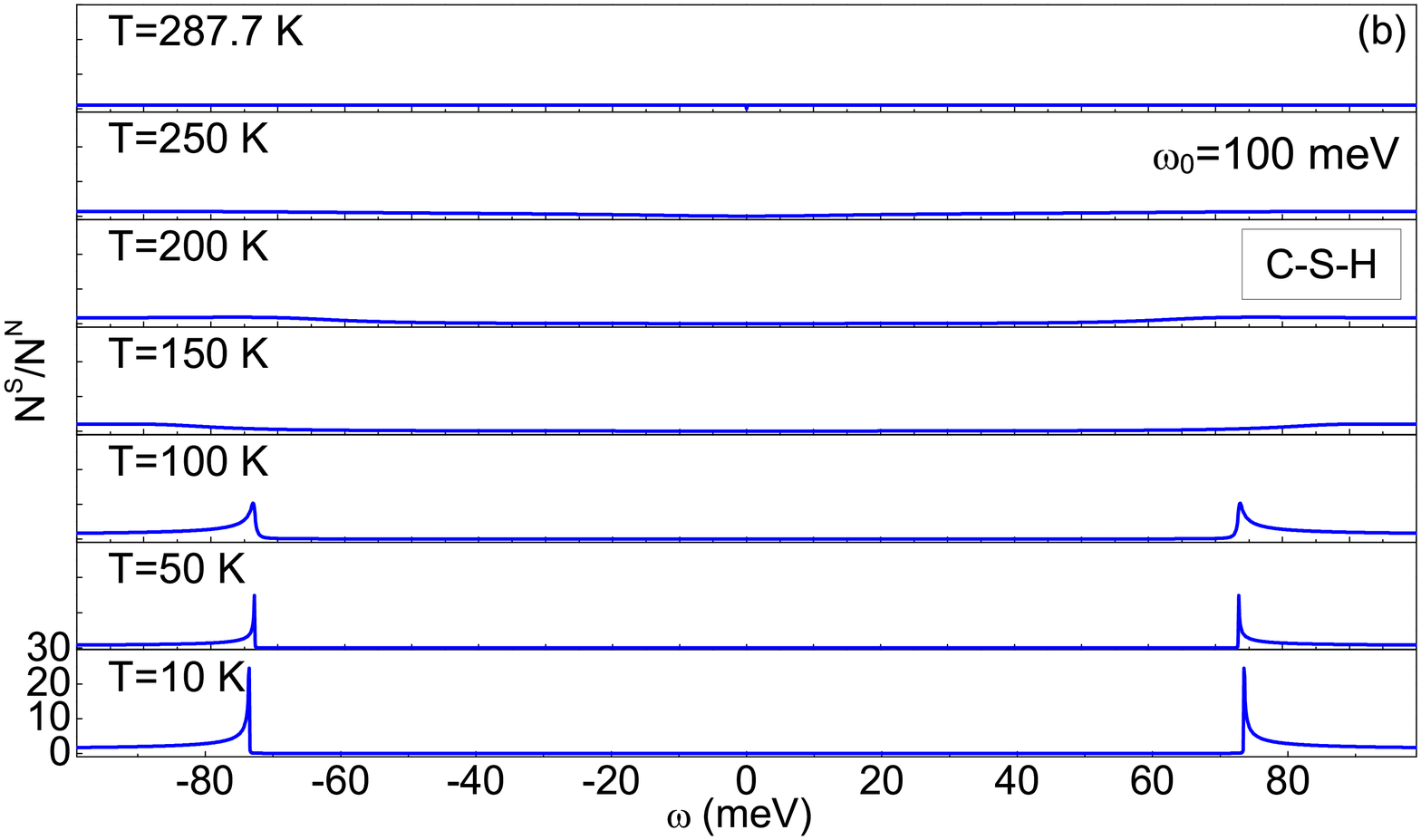}
\caption
{
  The quasiparticle density of states for C-S-H ($p=267$~GPa). The results were obtained for 
  $\omega_{0}=\omega_{D}$ and $\omega_{0}=100$~meV.
}
\label{Fig02-A}
\end{figure}

\item The free energy difference between the superconducting and the normal state  \textcolor{blue}{\cite{Carbotte1990A}}: 
\begin{eqnarray}
\frac{\Delta F}{\rho\left(\varepsilon_{F}\right)}=-\frac{2\pi}{\beta}\sum_{n=1}^{M}
\left(\sqrt{\omega^{2}_{n}+\Delta^{2}_{n}}- \left|\omega_{n}\right|\right)
\left(Z^{S}_{n}-Z^{N}_{n}\frac{\left|\omega_{n}\right|}
{\sqrt{\omega^{2}_{n}+\Delta^{2}_{n}}}\right), 
\label{r05-A} 
\end{eqnarray}  
where $\rho\left(\varepsilon_{F}\right)$ denotes the electron density of states at the Fermi level. The symbols $Z^{S}_{n}$ and $Z^{N}_{n}$ represent 
the wave function renormalization factor for supercondacting ($S$) and normal ($N$) state, respectively. Note that from the physical 
point of view, the negative values of $\Delta F$ prove thermodynamic stability of the superconducting condensate.
The first derivative of function given by \eq{r05-A} determines the entropy difference between superconducting and normal state 
($\Delta S$). The negative values of $\Delta S$ prove that the entropy of superconducting state is lower than the entropy of normal state due to the existence of Cooper pairs (see \fig{Fig03-A}).
\begin{figure}
\includegraphics[width=0.5\textwidth]{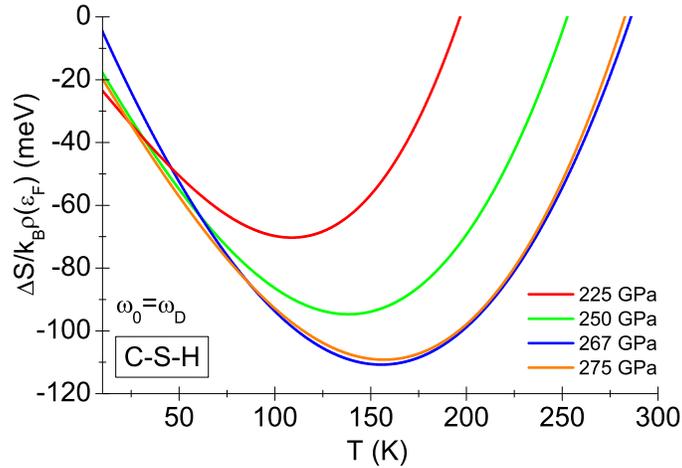}
\caption{The entropy difference between superconducting and normal state as a function of temperature for C-S-H.
         We took into account the selected pressure values and the case $\omega_{0}=\omega_{D}$.  
        }
\label{Fig03-A}        
\end{figure}

\item The thermodynamic critical field:
\begin{equation}
\frac{H_{C}}{\sqrt{\rho\left(\varepsilon_{F}\right)}}=\sqrt{-8\pi\left[\Delta F/\rho\left(\varepsilon_{F}\right)\right]}.
\label{r06-A}
\end{equation}

\item The specific heat difference between superconducting and normal state ($\Delta C=C^{S}-C^{N}$):
\begin{equation}
\frac{\Delta C\left(T\right)}{k_{B}\rho\left(\varepsilon_{F}\right)}=-\frac{1}{\beta}\frac{d^{2}\left[\Delta F/\rho\left(\varepsilon_{F}\right)\right]}
{d\left(k_{B}T\right)^{2}}.
\label{r07-A}
\end{equation}
The specific heat of normal state is the most convenient to estimate by the formula:
\begin{equation}
\frac{C^{N}\left(T\right)}{k_{B}\rho\left(\varepsilon_{F}\right)}=\frac{\gamma}{\beta},
\label{r08-A}
\end{equation}
where Sommerfeld constant is given by: $\gamma=\frac{2}{3}\pi^{2}\left(1+\lambda\right)$.

On the basis of results obtained for the thermodynamic critical field and the specific heat, it is also possible to calculate the values of dimensionless ratios: 
\begin{equation}
R_{H}=\frac{T_{C}C^{N}\left(T_{C}\right)}{H_{C}^{2}\left(0\right)},
\qquad {\rm and} \qquad
R_{C}=\frac{\Delta C\left(T_{C}\right)}{C^{N}\left(T_{C}\right)}.
\label{r09-A}
\end{equation}
\end{itemize}

\section{\label{Dod0B} Eliashberg formalism including the vertex correction}

We discuss formalism determining the effect of vertex corrections on thermodynamic properties of the superconducting state. 
Note that this type of analysis was carried out, among others, for the fullerene systems 
\textcolor{blue}{\cite{Pietronero1992A, Pickett1993A}}, the high-$T_{C}$ cuprates \textcolor{blue}{\cite{Uemura1991A, Uemura1992A, Ambrumenil1991A}}, 
the heavy fermion compounds \textcolor{blue}{\cite{Wojciechowski1996A}} and in the superconductors under high magnetic fields 
\textcolor{blue}{\cite{Goto1996A}}. In particular, the Eliashberg equations taking into account vertex corrections take the form \textcolor{blue}{\cite{Freericks1997A}}:
\begin{eqnarray}
\label{r01-B}
\varphi_{n}&=&\pi k_{B}T\sum_{m=-M}^{M}
\frac{\lambda_{n,m}-\mu_{m}^{\star}}
{\sqrt{\omega_m^2Z^{2}_{m}+\varphi^{2}_{m}}}\varphi_{m}\\ \nonumber
&-&
\frac{\pi^{3}\left(k_{B}T\right)^{2}}{4\varepsilon_{F}}
\sum_{m=-M}^{M}\sum_{m'=-M}^{M}
\frac{\lambda_{n,m}\lambda_{n,m'}}
{\sqrt{\left(\omega_m^2Z^{2}_{m}+\varphi^{2}_{m}\right)
       \left(\omega_{m'}^2Z^{2}_{m'}+\varphi^{2}_{m'}\right)
       \left(\omega_{-n+m+m'}^2Z^{2}_{-n+m+m'}+\varphi^{2}_{-n+m+m'}\right)}}\\ \nonumber
&\times&
\left[
\varphi_{m}\varphi_{m'}\varphi_{-n+m+m'}+2\varphi_{m}\omega_{m'}Z_{m'}\omega_{-n+m+m'}Z_{-n+m+m'}-\omega_{m}Z_{m}\omega_{m'}Z_{m'}
\varphi_{-n+m+m'}
\right],
\end{eqnarray}
and
\begin{eqnarray}
\label{r02-B}
Z_{n}&=&1+\frac{\pi k_{B}T}{\omega_{n}}\sum_{m=-M}^{M}
\frac{\lambda_{n,m}}{\sqrt{\omega_m^2Z^{2}_{m}+\varphi^{2}_{m}}}\omega_{m}Z_{m}\\ \nonumber
&-&
\frac{\pi^{3}\left(k_{B}T\right)^{2}}{4\varepsilon_{F}\omega_{n}}\sum_{m=-M}^{M}
\sum_{m'=-M}^{M}
\frac{\lambda_{n,m}\lambda_{n,m'}}
{\sqrt{\left(\omega_m^2Z^{2}_{m}+\varphi^{2}_{m}\right)
       \left(\omega_{m'}^2Z^{2}_{m'}+\varphi^{2}_{m'}\right)
       \left(\omega_{-n+m+m'}^2Z^{2}_{-n+m+m'}+\varphi^{2}_{-n+m+m'}\right)}}\\ \nonumber
&\times&
\left[
\omega_{m}Z_{m}\omega_{m'}Z_{m'}\omega_{-n+m+m'}Z_{-n+m+m'}+2\omega_{m}Z_{m}\varphi_{m'}\varphi_{-n+m+m'}-\varphi_{m}\varphi_{m'}\omega_{-n+m+m'}Z_{-n+m+m'}
\right].
\end{eqnarray}

The meaning of symbols in \eq{r01-B} and \eq{r02-B} were explained in \app{Dod0A}. 
Note that \eq{r01-B} and \eq{r02-B} are isotropic, which can use them to calculate the value 
of order parameter $\varphi_{n}$ and wave function renormalization factor $Z_{n}$ in self-consistent way 
only with respect to the Matsubara frequency $\omega_{n}$. This means that the self-consistent procedure does not apply 
to electron or phonon wave vector - the electron and phonon energies are averaged over the Fermi surface. 
It is worth emphasizing the fact that discussed equations are the same in form as equations 
that would be derived in local approximation \textcolor{blue}{\cite{Metzner1989A, Freericks1994A, Freericks1994B, Freericks1994C, Nicol1994A, Freericks1996A}}, where eigen energies are averaged over the entire Brillouin zone, not just the Fermi surface. 
Thus, the used approximation seems to be reasonable since the phonon-induced superconducting state in hydrogen 
containing systems is highly isotropic \textcolor{blue}{\cite{Durajski2016B}, \cite{Kostrzewa2018A}}. 
It is also worth noting that the literature gives the form of Eliashberg equations that take into account the vertex corrections explicitly 
dependent on wave vector {\bf k} \textcolor{blue}{\cite{Grimaldi1995A, Pietronero1995A}}. Nevertheless, due to great mathematical difficulties, 
it was not possible to obtain their self-consistent solutions ($\Delta_{n,{\bf k}}$ and $Z_{n,{\bf k}}$). 

The system of equations \eq{r01-B} and \eq{r02-B} was originally used to investigate the properties of superconducting state induced in 
lead \textcolor{blue}{\cite{Freericks1997A}}. Later, the discussed model was successfully applied to analysis of superconducting state 
in hydrogen-rich systems such as H$_{5}$S$_{2}$ \textcolor{blue}{\cite{Kostrzewa2018A}}, PH$_{3}$ and H$_{3}$S 
\textcolor{blue}{\cite{Durajski2016B}}. 
In these compounds, the critical temperature is equal to: $36$~K, $80$~K, and $200$~K for H$_{5}$S$_{2}$, PH$_{3}$, 
and H$_{3}$S, respectively. Let us note that using the equations \eq{r01-B} and \eq{r02-B} for above-mentioned materials 
one can get the results which are consistent with experimental data. Moreover, the equations \eq{r01-B} and \eq{r02-B} were used to study of superconducting state in low-dimensional systems such as LiC$_{6}$ \textcolor{blue}{\cite{Profeta2012A, Ludbrook2015A, Zheng2016A, Szczesniak2019A}} and Li-hBN \textcolor{blue}{\cite{Shimada2017A, Szewczyk2020A}}. In this cases, the dimensionless ratio $\lambda \omega_{D}/\varepsilon_{F}$ reaches high values of $0.09$ for LiC$_{6}$ and $0.46$ for Li-hBN. This means that in LiC$_{6}$ and Li-hBN, the analysis of superconducting state should not be carried out within the CEE formalism.

From the numerical point of view, the Eliashberg equations taking into account vertex corrections were solved for the large number of Matsubara frequencies ($M=2000$). This ensured stability of solutions in the temperature range from $T_{0}$ to $T_{C}$. 
For C-S-H and LaH$_{10}$ systems these are $10$~K and $15$~K, respectively.

\begin{figure}
\includegraphics[width=0.48\columnwidth]{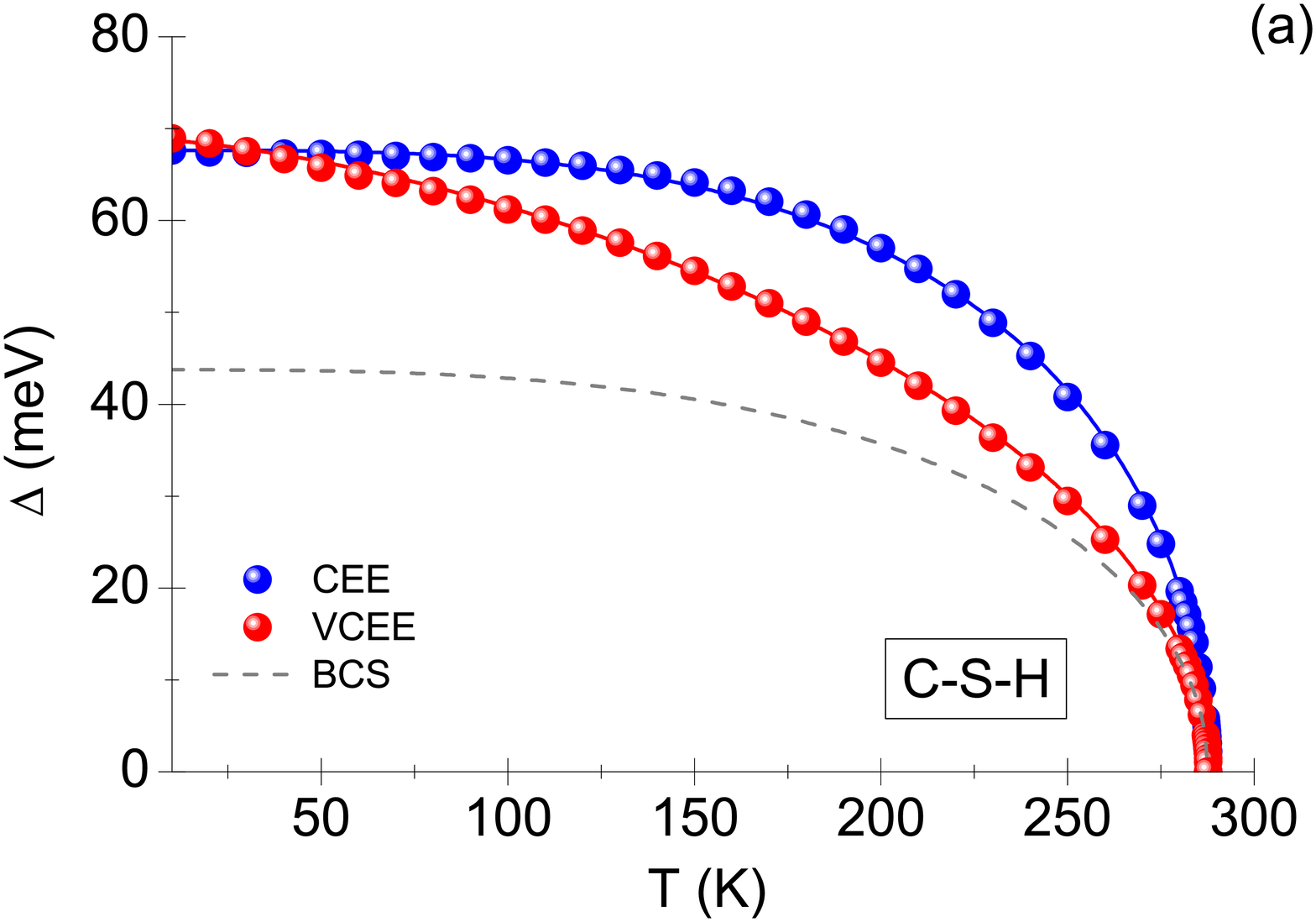}
\includegraphics[width=0.48\columnwidth]{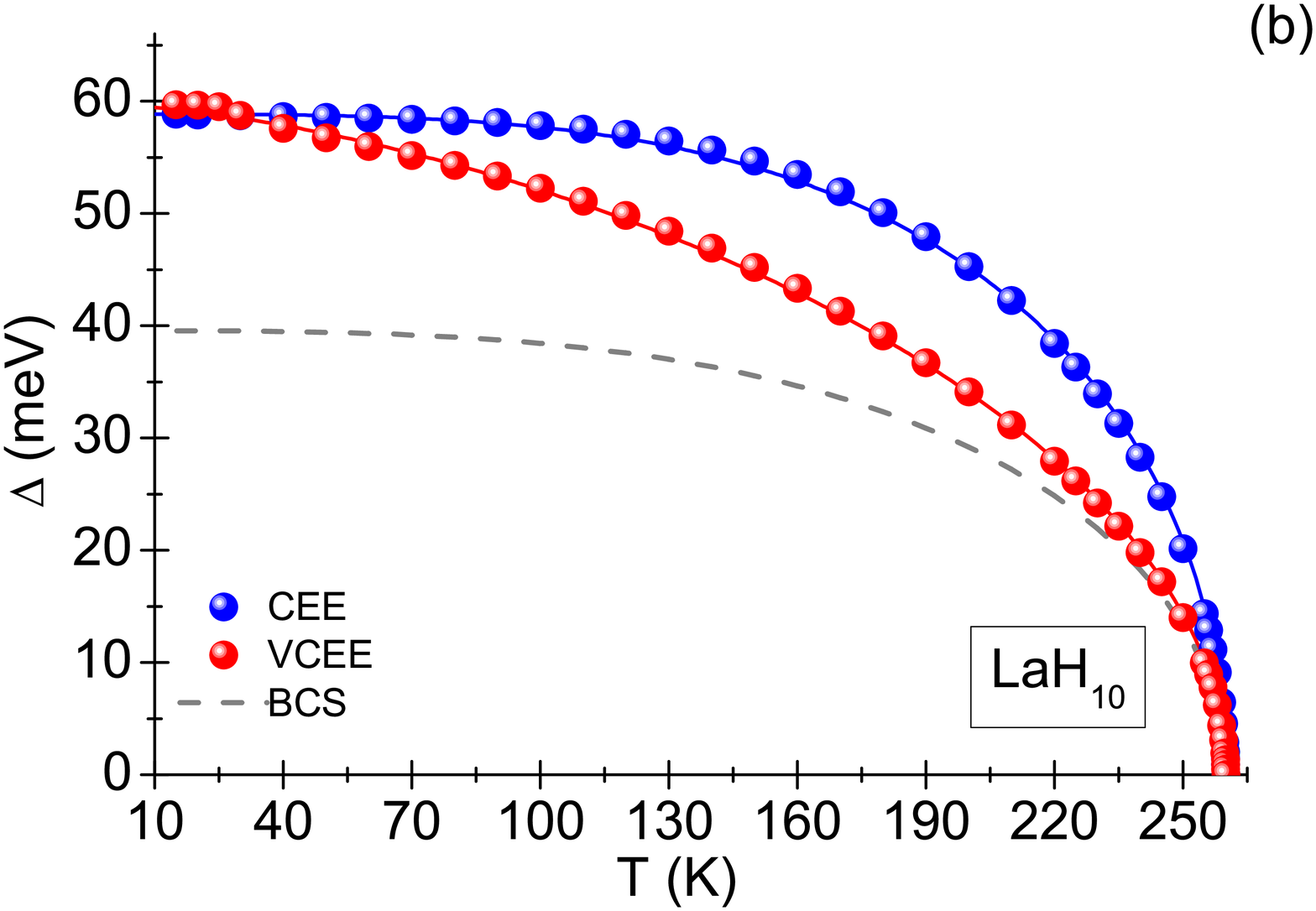}
\caption{
    The order parameter as a function of temperature obtained using the classical formalism of Eliashberg 
    equations CEE and Eliashberg 
    equations with the vertex corrections VCEE. The $\mu^{\star}=0.1$ was assumed. 
    Figure (a) the results obtained for the C-S-H system.
    Figure (b) the results for ${\rm LaH_{10}}$ superconductor. 
    The colored spheres represent numerical results. 
    Solid lines were obtained with \eq{r01-IIb}. 
    The dashed lines are predictions of BCS theory. 
}
\label{Fig02-IIIa}
\end{figure}

\begin{figure}
\includegraphics[width=0.24\columnwidth]{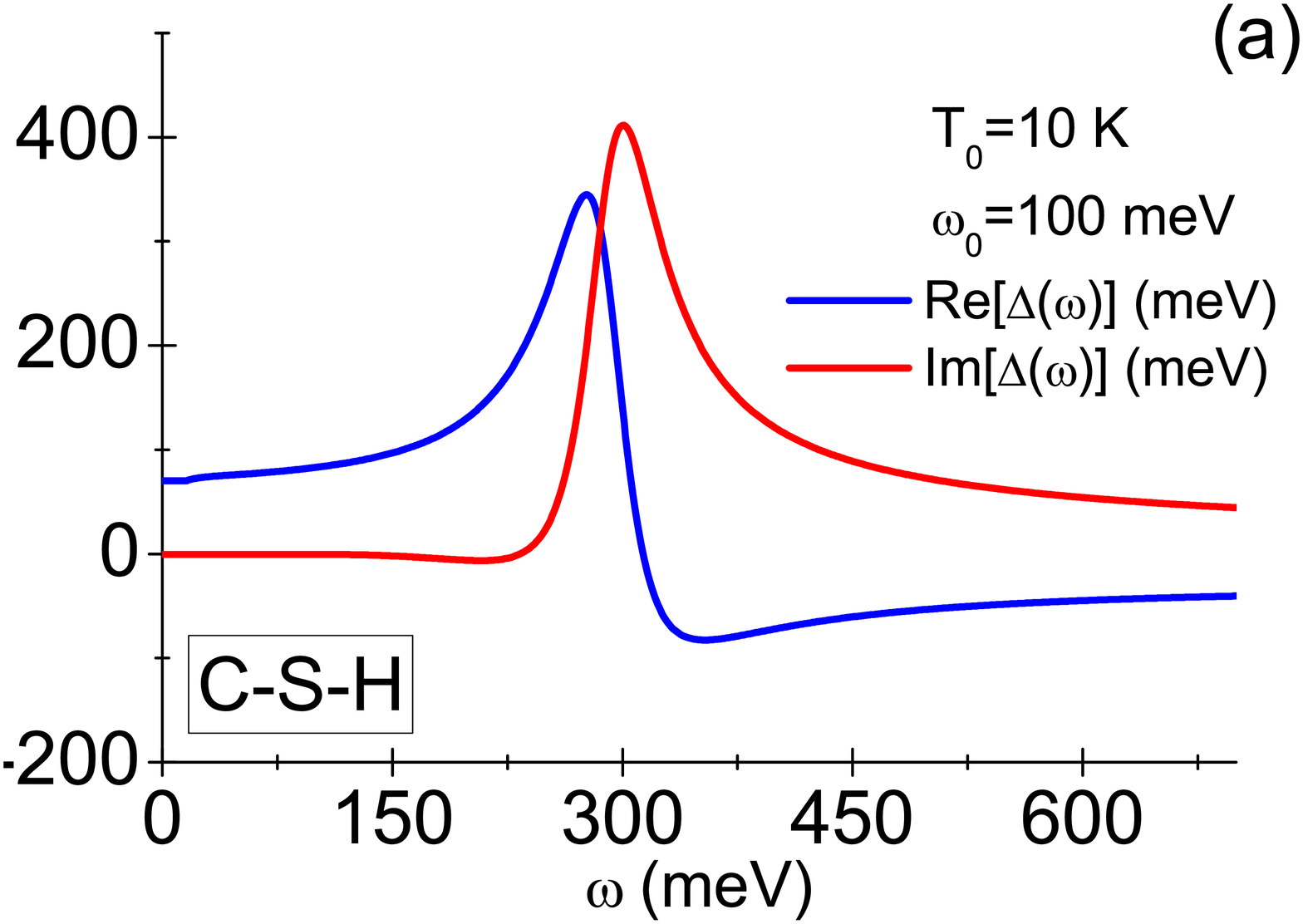}
\includegraphics[width=0.24\columnwidth]{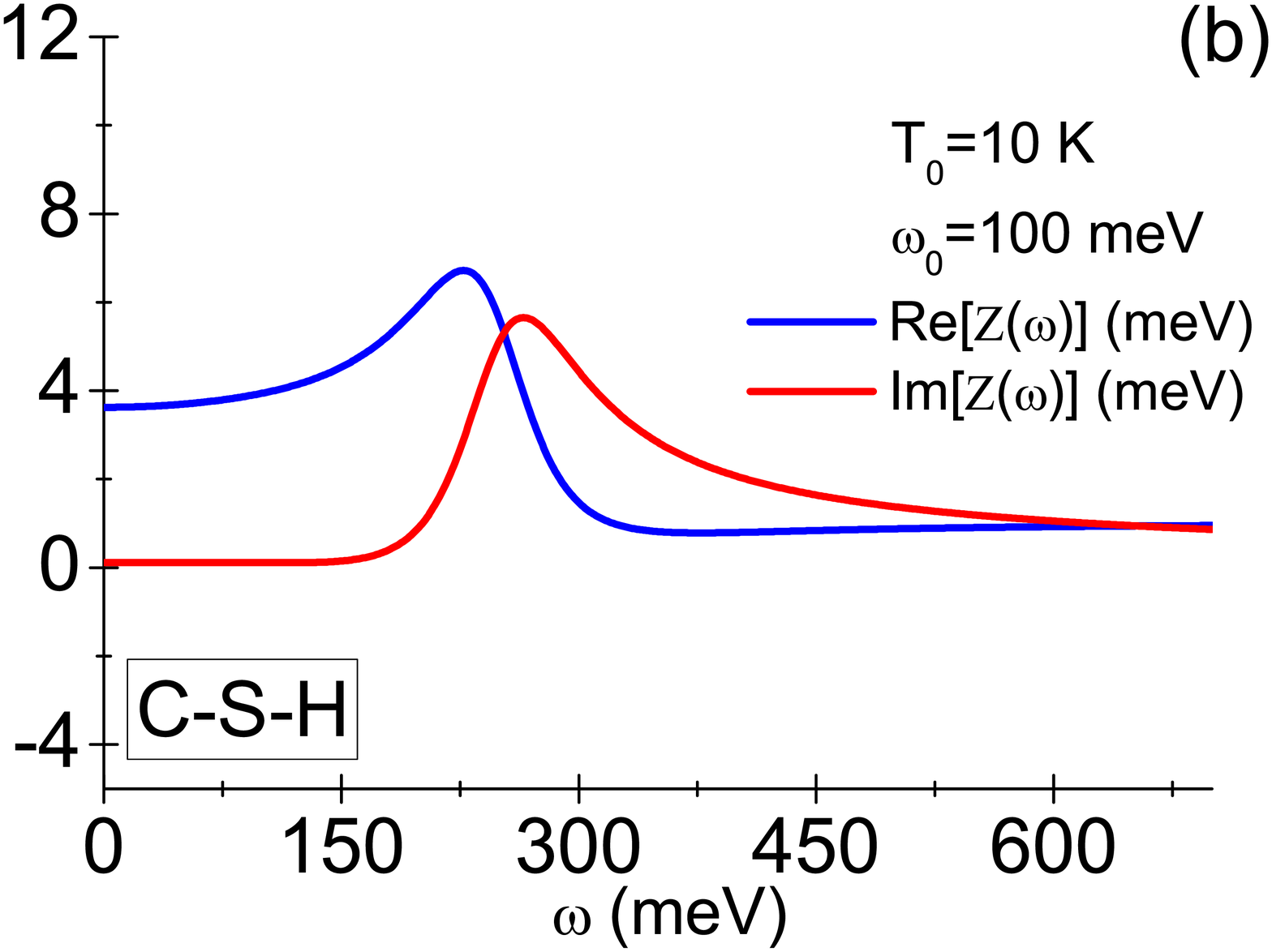}
\includegraphics[width=0.24\columnwidth]{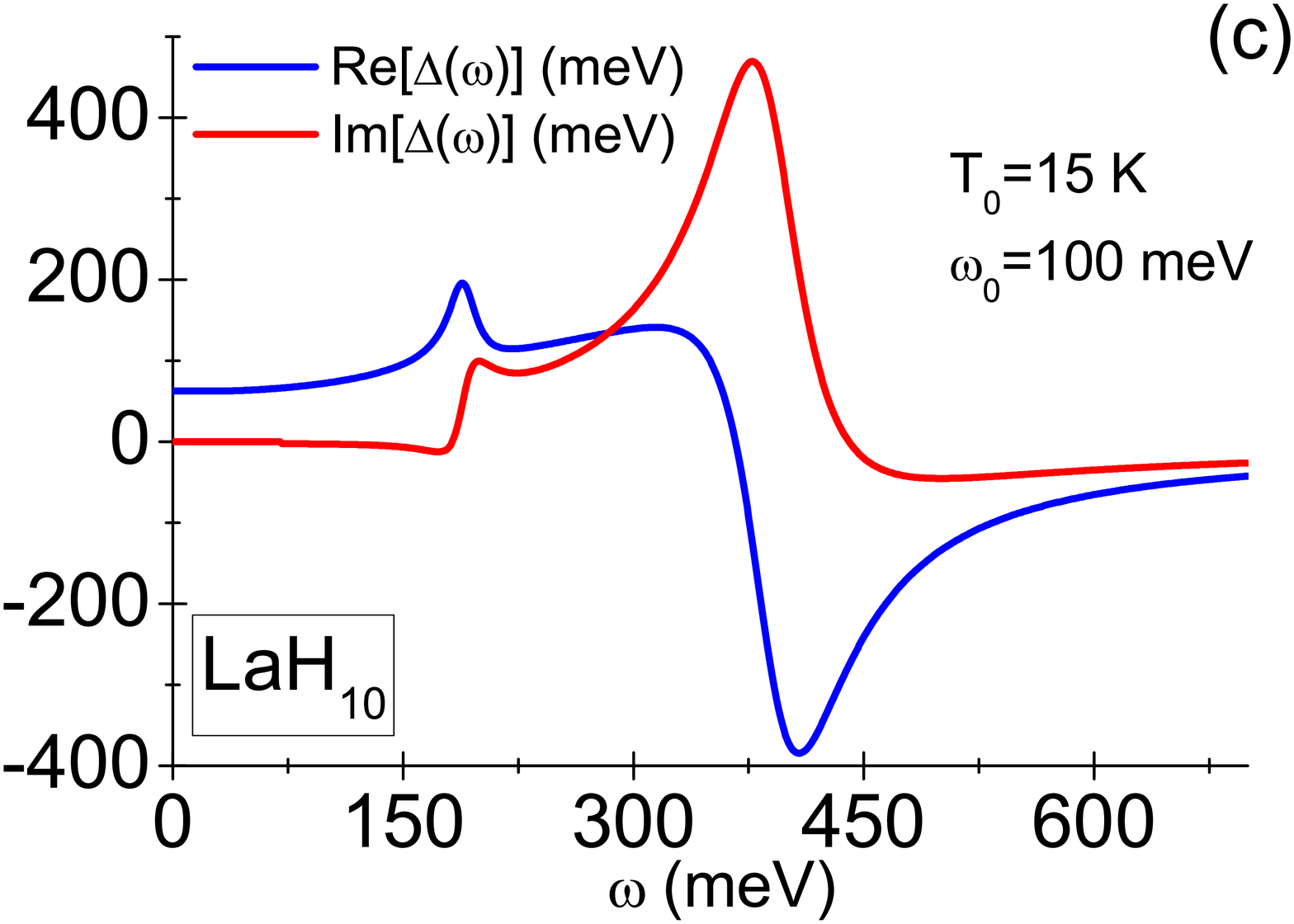}
\includegraphics[width=0.24\columnwidth]{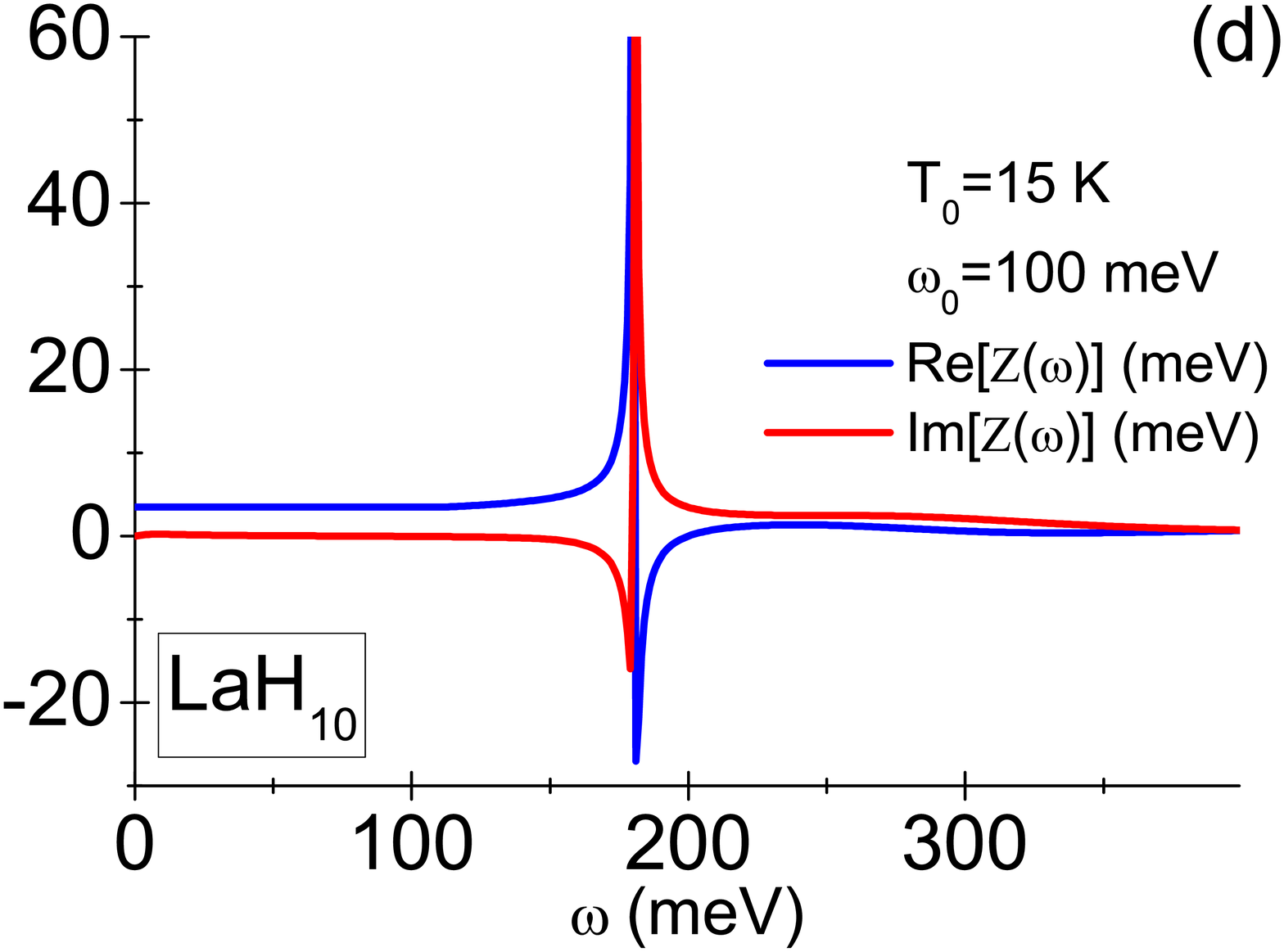}
\caption{The real and imaginary part of order parameter (a, c), and wave function renormalization factor (b, d) on real axis for $T=T_{0}$. 
         Results obtained for C-S-H and LaH$_{10}$ superconductors.}
\label{Fig01-B}        
\end{figure}

The physical values of order parameter presented in \fig{Fig02-IIIa} and \fig{Fig01-IV} were obtained by using the analytical continuation method 
($\Delta_{n}\rightarrow \Delta(\omega)$) \textcolor{blue}{\cite{Beach2000A}}. The sample functions $\Delta(\omega)$ and $Z(\omega)$ 
for C-S-H and LaH$_{10}$ superconductors taking into account vertex corrections  are presented in \fig{Fig01-B}. The characteristic maxima and minima of discussed functions correspond to frequency regions with the particularly high electron-phonon coupling.

%

\section{\label{II} Intermediate values of C-S-H electron-phonon coupling constant}

\subsection{Electron and phonon properties}

The target structure that we selected for our analysis from the paper \textcolor{blue}{\cite{Snider2020A}} was optimized by Kohn-Sham density function theory (DFT) \textcolor{blue}{\cite{Parr1989}}, within the projector augmented wave (PAW) method and with the generalized gradient approximation 
of Perdew-Burke-Ernzerhof (GGA-PBE) to the exchange-correlation functional \textcolor{blue}{\cite{Perdew1996}} as implemented in the Quantum-Espresso \textit{ab initio} simulation package \textcolor{blue}{\cite{Giannozzi2009, Giannozzi2017}}.

\begin{figure}
\includegraphics[width=0.55\textwidth]{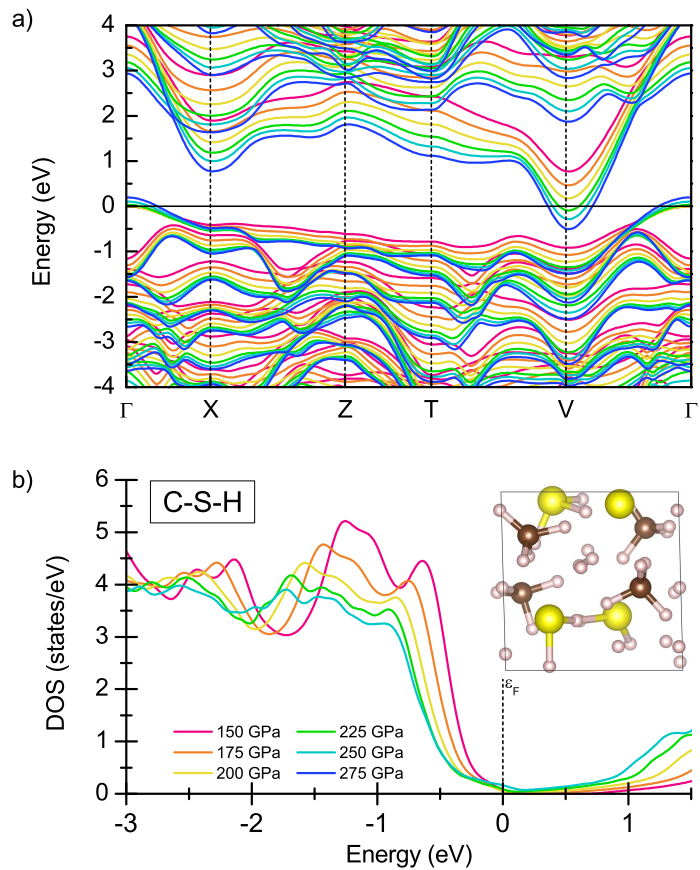}
\caption{ (a) Electron structure; and (b) electron density of states for C-S-H system.
          The inset shows the crystal structure of carbonaceous sulfur hydride at $267$~GPa.} 
\label{Fig01-IIa}        
\end{figure}

The kinetic energy cut-off for the wave function is set to $80$~Ry and the kinetic energy cut-off for charge density is $800$~Ry. 
The Brillouin zone is sampled utilizing a $6\times6\times6$ \textbf{k}-mesh in the Monkhorst-Pack scheme to reach the convergence of better 
than $1$~meV per atom. During the geometric optimization, both lattice constants and atomic positions are fully relaxed by using 
the Broyden-Fletcher-Goldfarb-Shanno (BFGS) quasi-Newton algorithm \textcolor{blue}{\cite{Billeter2003}} until the residual forces acting on the atoms remain smaller than $0.001$~eV/{\AA} and the total energy change is smaller than $10^{-5}$~eV. The underlying structure relaxation indicates that the stoichiometry (H$_2$S)(CH$_4$)H$_2$ is a mixture of CH$_4$, H$_2$S and hydrogen molecules in the host framework (see inset in \fig{Fig01-IIa}~(b)).
Moreover, the pressure-dependences optimization shows that the unit-cell volume of carbonaceous sulfur hydride decreases from $121.87$~\AA$^3$ 
at $150$ GPa to $98.61$~\AA$^3$ at $275$ GPa.

As shown in \fig{Fig01-IIa}, the calculated band structure of carbonaceous sulfur hydride exhibits metallic character above $225$~GPa. 
As a result, we observe the non-zero value of total density of states (DOS) at the Fermi level. It should be noted, that the obtained results agree with the pressure dependence of $T_{C}$ reported by Snider \textit{et al.} in paper \textcolor{blue}{\cite{Snider2020A}} where, the sharp increase in 
$T_{C}$ is observed above $220$~GPa. This can suggest existence of pressure-induced phase transition like in the case of H$_3$S 
\textcolor{blue}{\cite{Drozdov2015A, Mozaffari2019A}}.

\begin{figure}
\includegraphics[width=0.8\textwidth]{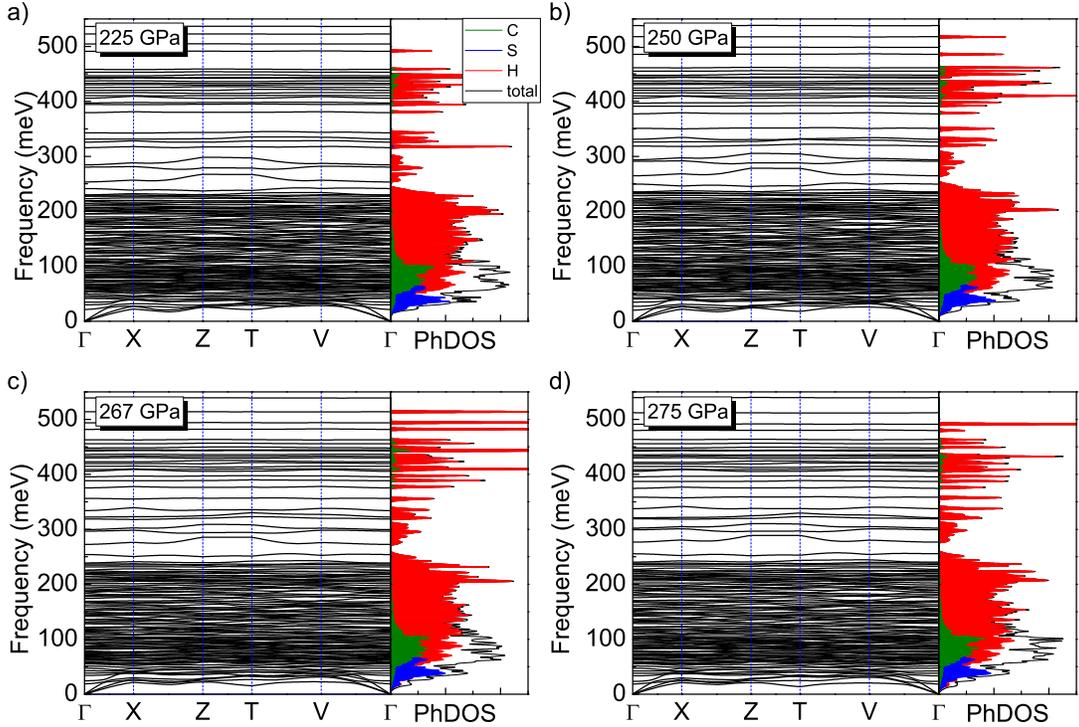}
\caption{Phonon dispersion curves and phonon density of states (PhDOS) projected onto C, S and H atoms for carbonaceous sulfur hydride 
         at $225$, $250$, $267$, and $275$~GPa.
        } 
\label{Fig02-IIa}        
\end{figure}

The phonon properties were computed by using the PHONOPY program \textcolor{blue}{\cite{Togo2015A}}. The phonon dispersion along 
$\Gamma$-K-Z-T-V-$\Gamma$ high-symmetry line in the first Brillouin zone is plotted in \fig{Fig02-IIa}. No negative phonon branches were found, indicating that the investigated system is dynamically stable in the pressure range from $225$ to $275$~GPa. To know more about the atomic vibration information, the atom-projected phonon DOS curves are also plotted. We find that since the S and C atoms are heavier than the H atom, the vibration modes at the low-energy range are mainly contributed by S and C atoms. The vibration of H atoms contributes to the whole range of energy and is the only contribution to the modes for energy in the range of $\sim$200-370 meV. Interestingly, at higher energy, the phonon modes from C appear again. This result highlights that the H atoms play a dominant role in superconductivity, but C vibrations at high frequencies, are also important because increase the maximal phonon frequency. Which in turn, according to the BCS theory, is one of the reasons responsible for the high critical temperature.
%

\subsection{Thermodynamic properties of superconducting state at $p\in\{225, 250, 267, 275\}$~GPa}

\begin{table*}
\caption{\label{Tab01-IIb} Basic thermodynamic parameters characterizing the superconducting state in C-S-H system. 
                           Theoretical and experimental results for the pressure of $267$~GPa.}
\begin{ruledtabular}
\begin{tabular}{|c|c|c|c|c|}
                         &                                     &                                  &                                   &            \\
Thermodynamic parameter  & CEE model ($\omega_{0}=\omega_{D}$) & CEE model ($\omega_{0}=100$~meV) & VCEE model ($\omega_{0}=100$~meV) &  Exp.      \\
                         &                                     &                                  &                                   &            \\
\hline
                         &                                     &                                  &                                   &            \\
$\omega_{D}$~(K)         & 6258.7                              & -                                & -                                 & -          \\ 
                         &                                     &                                  &                                   &            \\
$\mu^{\star}$            & 0.1                                 & 0.1 (0.2)                        & 0.1                               & -          \\ 
                         &                                     &                                  &                                   &            \\
$\lambda$                & 0.75                                & 3.26  (3.95)                     & 3.31                              & -          \\
                         &                                     &                                  &                                   &            \\
$T_{C}$~(K)              & 287.7                               & 287.7 (287.7)                    & 287.7                             & 287.7      \\
                         &                                     &                                  &                                   &            \\
$\Delta\left(0\right)$~(meV) &   46.05                         & 67.64 (69.28)                    & 68.94                             & -          \\
                         &                                     &                                  &                                   &            \\$R_{\Delta}$             &       3.71                          & 5.46 (5.59)                      & 5.56                              & -          \\
                         &                                     &                                  &                                   &            \\
$Z\left(T_{0}\right)$    &       1.74                          & 3.56 (4.09)                      & 3.48                              & -          \\
                         &                                     &                                  &                                   &            \\
$Z\left(T_{C}\right)$    &       1.75                          & 4.26 (4.95)                      & 2.69                              & -          \\
                         &                                     &                                  &                                   &            \\
$H_{C}\left(0\right)/\sqrt{\rho\left(\varepsilon_{F}\right)}$~(meV) &     210.90   & 393.97       & -                                 & -          \\
                         &                                     &                                  &                                   &            \\
$R_{H}$                  &       0.159                         & 0.177                            & -                                 & -          \\
                         &                                     &                                  &                                   &            \\
$\Delta C\left(T_{C}\right)/k_{B}\rho\left(\varepsilon_{F}\right)$~(meV) &    524.21     & 2650.32& -                                 & -          \\
                         &                                     &                                  &                                   &            \\
$R_{C}$                  &       1.84                          & 2.37                             & -                                 & -          \\
                         &                                     &                                  &                                   &           
\end{tabular}
\end{ruledtabular}
\end{table*}

\begin{table*}
\caption{\label{Tab02-IIb} Basic thermodynamic parameters characterizing the superconducting state in C-S-H system. 
                           Theoretical and experimental results for the pressure of $225$~GPa, $250$~GPa, and $275$~GPa. 
                           It has been assumed $\omega_{0}=\omega_{D}$.}
\begin{ruledtabular}
\begin{tabular}{|c|c|c|c|c|c|c|}
                         &                  &                   &                 &                   &                 &                          \\
Parameter                & CEE ($225$~GPa)  & Exp.              & CEE ($250$~GPa) & Exp.              & CEE ($275$~GPa) & Exp.                     \\
                         &                  &                   &                 &                   &                 &                          \\
\hline
                         &                  &                   &                 &                   &                 &                          \\
$\omega_{D}$~(K)         & 6229.5           & -                 & 6245.5          & -                 & 6264.2          &  -                       \\ 
                         &                  &                   &                 &                   &                 &                          \\
$\mu^{\star}$            & 0.1              & -                 & 0.1             & -                 & 0.1             &  -                       \\ 
                         &                  &                   &                 &                   &                 &                          \\
$\lambda$                & 0.65             & -                 & 0.71            & -                 & 0.75            &  -                       \\
                         &                  &                   &                 &                   &                 &                          \\
$T_{C}$~(K)              & 200              & 200               & 255             & 255               & 286             & 286                      \\
                         &                  &                   &                 &                   &                 &                          \\
$\Delta\left(0\right)$~meV & 31.53          & -                 & 40.60           & -                 & 45.68           & -                        \\
                         &                  &                   &                 &                   &                 &                          \\$R_{\Delta}$             &  3.72            & -                 & 3.68            & -                 & 3.93            & -                        \\
                         &                  &                   &                 &                   &                 &                          \\
$Z\left(T_{0}\right)$    &  1.64            & -                 & 1.70            & -                 & 1.73            & -                        \\
                         &                  &                   &                 &                   &                 &                          \\
$Z\left(T_{C}\right)$    &  1.65            & -                 & 1.71            & -                 & 1.75            & -                        \\
                         &                  &                   &                 &                   &                 &                          \\
$H_{C}\left(0\right)/\sqrt{\rho\left(\varepsilon_{F}\right)}$~meV & 144.74 & -    & 186.61 & -        & 212.18          & -                        \\
                         &                  &                   &                 &                   &                 &                          \\
$R_{H}$                  &   0.154          & -                 & 0.157           & -                 & 0.154           & -                        \\
                         &                  &                   &                 &                   &                 &                          \\
$\Delta C\left(T_{C}\right)/k_{B}\rho\left(\varepsilon_{F}\right)$~meV &  372.48  & - & 470.15 & -    & 538.99          & -                        \\
                         &                  &                   &                 &                   &                 &                          \\
$R_{C}$                  &   1.99           & -                 & 1.90            & -                 & 1.94            & -                        \\
                         &                  &                   &                 &                   &                 & 
\end{tabular}
\end{ruledtabular}
\end{table*}

\begin{figure}
\includegraphics[width=0.48\textwidth]{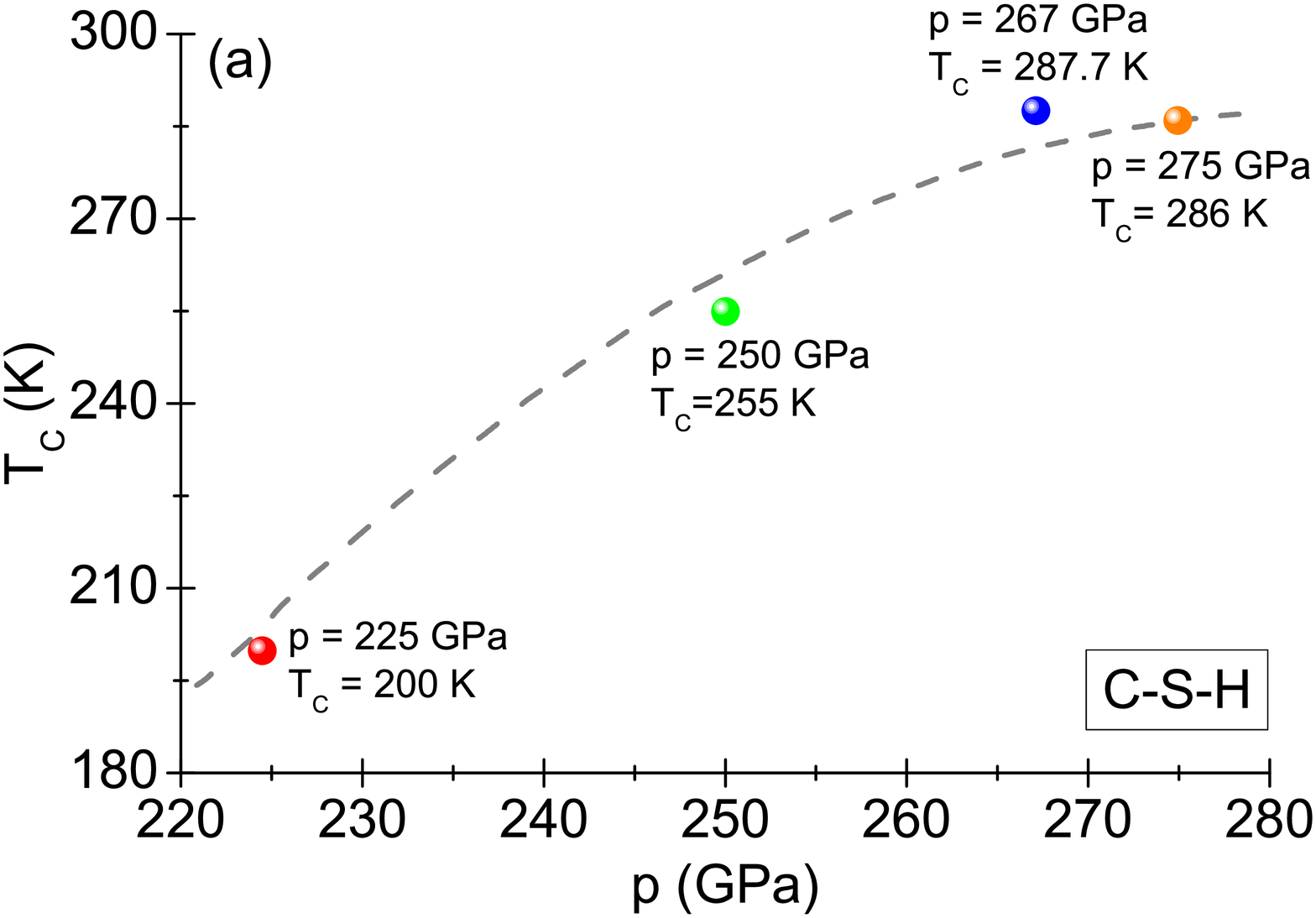}
\includegraphics[width=0.48\textwidth]{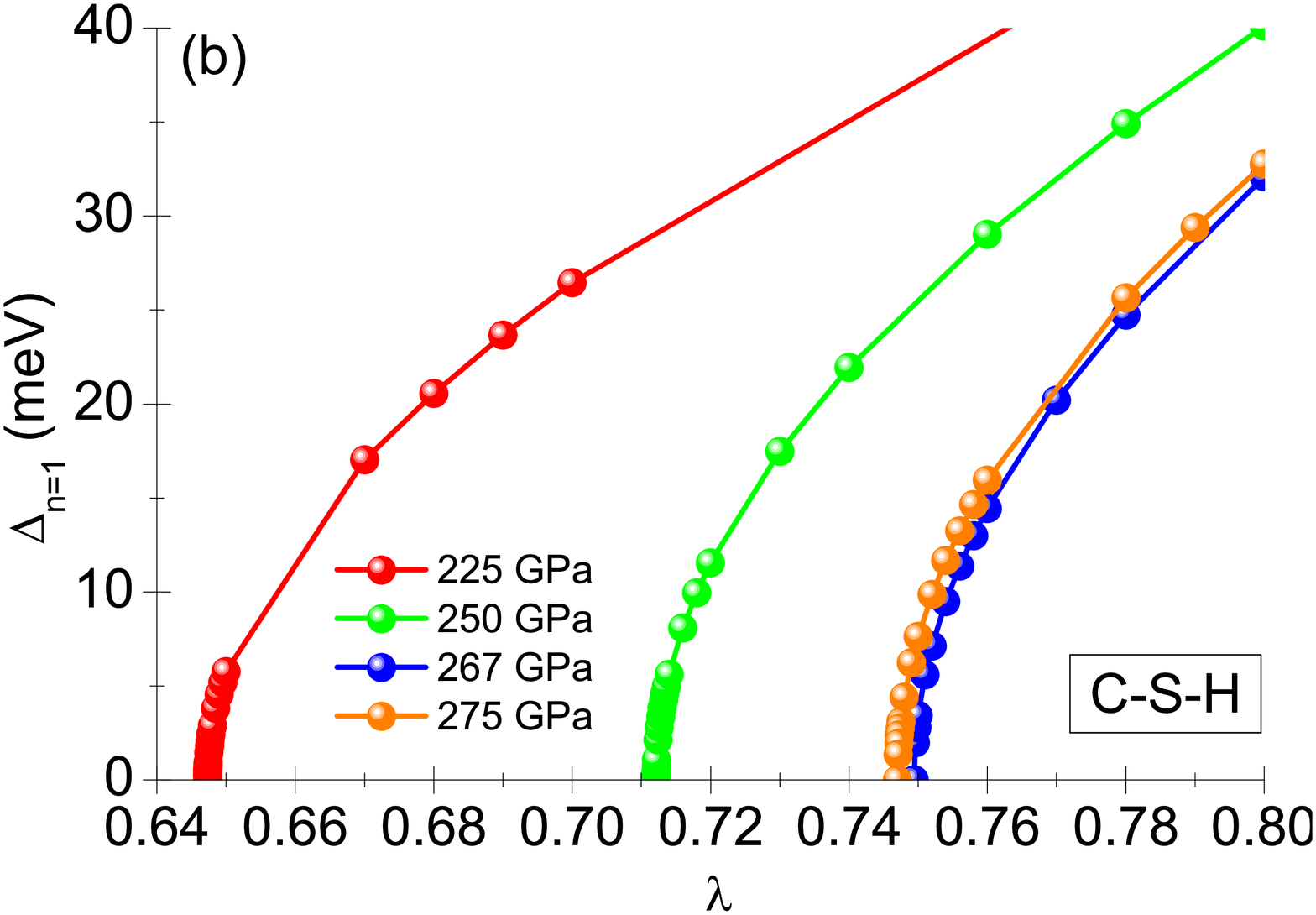}
\caption{(a) Critical temperature of superconducting state in the C-S-H system for selected values of pressure. 
             The blue, red and green points represent experimental data contained in the paper \textcolor{blue}{\cite{Snider2020A}}. 
             The orange point was obtained by using approximation curve (gray dashed line).
         (b) Maximum value of order parameter as a function of electron-phonon coupling constant. 
             The characteristic frequency $\omega_{0}$ is equal to Debye frequency. The results were obtained in the CEE approach ($\mu^{\star}=0.1$).}  
\label{Fig01-IIb}        
\end{figure}

\begin{figure}
\includegraphics[width=0.48\columnwidth]{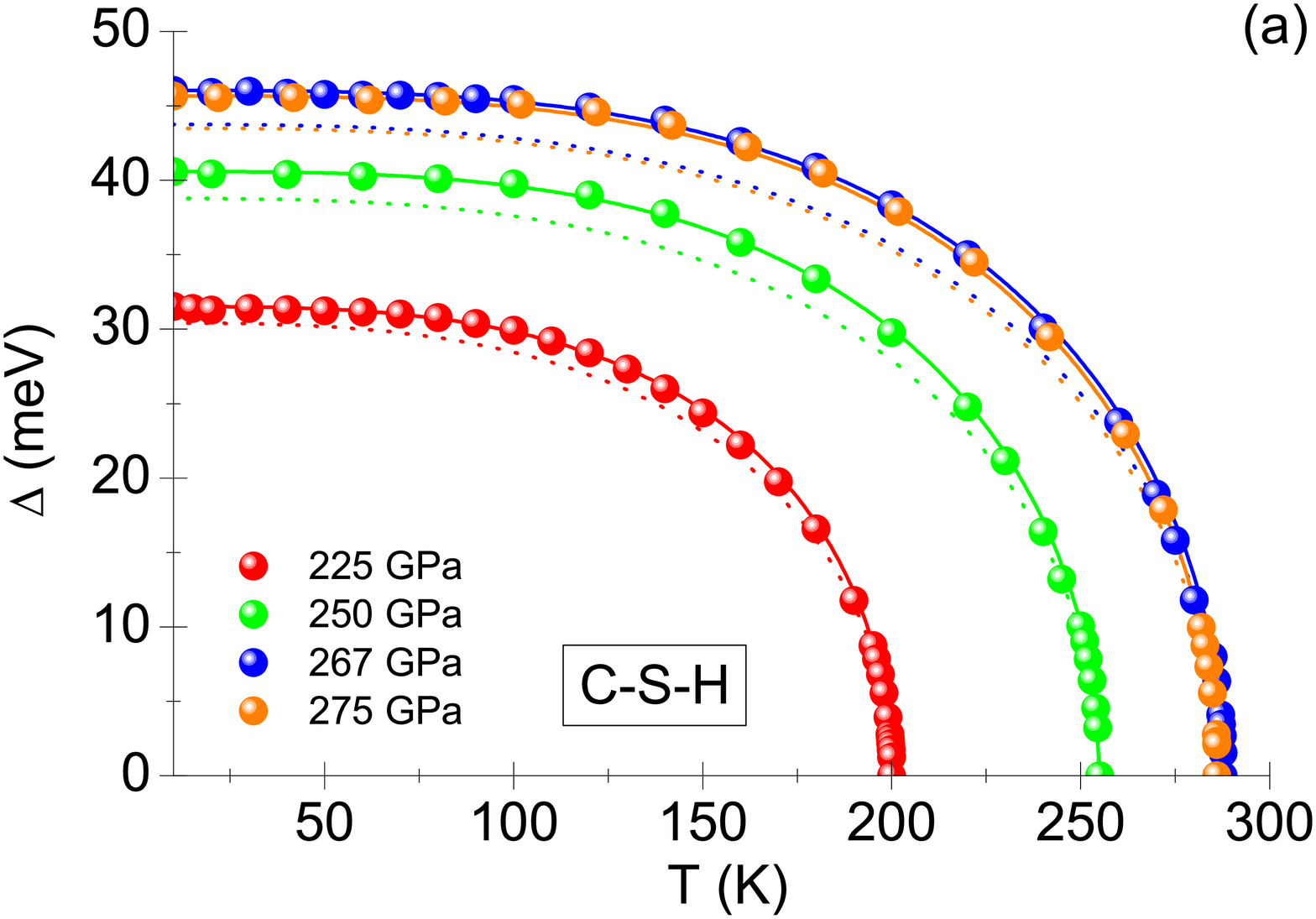}
\includegraphics[width=0.48\columnwidth]{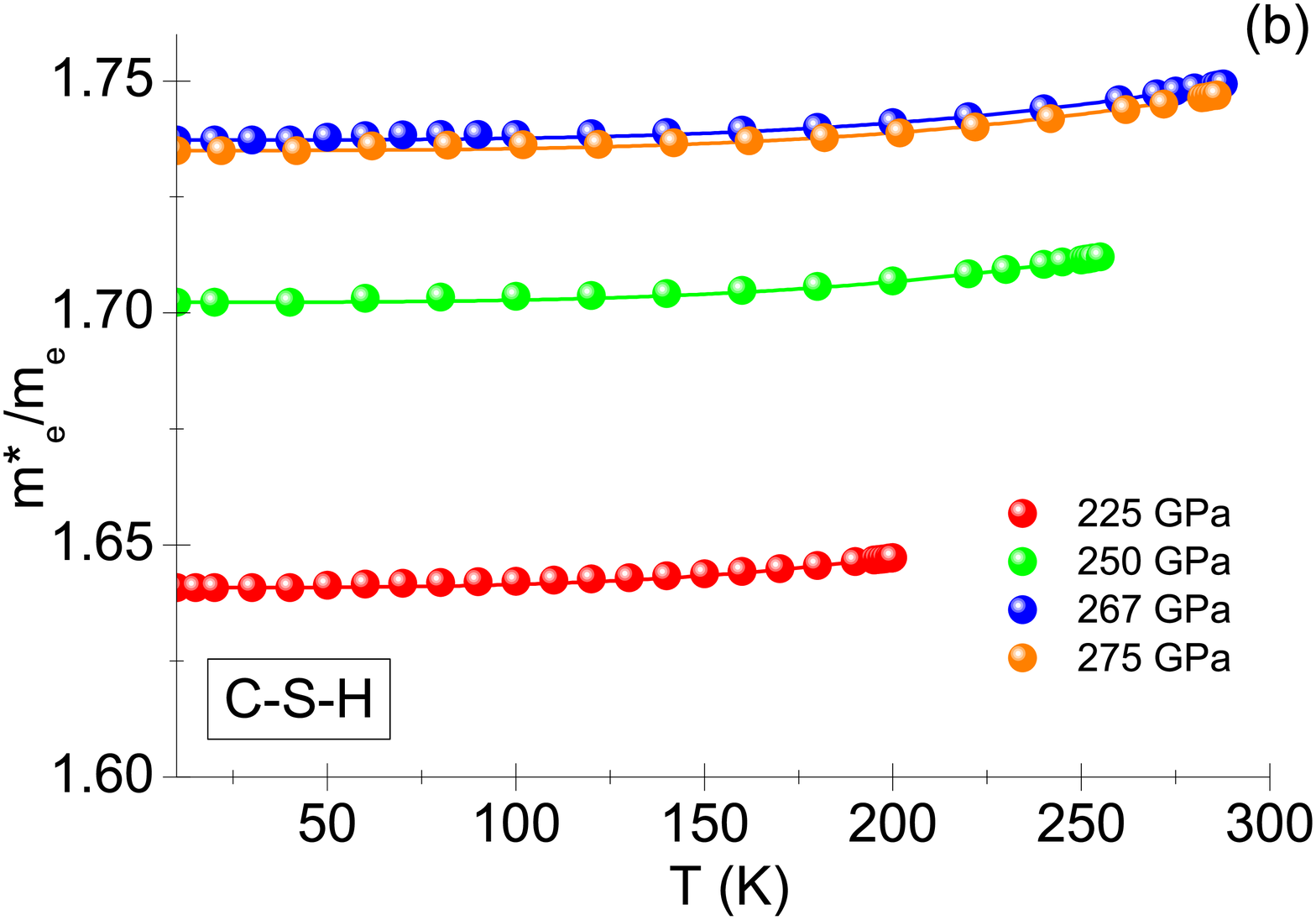}
\includegraphics[width=0.48\columnwidth]{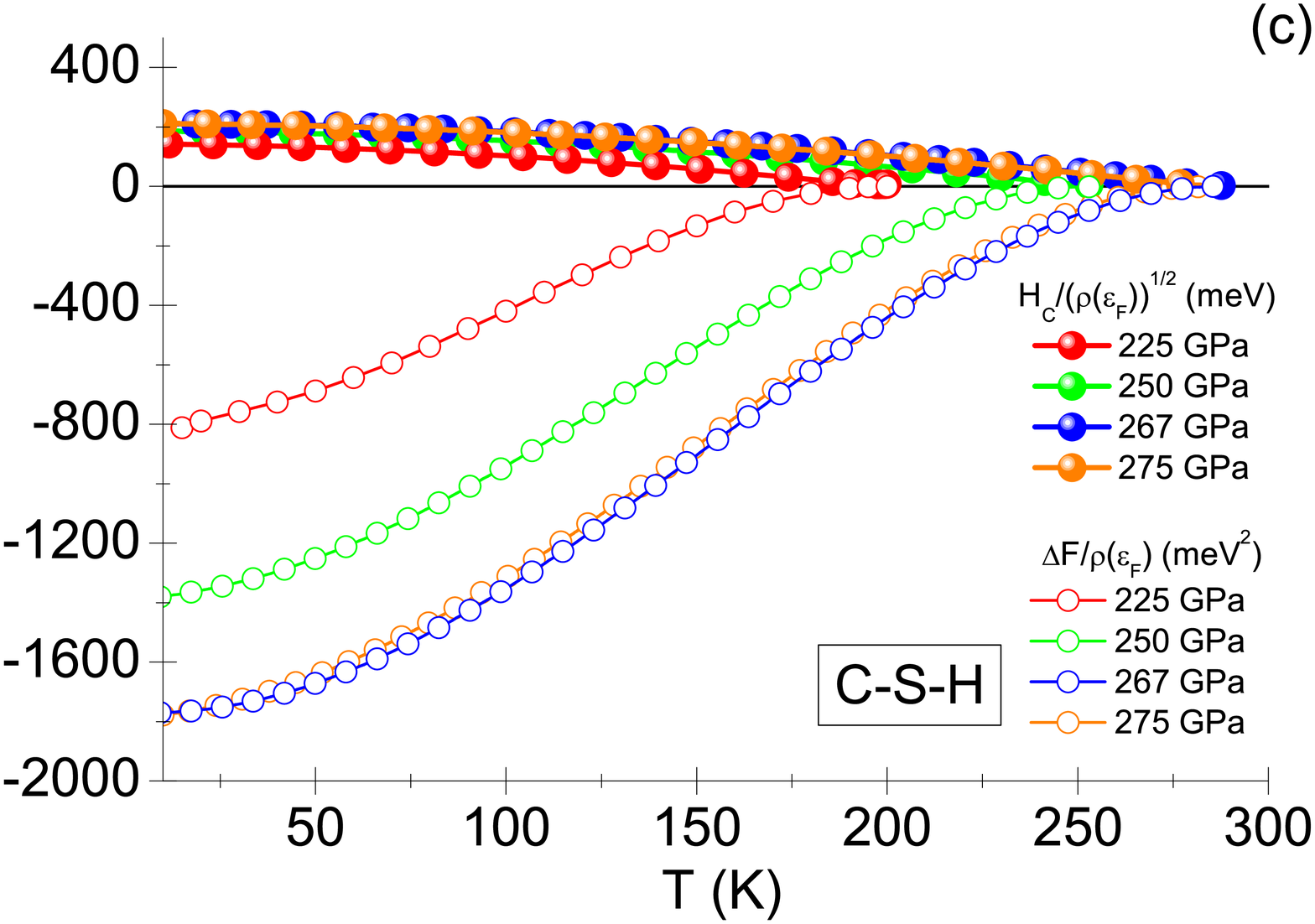}
\includegraphics[width=0.48\columnwidth]{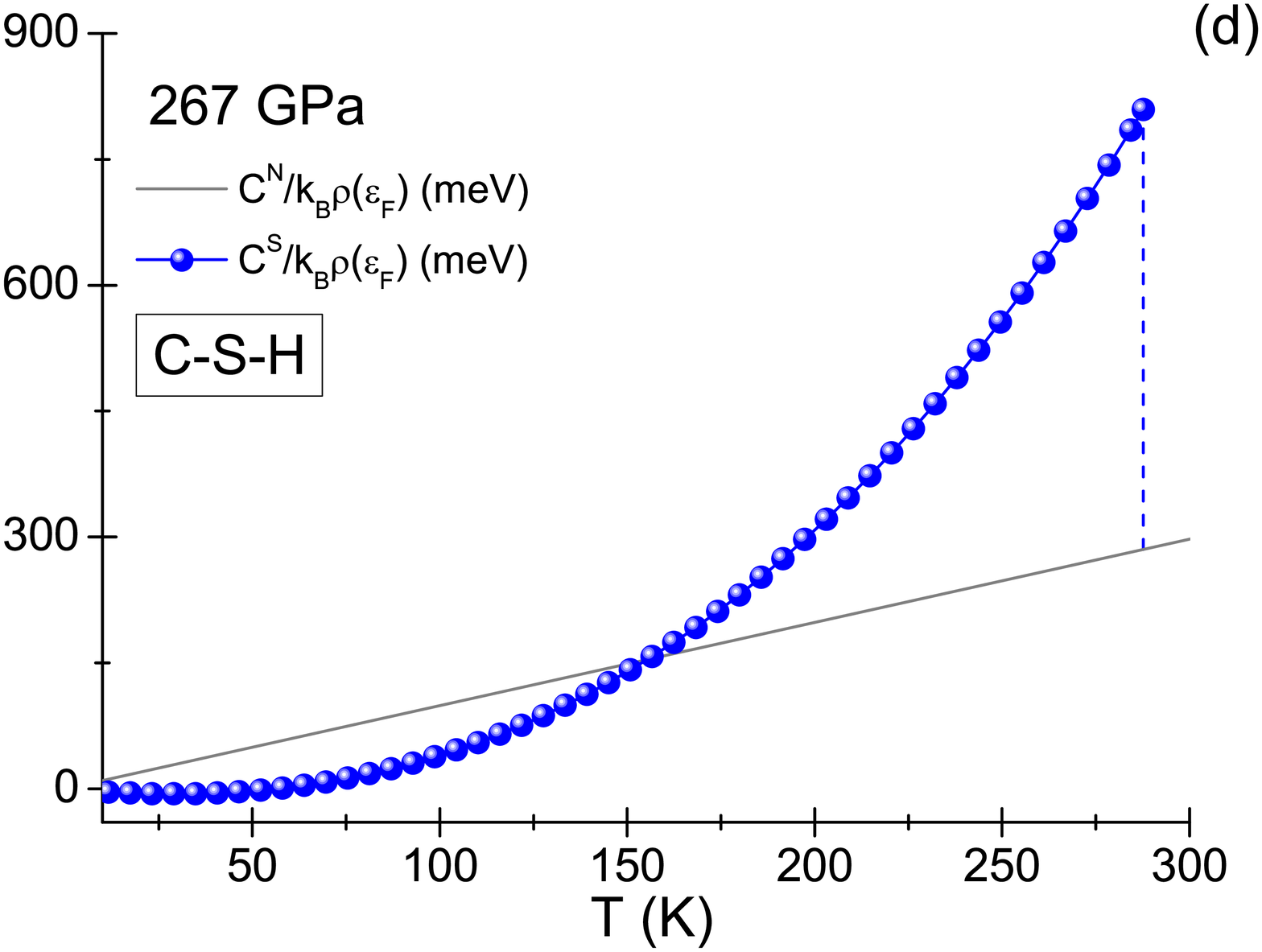}
\caption
{
(a) Order parameter $\Delta$ as a function of temperature obtained by using the classical Eliashberg formalism ($\mu^{\star}=0.1$). 
    The numerical results are represented by circles. The solid lines have been obtained using \eq{r01-IIb}. 
    The dashed lines represent the predictions of BCS theory.
(b) Ratio of the electron effective mass to the electron band mass as a function of temperature. The numerical results are represented 
    by circles. The solid lines have been obtained using \eq{r02-IIb}.
(c) Lower panel: the free energy difference between superconducting and normal state as a function of temperature. 
    Upper panel: the influence of temperature on thermodynamic critical field. 
(d) The specific heat of superconducting and normal state as a function of temperature ($p=267$~GPa).   
}
\label{Fig02-IIb}
\end{figure}

In the first step, we assumed that characteristic frequency $\omega_{0}$ in the electron-phonon pairing kernel (\eq{r03-A}) 
is equal to Debye frequency $\omega_{D}$. The values of $\omega_{D}$  were read from the data obtained by DFT method (see also \tab{Tab01-IIb} 
and \tab{Tab02-IIb}). For the crystal structure considered in the study, the Debye frequencies assume high values of $\sim 6200$~K. 
This result is caused by existence of the quasi-free hydrogen molecules present in the C-S-H structure, which was illustrated on the 
phonon density of state graph (\fig{Fig02-IIa}) obtained for the pressure values $225$~GPa, $250$~GPa, $267$~GPa, and $275$~GPa. 

The basic thermodynamic parameters of superconducting state in C-S-H system were determined within the framework of 
classic Eliashberg equations (see \app{Dod0A}). We solved the Eliashberg equations using the numerical methods that we 
developed in the paper \textcolor{blue}{\cite{Szczesniak2006A}}. For $M=2200$ equations, the physically correct solutions can be obtained 
in temperature range from $T_{0}=10$~K to $T_{C}$. Note that there is the restriction for solutions of Eliashberg equations in the low temperatures 
due to the fact that temperature zero Kelvin corresponds infinite number of the Matsubara frequencies ($M=+\infty$).

The depairing electron correlations were considered parametrically using the Coulomb pseudopotential $\mu^{\star}$ 
\textcolor{blue}{\cite{Morel1962A}}. We assumed in the numerical calculations $\mu^{\star}=0.1$.

The values of electron-phonon coupling constant $\lambda$ were calculated on the basis of experimental critical temperature values (\fig{Fig01-IIb}~(a)) 
by the equation $\left[\Delta_{n=1}\right]_{T=T_{C}}=0$ (see also \fig{Fig01-IIb}~(b)). It turns out that for the pressures considered in the study, the electron-phonon coupling constant range from $0.65$ to $0.75$ - the exact results can be found in \tab{Tab01-IIb} and \tab{Tab02-IIb}. 
From the physical point of view, this means that superconducting state in C-S-H system is induced by electron-fonon interaction characterized by the intermediate value of coupling constant. This means that the high value of critical temperature for C-S-H results primarily from the high value of Debye frequency.  

In \fig{Fig02-IIb}~(a) we have presented the temperature dependence of order parameter for pressure values of $225$~GPa, $250$~GPa, $267$~GPa, and $275$~GPa. The physical values of $\Delta$ have been calculated using the method of analytical continuation characterized in \app{Dod0A}.
It can be seen that due to relatively low values of electron-phonon coupling constant, the functions of order parameter do not differ 
substantially from the curves of BCS theory \cite{Bardeen1957A, Bardeen1957B}. In particular, the CEE numerical results can be parameterized 
by the formula:
\begin{equation}
\label{r01-IIb}
\Delta\left(T\right)=\Delta\left(0\right)\sqrt{1-\left(\frac{T}{T_{C}}\right)^{\eta}}, 
\end{equation}
where $\Delta\left(0\right)=\Delta\left(T_{0}\right)$ and $\eta=3.25$. As part of the BCS theory, we get $\eta=3$ and
$\Delta\left(0\right)=1.765k_{B}T_{C}$ \cite{Eschrig2001A}. In the case of numerical results, the ratio of order parameter 
to critical temperature ranges from $3.7$ to $3.9$ (\tab{Tab01-IIb} and \tab{Tab02-IIb}). For the BCS theory, we get the value $3.53$ 
\cite{Bardeen1957A, Bardeen1957B}.

The Eliashberg formalism allows to calculate the ratio 
of electron effective mass ($m^{\star}_{e}$) to electron band mass $m_{e}$. The numerical results obtained with the Eliashberg equations are collected in \fig{Fig02-IIb}~(b). They can be parameterized with the formula:
\begin{equation}
\label{r02-IIb}
m_{e}^{\star}/m_{e}=\left[Z\left(T_{C}\right)-Z\left(T_{0}\right)\right](T/T_{C})^{\eta}+Z\left(T_{0}\right),   
\end{equation}
where the values of $Z\left(0\right)=Z\left(T_{0}\right)$ and $Z\left(T_{C}\right)=1+\lambda$ can be found in the  \tab{Tab01-IIb} or \tab{Tab02-IIb}. 
Note that in the case of BCS theory, we get  $m_{e}^{\star}=m_{e}$.

In the next step, we have calculated numerically temperature dependence of free energy difference between the superconducting and the normal state $\Delta F\left(T\right)$, the thermodynamic critical field $H_{C}\left(T\right)$, and the specific heat in superconducting $C^{S}\left(T\right)$ and normal 
$C^{N}\left(T\right)$ state. The results have been presented in \fig{Fig02-IIb}~(c) and (d). The characteristic values of  discussed functions have been summarized in \tab{Tab01-IIb} and \tab{Tab02-IIb}.

Based on obtained data, the following dimensionless ratios can be calculated: $R_{H}=T_{C}C^{N}\left(T_{C}\right)/H_{C}^{2}\left(0\right)$ and 
$R_{C}=\Delta C\left(T_{C}\right)/C^{N}\left(T_{C}\right)$. On the basis of \tab{Tab01-IIb} and \tab{Tab02-IIb}, we can conclude that their values do not differ substantially from those predicted by BCS theory ($0.168$ and $1.43$) \textcolor{blue}{\cite{Bardeen1957A, Bardeen1957B}}.

\section{\label{III} High value of C-S-H electron-phonon coupling constant}

\subsection{The thermodynamic properties of C-S-H superconducting state at $p=267$~GPa}

The calculations based on use of genetic evolutionary algorithms and DFT method \textcolor{blue}{\cite{Hu2020A}} 
suggest also the different scenario than proposed for C-S-H system in the paper by Sinder {\it et al.} 
\textcolor{blue}{\cite{Snider2020A}}. The authors of article \textcolor{blue}{\cite{Hu2020A}} showed that the replacement 
of small amount of sulfur atoms by carbon in compounds like ${\rm C_{1}S_{15}H_{48}}$ and ${\rm C_{1}S_{17}H_{54}}$ results 
in stronger electron-phonon coupling and higher averaged phonon frequency that increases with the pressure. 
As a result, the critical temperature reaches the room temperature value at $\sim 270$~GPa. 
Additionally, the calculated superconducting transition temperature of ${\rm C_{1}S_{15}H_{48}}$ and ${\rm C_{1}S_{17}H_{54}}$ 
as a function of pressure shows the good agreement with experimental measurements for C-S-H \textcolor{blue}{\cite{Snider2020A}}.

Under the approach that we consider in this paper, for the pressure of $267$~GPa, the Hu {\it et al.} results \textcolor{blue}{\cite{Hu2020A}} 
can be reproduced by taking $\omega_{0}\sim 100$~meV. 
Using the experimental data obtained for critical temperature (\fig{Fig01-IIb}~(a)) and equation $[\Delta_{n=1}]_{T=T_{C}}=0$, 
the calculated values of electron-phonon coupling constant in CEE model are $3.26$ and $3.95$ for $\mu^{\star}=0.1$ 
and $\mu^{\star}=0.2$, respectively. In paper \textcolor{blue}{\cite{Hu2020A}} estimated value of $\lambda$ 
for ${\rm C_{1}S_{17}H_{54}}$ is approximately $2.8$. The above-mentioned results clearly suggest that the high value of critical temperature 
in C-S-H system is induced by high value of the electron-phonon coupling constant and high value of the logarithmic phonon frequency 
$\omega_{\rm ln}$ of about $1550$~K (the value of Debye frequency is about $3597$~K \textcolor{blue}{\cite{Hu2020A}}).

\begin{figure}
\includegraphics[width=0.5\columnwidth]{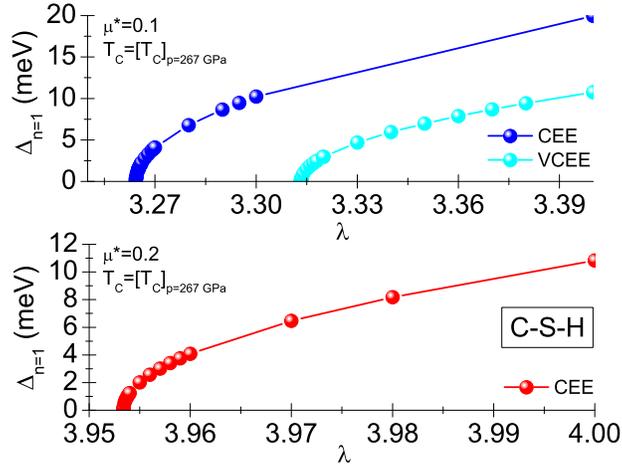}
\caption{
         Maximum value of order parameter as a function of electron-phonon coupling constant for the characteristic 
         frequency of $100$~meV. Assumed $T=[T_{C}]_{p=267\hspace*{1mm}{\rm GPa}}$ 
         and $\mu^{\star}\in\{0.1, 0.2\}$. The results were obtained under CEE and VCEE model.
         }
\label{Fig01-IIIa}
\end{figure}

Taking into account the present case, in the area of strong electron-phonon coupling, the importance of vertex corrections for the electron-phonon interaction should be additionally examined. 
Our remark is due to the fact that their influence on obtained results is related to the value of dimensionless ratio 
$\lambda\omega_{D}/\varepsilon_{F}$, which explicitly depends on the  value of electron-phonon coupling constant 
\textcolor{blue}{\cite{Pietronero1992A, Pietronero1995A, Grimaldi1995A}}. For C-S-H system its value is equal to $0.02$, 
which means that in the static limit vertex corrections do not significantly modify the results obtained under the classical Eliashberg formalism. Nevertheless, when one considers the full dependence of order parameter on Matsubara frequency (dynamic effects) then the answer to question about the significance of vertex corrections requires the self-consistent solution of appropriately modified Eliashberg equations \textcolor{blue}{\cite{Freericks1997A}} (see VCEE schema discussed in detail in \app{Dod0B}). 
Our calculations showed that in VCEE scheme, for the Coulomb pseudopotential $0.1$, the value of electron-phonon coupling constant is slightly increased 
($\lambda=3.31$) compared to the result obtained under CEE scheme ($\lambda=3.26$). 
The full order parameter dependencies on $\lambda$ for both CEE and VCEE schemas are shown in \fig{Fig01-IIIa}. 

In \fig{Fig02-IIIa}~(a) we plotted the temperature dependence of order parameter determined using the classical Eliashberg equations and equations with 
vertex corrections. We obtained the physical values of order parameter using the method of analytical continuation the solutions of Eliashberg equations from imaginary axis \textcolor{blue}{\cite{Beach2000A}} (see also \app{Dod0A} and \app{Dod0B}). It can be seen that due to significant strong-coupling effects, the obtained curves differ very clearly from BCS curve \textcolor{blue}{\cite{Bardeen1957A, Bardeen1957B}}. In particular, the numerical results can be parameterized using the formula \eq{r01-IIb}. In the CEE approach we get $\eta=3.35$, while for VCEE the exponent $\eta$ is equal to $1.5$. \fig{Fig02-IIIa}~(a) allows to evaluate the impact of vertex corrections on the values of order parameter. In particular, it can be seen that vertex corrections clearly underestimate the $\Delta\left(T\right)$ values in temperature range from about $50$~K to about $275$~K. Outside the indicated range, their importance is negligible, which means that near zero Kelvin or near $T_{C}$, the thermodynamic properties of C-S-H system, with high accuracy, 
can be calculated within classical Eliashberg formalism. Note that in the case of ${\rm LaH_{10}}$ superconductor ($p=190$~GPa and $T_{C}=260$~K \textcolor{blue}{\cite{Somayazulu2019A}}), the very similar scenario as for C-S-H is realized, as presented in \fig{Fig02-IIIa}~(b). The method for obtaining results for ${\rm LaH_{10}}$ was discussed in \app{Dod0E}. The $\eta$ exponent values are $3.4$ and $1.5$ for the CEE and VCEE formalism, respectively.

\begin{figure}
\includegraphics[width=0.3\columnwidth]{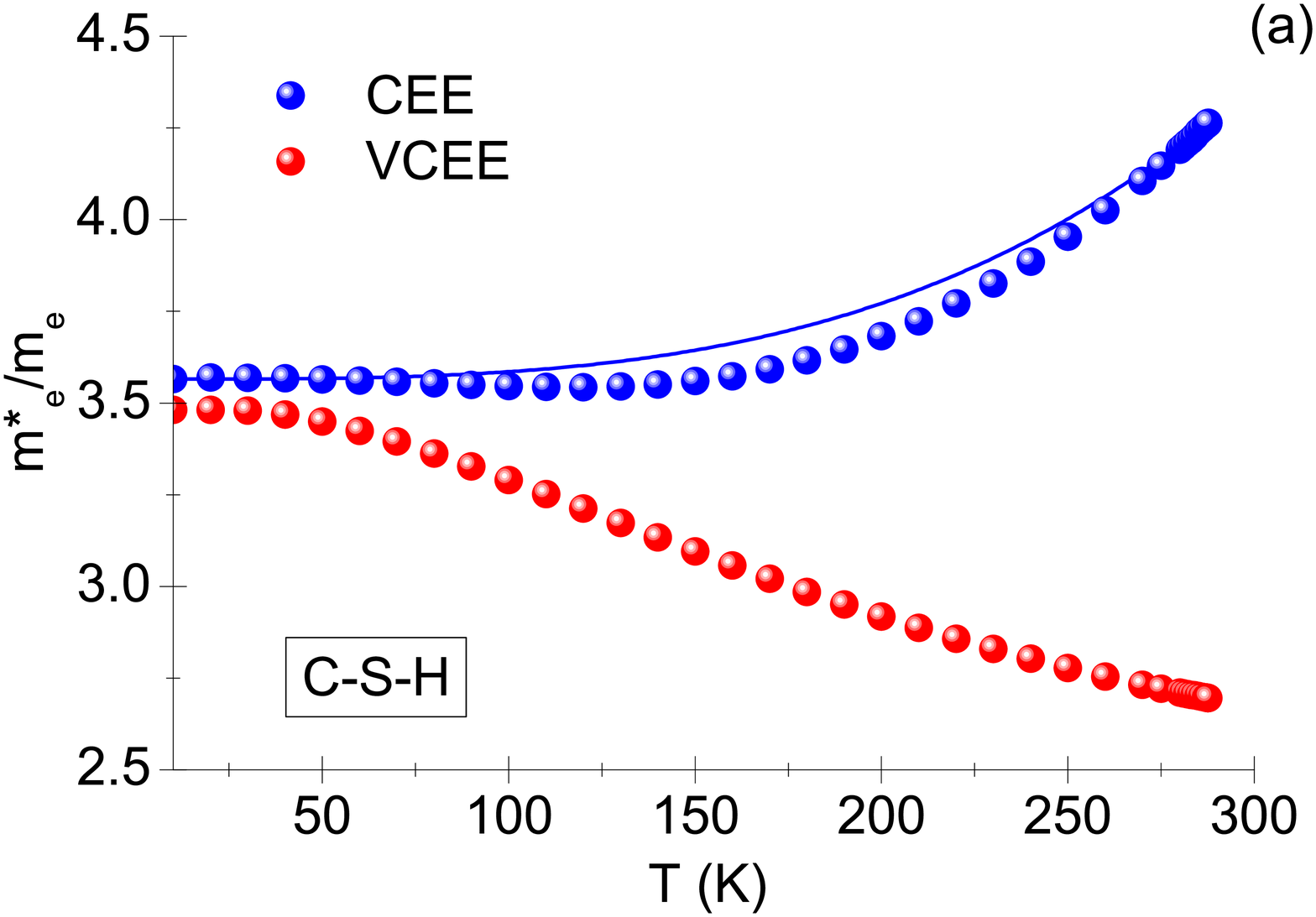}
\includegraphics[width=0.3\columnwidth]{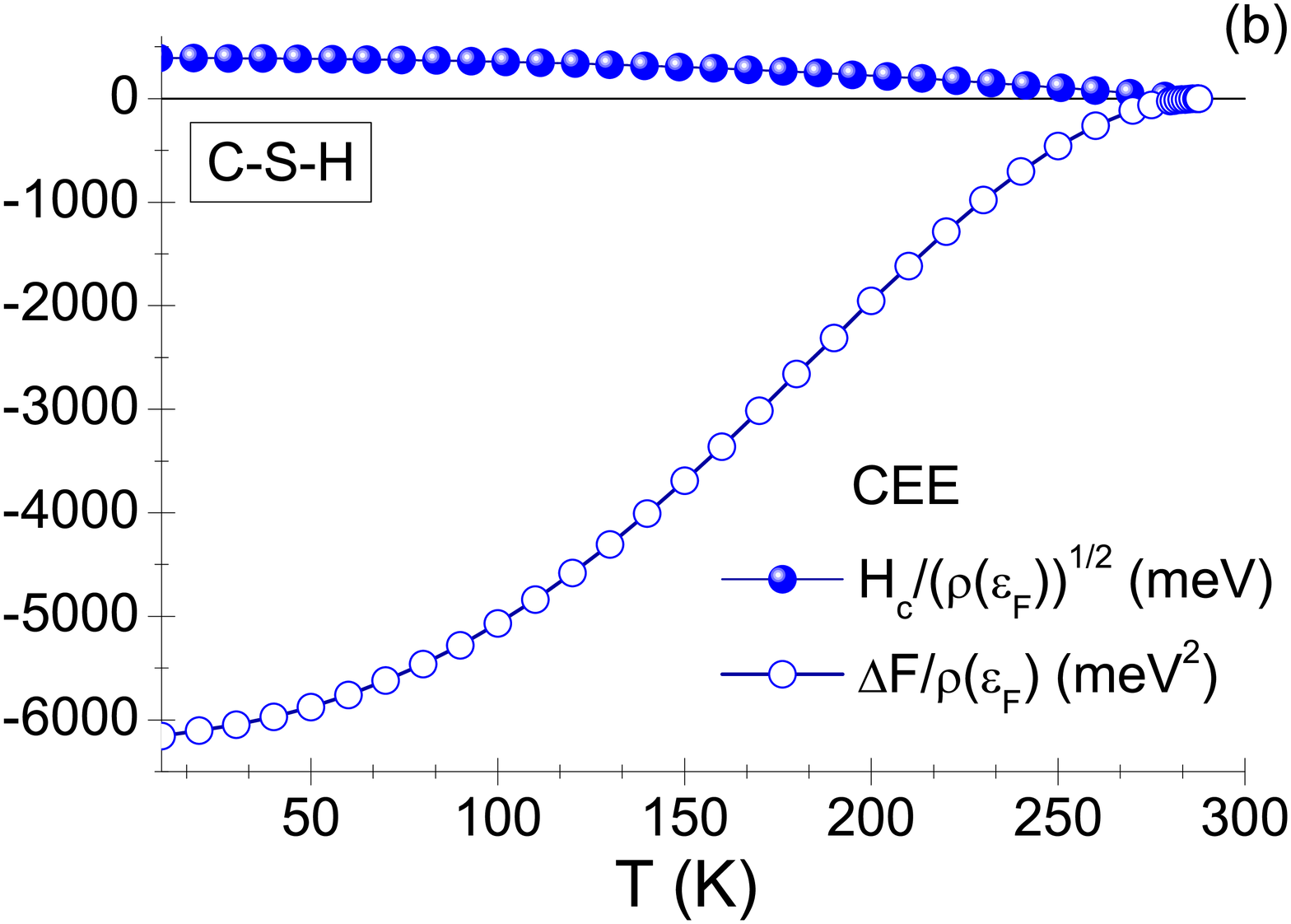}
\includegraphics[width=0.3\columnwidth]{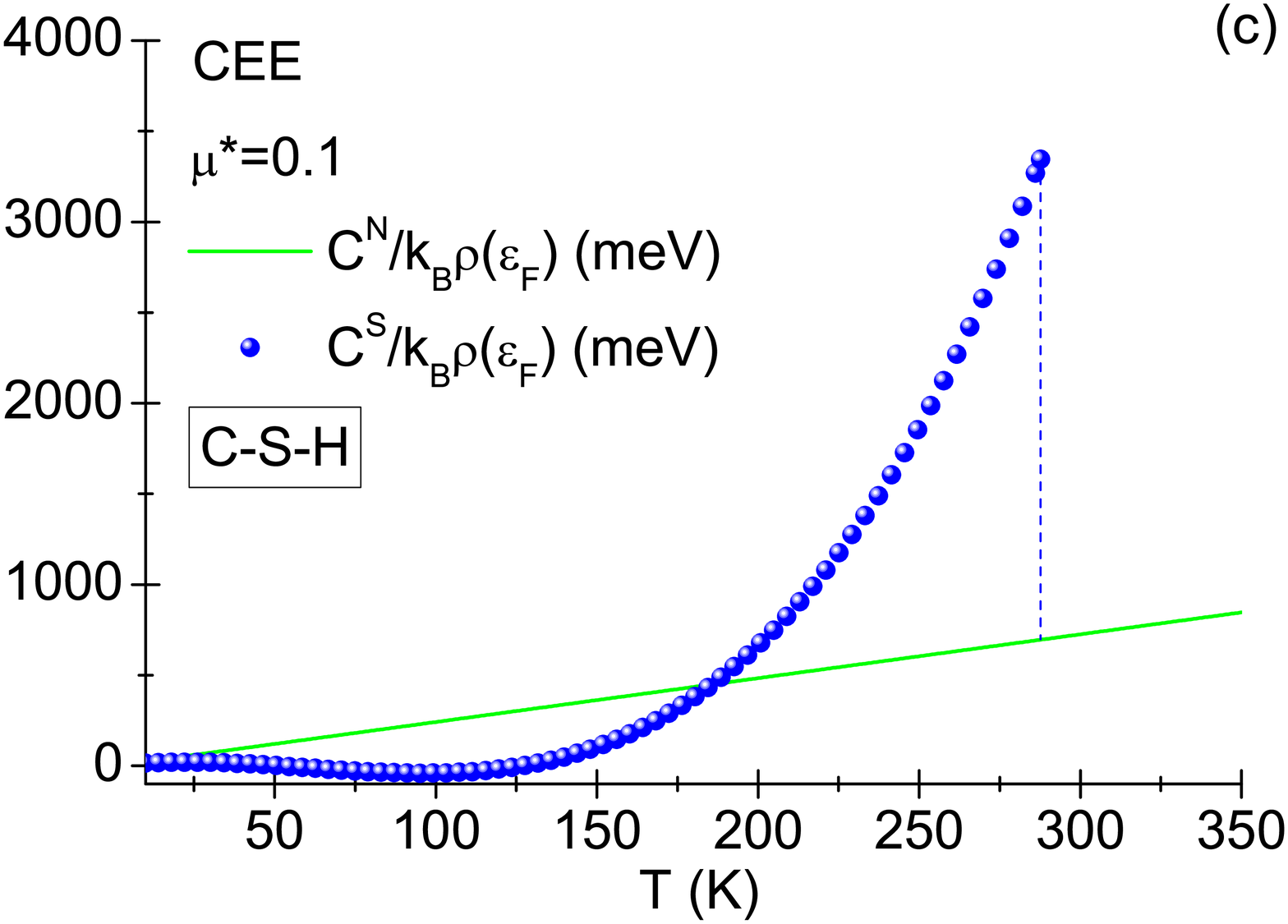}
\caption{  
(a) The ratio of effective mass to band mass of electron as a function of temperature. 
    The solid line was obtained with \eq{r02-IIb}. 
(b) Upper panel: the thermodynamic critical field as a function of temperature.
    Lower panel: the free energy difference between superconducting and normal states as a function of temperature.
(c) The specific heat of superconducting and normal states as a function of temperature.   
}
\label{Fig03-IIIa}
\end{figure}

In \fig{Fig03-IIIa}~(a), we plotted the temperature dependence of ratio: $m^{\star}_{e}/m_{e}$ - the effective mass of electron to the band mass of electron. Due to the very high value of electron-phonon coupling constant, also the ratio $m^{\star}_{e}/m_{e}$ takes 
high values. In particular, under classical Migdal-Eliashberg scheme, we obtained $m^{\star}_{e}/m_{e}=3.56$ ($T=T_{0}$). 
On the other hand, for $T=T_{C}$, the value of $m^{\star}_{e}/m_{e}$ is equal to $4.26$. This result is consistent with exact analytical 
result: $m^{\star}_{e}/m_{e}=1+\lambda$ \textcolor{blue}{\cite{Carbotte1990A}}, which confirms the high quality of presented numerical results.  

The results collected in \fig{Fig03-IIIa}~(a) prove also that the vertex corrections clearly change the temperature dependence of effective mass 
of the electron. Importantly, with the increasing temperature, the ratio $m^{\star}_{e}/m_{e}$ significantly decreases reaching the value of $2.69$ 
for critical temperature. 

As we have shown in area of the low temperatures ($T\sim T_{0}$) and in vicinity of the critical temperature, 
the vertex corrections slightly affect the values of order parameter. This means that in the temperature ranges of interest, 
the thermodynamic parameters of superconducting state can be calculated within the CEE scheme. 
All the formulas needed for this purpose have been collected and discussed in \app{Dod0A}. 

In \fig{Fig03-IIIa}~(b) we presented the influence of temperature on the value of free energy difference between 
superconducting and normal states ($\Delta F$). From this, the temperature dependence of thermodynamic critical field $H_{C}$ 
can be determined. For $T=T_{0}$ the value of thermodynamic critical field 
is equal to $H_{C}\left(0\right)/\sqrt{\rho\left(\varepsilon_{F}\right)}=393.97$~meV (see also \tab{Tab01-IIb}). 
This means that the dimensionless ratio $R_{H}$ is equal to $0.177$. Recall that for all superconducting systems, 
the BCS model predicts $R_{H}=0.168$ \textcolor{blue}{\cite{Bardeen1957A, Bardeen1957B}}. 

The forms of specific heat curves for superconducting and normal states are plotted in \fig{Fig03-IIIa}~(c). 
The specific heat jump occurring at critical temperature is characterized by the value 
$\Delta C\left(T_{C}\right)/k_{B}\rho\left(\varepsilon_{F}\right)=2650.32$~meV. 
Hence, the dimensionless ratio $R_{C}$ is equal to $2.37$. The BCS model predicts $R_{C}=1.43$ 
\textcolor{blue}{\cite{Bardeen1957A, Bardeen1957B}}.

\section{\label{Dod0D} Characterization of selected hydrogen-rich compounds in terms of superconducting state}

The literature that describing the superconducting state in hydrogen-rich compounds is very extensive (see \textcolor{blue}{\cite{Szczesniak2013A, Durajski2020A}} and references therein). In \tab{Tab01-D}, we have collected the most important information on experimental and theoretical results obtained so far. In particular, we have taken into account superconductors containing carbon and sulphur due to the fact that these elements are present 
in C-S-H system. Additionally, we have reported results for $\rm{LaH_{10}}$, $\rm{YH_{10}}$, and $\rm{YH_{6}}$, due to the very high value 
of critical temperature in these compounds. 

Note that research on sulphur in the context of superconductivity dates back to $1997$, when Struzhkin 
{\it et al.} \textcolor{blue}{\cite{Struzhkin1997A}} experimentally demonstrated the disappearance of electrical resistance 
in temperature range from $10$~K to $17$~K, and the pressure of $93$-$157$~GPa. Thus, they established the record value of 
critical temperature for pure element, for that moment. Subsequently, the theoretical analysis carried out in $2014$ by 
Li {\it et al.} \textcolor{blue}{\cite{Li2014A}} suggested the existence of superconducting state with much higher value of 
critical temperature ($T_{C}=80$~K) in $\rm{H_{2}S}$ for the pressure $160$~GPa. The experimental verification 
of Li {\it et al.} \textcolor{blue}{\cite{Li2014A}} result by Drozdov {\it et al.} \textcolor{blue}{\cite{Drozdov2015A}} 
unexpectedly showed that the value of critical temperature was almost twice as high ($T_{C}=$150~K). 

In addition, in the same experiment \textcolor{blue}{\cite{Drozdov2015A}}, the superconducting properties of $\rm{H_{3}S}$ were 
investigated, confirming the previous theoretical predictions of Duan {\it et al.} \textcolor{blue}{\cite{Duan2014}} about 
superconducting state with the record critical temperature (at that time) of as much as $203$~K. 
Referring to the \tab{Tab01-D}, we can see that the superconducting properties of sulfur and hydrogen compounds have been 
tested for the wide pressure range, however, no higher critical temperature value than $T_{C}=203$~K has been found.  
While extending the research, promising results were obtained by adding carbon to sulfur and hydrogen. 
For example, the theoretically analyzed $\rm{CSH_{7}}$ compound can presumably become superconducting at temperature 
of about $200$~K \textcolor{blue}{\cite{Cui2020A}}. On the other hand, the C-S-H system, widely discussed in this paper, 
seems to be the undisputed record holder, which according to the experimental data reaches the $T_C$ value of $287.7$~K 
\textcolor{blue}{\cite{Snider2020A}}.

The $\rm{YH_{6}}$, $\rm{LaH_{10}}$, and $\rm{YH_{10}}$ compounds were indicated as potential high-temperature superconductors based 
on the results of theoretical considerations ($\rm{YH_{6}}$ in 2015 \textcolor{blue}{\cite{Li2015A}}, 
$\rm{LaH_{10}}$ and $\rm{YH_{10}}$ two years later \textcolor{blue}{\cite{Liu2017}}). Although, according to calculations, 
$\rm{YH_{10}}$ compound has the highest critical temperature among mentioned superconductors (see \tab{Tab01-D}), this result 
has not been confirmed experimentally so far. The situation is different for the $\rm{LaH_{10}}$ compound, for which in $2019$ 
it was experimentally demonstrated that its critical temperature value is equal to $260$~K ($p=190$~GPa) \cite{Somayazulu2019A}.
This result agrees quite well with previous theoretical predictions \textcolor{blue}{\cite{Liu2017}}. For $\rm{YH_{6}}$ compound, 
the latest experimental studies \textcolor{blue}{\cite{Troyan2020A}} show that theoretical calculations have overestimated 
the critical temperature (by about $40$~K) \cite{Li2015A}. In our opinion, this inconsistency is seeming. 
The detailed analysis of this issue is provided in \app{Dod0F}.

It should be noted that all of mentined hydrogen-rich compounds for which $T_{C}$ exceeds $200$~K are characterized by the high 
value of electron-phonon coupling constant (see \tab{Tab01-D}), what suggests the similar scenario for C-S-H system.  

\begin{table*}
\caption{\label{Tab01-D} Selected elements or compounds containing hydrogen in which the superconducting state has been confirmed or is expected to exist.
}
\begin{ruledtabular}
\begin{tabular}{|c|c|c|c|c|c|c|c|c|}
                 &     &                     &       &       &     &     &     &           \\
Stoichiometry & $p$~(GPa) & Structure & $\lambda$ & $\omega_{{\rm ln}}/k_{B}$~(K)  & $\mu^{\star}$ & $T^{\rm CEE}_{C}$~(K) & $T^{\rm Exp.}_{C}$~(K) &  Theor./Exp. \\
                 &               &        &       &       &     &     &     &                                          \\
\hline
                 &     &                     &       &       &     &     &     &                                          \\
${\rm CSH_{7}}$  & 100 &  {\it Cm}   &  1.20  &  1091     & 0.1 & 108 & -   & Theor. \textcolor{blue}{\cite{Cui2020A}} \\
${\rm CSH_{7}}$  & 150 &  {\it R3m}  &  2.47  &  925      & 0.1 & 194 & -   & Theor. \textcolor{blue}{\cite{Cui2020A}} \\
${\rm CSH_{7}}$  & 200 &  {\it R3m}  &  1.35  &  1379     & 0.1 & 158 & -   & Theor. \textcolor{blue}{\cite{Cui2020A}} \\
${\rm CSH_{7}}$  & 150 &  {\it Pmna} &  3.06  &  672     & 0.1 & 170 & -   & Theor. \textcolor{blue}{\cite{Cui2020A}} \\
${\rm CSH_{7}}$  & 200 &  {\it Pmna} &  1.06  &  1622     & 0.1 & 138 & -   & Theor. \textcolor{blue}{\cite{Cui2020A}} \\
                 &     &             &        &              &     &     &     &                                          \\
\hline
           &     &             &        &              &     &     &       &                                          \\
C-S-H      & 225 &  -          &  -     &  -        & -   & -   & 200   & Exp. \textcolor{blue}{\cite{Snider2020A}}\\
C-S-H      & 250 &  -          &  -     &  -        & -   & -   & 255   & Exp. \textcolor{blue}{\cite{Snider2020A}}\\
C-S-H      & 267 &  -          &  -     &  -       & -   & -   & 287.7 & Exp. \textcolor{blue}{\cite{Snider2020A}}\\
C-S-H      & 275 &  -          &  -     &  -        & -   & -   & 286   & Exp. \textcolor{blue}{\cite{Snider2020A}}\\
           &     &             &        &       &           &     &       &                                          \\                 
\hline
                 &     &                      &                         &   &      &       &     &                                               \\
$\rm H_{2}S$     & 160 & $P1\rightarrow Cmca$ & $\sim 1.2$ & $\sim 1000$    & 0.13 & 82    & -   & Theor. \textcolor{blue}{\cite{Li2014A}}       \\                 
$\rm H_{2}S$     & 150 &  -                   & -          &  -           & -    & -     & 150 & Exp. \textcolor{blue}{\cite{Drozdov2015A}}    \\
$\rm H_{2}S$     & 130-180&  $P1, Cmca$            & 1.28       &  960           & 0.15 & 31-88 &  -  & Theor. \textcolor{blue}{\cite{Durajski2015A}} \\                 
                 &     &                                  &             &   &      &       &     &                                               \\
\hline
                 &                                    &        &       &       &     &            &     &                                            \\
$\rm H_{3}S$     & 200       & {\it Im$\overline{3}$m} & 2.19   & 1335       & 0.1 & 204        & -   & Theor. \textcolor{blue}{\cite{Duan2014}}   \\
$\rm H_{3}S$     & 157 & {\it Im$\overline{3}$m} & 1.94      &   1321      & 0.16& 216 & -   & Theor. \textcolor{blue}{\cite{Errea2016A}} \\                 
$\rm H_{3}S$     & 155       &  -                      & -      & -          & -   & -          & 203 & Exp.\textcolor{blue}{\cite{Drozdov2015A}}  \\
$\rm H_{3}S$     & 150 & $R3m$ & 2.067 & 1056     & 0.123 & 203 & - & Theor. \textcolor{blue}{\cite{Durajski2016A}} \\
$\rm H_{3}S$     & 250 & {\it Im$\overline{3}$m} & 1.31  & 1485  & 0.13  & 164 & - & Theor. \textcolor{blue}{\cite{Durajski2017A}} \\
$\rm H_{3}S$     & 350 & {\it Im$\overline{3}$m} & 1.22  & 1301  & 0.13  & 129 & - & Theor. \textcolor{blue}{\cite{Durajski2017A}} \\
$\rm H_{3}S$     & 450 & {\it Im$\overline{3}$m} & 1.26  & 1464  & 0.13  & 146 & - & Theor. \textcolor{blue}{\cite{Durajski2017A}} \\
$\rm H_{3}S$     & 500 & {\it Im$\overline{3}$m} & 1.32  & 1454  & 0.13  & 156 & - & Theor. \textcolor{blue}{\cite{Durajski2017A}} \\
                 &     &   &       &         &       &     &   &                                               \\ 
\hline
                 &     &                   &            &   &            &           &     &                                               \\
$\rm LaH_{10}$   & 300 & $Fm\overline{3}m$ & 1.78 & 1488      & 0.1 (0.13) & 254 (241)& -   & Theor. \textcolor{blue}{\cite{Liu2017}}       \\
$\rm LaH_{10}$   & 250 & sodalite-like fcc & 2.2 & 1253      & 0.1 (0.13) & 274 (257) & -   & Theor. \textcolor{blue}{\cite{Liu2017}}       \\
$\rm LaH_{10}$   & 190 & -                 & -   & -    & -          & -         & 260 & Exp. \textcolor{blue}{\cite{Somayazulu2019A}} \\                 
$\rm LaH_{10}$   & 170 & -                 & -   & -     & -          & -         & 250 & Exp. \textcolor{blue}{\cite{Drozdov2019A}}    \\                 
$\rm LaH_{10}$   & 150 & {\it R$\overline{3}$m}                & 2.2 & -      & 0.1        & 215       & -   & Theor. \textcolor{blue}{\cite{Kostrzewa2020A}}\\
$\rm LaH_{10}$   & 190 & {\it Fm$\overline{3}$m}                 & 2.8 & -      & 0.1        & 260       & -   & Theor. \textcolor{blue}{\cite{Kostrzewa2020A}}\\
$\rm LaH_{10}$   & 170 & $Fm\overline{3}m$ & 3.94& 801      & 0.2        & 259       &   -  & Theor. \textcolor{blue}{\cite{Kruglov2020A}}  \\
$\rm LaH_{10}$   & 150 & $R\overline{3}m$  & 2.77& 833      & 0.2        & 203       &    - & Theor. \textcolor{blue}{\cite{Kruglov2020A}}  \\
                 &     &                   &     &          &            &           &     &                                               \\
\hline
                 &     &             &        &             &     &     &     &                                                    \\
$\rm YH_{10}$    & 250 & {\it Fm$\overline{3}$m} & 2.56 & 1282  & 0.1 (0.13) & 326 (305) & -  & Theor.\textcolor{blue}{\cite{Liu2017}} \\     
$\rm YH_{10}$    & 300 & {\it Im$\overline{3}$m} & 2.06 & 1511  & 0.1 (0.13) & 308 (286) & -  & Theor.\textcolor{blue}{\cite{Liu2017}} \\  
$\rm YH_{10}$    & 250 & sodalite-like fcc& 2.67 & 1102  &0.1 & 291 & -  & Theor.\textcolor{blue}{\cite{Tanaka2017}} \\          
$\rm YH_{10}$    & 300 & sodalite-like fcc & 2.00 & 1450 & 0.1& 275 & -  & Theor.\textcolor{blue}{\cite{Tanaka2017}} \\            
                 &     &             &              &       &     &     &     &                                                    \\ 

\hline
                 &     &             &        &             &     &     &     &                                             \\
$\rm YH_{6}$     & 120   & {\it Im$\overline{3}$m}          & 2.93      & 1080          & 0.1 (0.13)   & 264 (251)   & -   & Theor.\textcolor{blue}{\cite{Li2015A}}      \\                       
$\rm YH_{6}$     & 166 & {\it Im$\overline{3}$m}           & -         & -     & -   & -   & 224 & Exp. \textcolor{blue}{\cite{Troyan2020A}}   \\
$\rm YH_{6}$     & 165   & {\it Im$\overline{3}$m}           & 1.71     & 1333         & 0.1 (0.15)   & 247 (236)   & -   & Theor.\textcolor{blue}{\cite{Troyan2020A}}  \\ 
$\rm YH_{6}$     & 300   & {\it Im$\overline{3}$m}           & 1.73     & 1612          & 0.11   & 290   & -   & Theor.\textcolor{blue}{\cite{Heil2019A}}  \\                  
                 &                &        &       &       &     &     &     &                                             \\                 
\hline
                 &     &             &             &       &     &     &     &                                          \\
  S              &   93-157  &        -     & -       &    -   &   -        & -    &   10-17  & Exp.\textcolor{blue}{\cite{Struzhkin1997A}}           \\
  S              &   160       &    $\beta${\it-Po}   &  0.75    &437       & 0.127    & 17    &  -   & Theor.\textcolor{blue}{\cite{Durajski2012A}}         \\
                 &     &                    &       &       &     &     &     &                                          \\
\hline
                 &     &             &             &       &     &     &     &\\
H                & 480  &  $I4_1\slash amd$                 &   2.17    &   1870    &  0.1 (0.13)   &  284 (266)   &    - & Theor. \textcolor{blue}{\cite{Yan2011}  }                                      \\
H                & 539   &   $I4_1\slash amd$                 &   2.01    &   2016    &  0.1 (0.13)   &  291 (272)   &    - & Theor.\textcolor{blue}{\cite{Yan2011}  }                                      \\    
H                & 608   &   $I4_1\slash amd$                 &   1.89    &  2106    &  0.1 (0.13)   &  291 (270)   &    - & Theor.\textcolor{blue}{\cite{Yan2011}  }                                      \\    
H                & 802   &  $I4_1\slash amd$                 &   1.68    &   2239    &  0.1 (0.13)   &  282 (260)   &    - & Theor.\textcolor{blue}{\cite{Yan2011}  }                                      \\    
H                & 802  &        $I4_1\slash amd$            &   1.7    &  -     &   0.1 (0.2)  &  332.7 (259.4)   &   -  &  Theor.\textcolor{blue}{\cite{Durajski2014A}   }   \\
H                & 2000  &        fcc            &   7.32    &  1035     &   0.1 (0.5)  &   631 (413)  &   -  &  Theor.\textcolor{blue}{\cite{Szczesniak2009A}   }                                  \\                     
$\rm H_{2} $     & 494 &  2D model   &  0.628 &  7955.064  &  0.17385  & 84.49 &  -& Theor. \textcolor{blue}{\cite{Kostrzewa2021A}} \\
$\rm H_{2} $     & 686 &  2D model   &  0.573 &  10985.66  &  0.17997  & 64.66 &  -& Theor. \textcolor{blue}{\cite{Kostrzewa2021A}} \\
                 &     &             &        &            &           &       &   &                                                \\                 
\end{tabular}
\end{ruledtabular}
\end{table*}
%

\section{\label{Dod0E} Detailed characteristics of superconducting state in ${\rm LaH_{10}}$ compound}

\begin{figure}
\includegraphics[width=0.50\columnwidth]{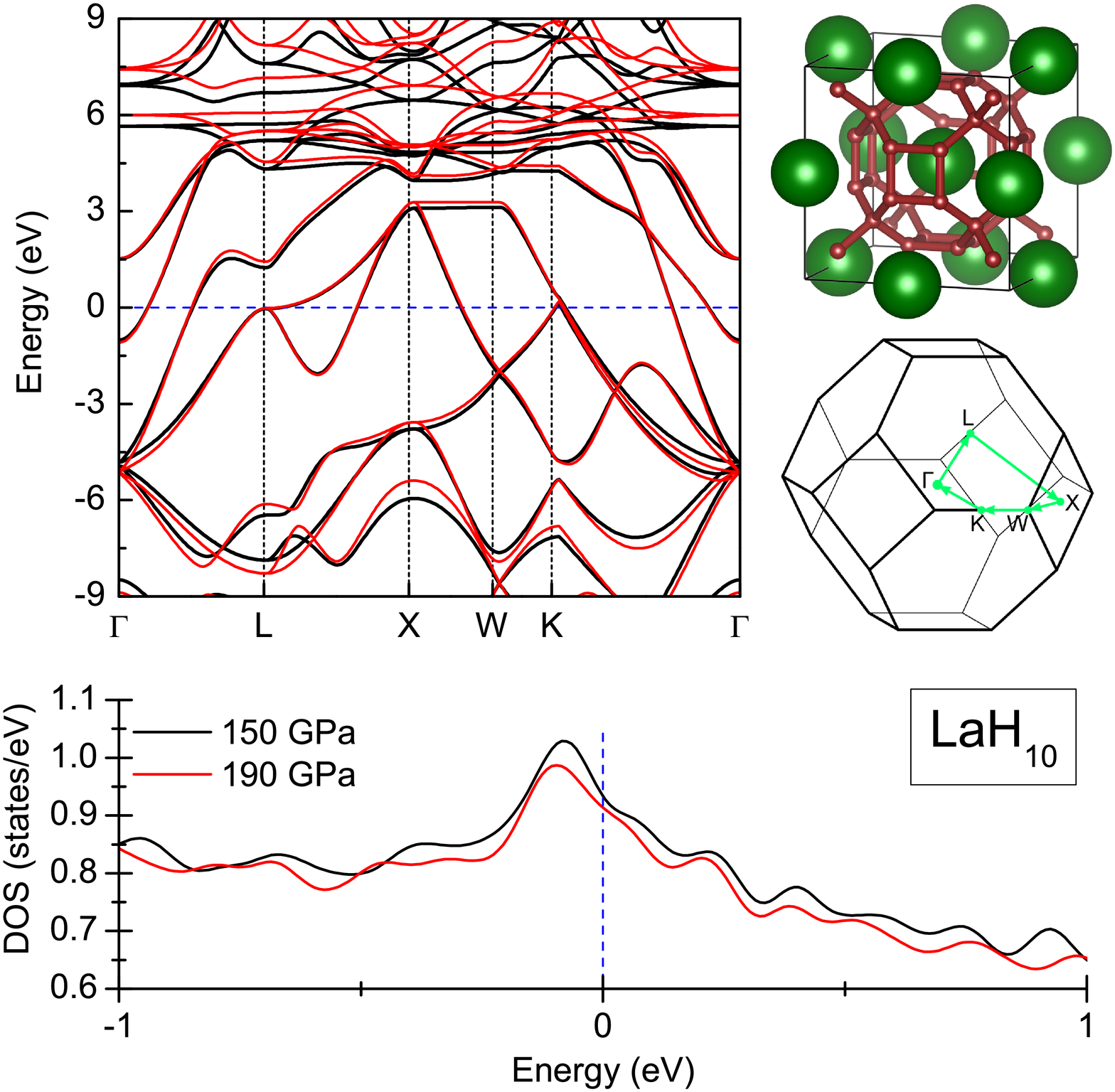}
\caption
{
         Band structures along the $\Gamma$-L-X-W-K-$\Gamma$ high-symmetry line and DOS of LaH$_{10}$ at $150$ GPa (black lines) and $190$ GPa 
         (red lines) with their Fermi energies set to be zero. Moreover, the optimized crystal structure and the Brillouin zone for 
         cubic (Fm$\overline{3}$m) clathrate-type structure with special {\bf k}-point paths were plotted.
}
\label{Fig01-E}
\end{figure}

For LaH$_{10}$ the calculations of atomic structure relaxation, electronic structure, and phonon properties were performed based on the DFT method 
within the generalized gradient approximation of the Perdew-Burke-Ernzerhof exchange-correlation functional. The Brillouin zone was sampled using a $24\times 24\times 24$ {\bf k}-points grid according to the Monkhorst-Pack scheme. On the base of convergence tests, the kinetic energy cut-off for the wave functions and charge density were taken as $80$~Ry and $1000$~Ry, respectively. The cubic clathrate-type structure of investigated system with the space group {\it Fm$\overline{3}$m} consists of the cage of $32$ H atoms surrounding a La atom. The optimized lattice parameters $a=b=c$ 
of LaH$_{10}$ at $150$~GPa and $190$~GPa are $5.0964$ and $4.9834$~\AA, repectively. We found that LaH$_{10}$ has a DOS that reaches $0.91 - 0.93$ states/eV at the Fermi level, because of the presence of a van Hove singularity in the vicinity, as shown in \fig{Fig01-E}. By increasing the pressure, the DOS peak can be modulated. 

Thermodynamic parameters of the ${\rm LaH_{10}}$ superconductor, subjected to external pressure of $190$~GPa, were determined taking into account 
the Eliashberg equations on imaginary axis (see \app{Dod0A} and \app{Dod0B}). In the first step, we calculated the value of electron-phonon coupling constant. We used the condition: $\left[\Delta_{n=1}\right]_{T=T_{C}}=0$, where $T_{C}=260$~K is experimental result taken from the paper \cite{Somayazulu2019A}. For classical Eliashberg equations, we obtained: $\lambda=2.82$, for Eliashberg equations taking into account the vertex corrections, we had: $\lambda=2.77$. Based on discussed data, we conclude that the ${\rm LaH_{10}}$ superconductor is the system characterized by 
strong electron-phonon coupling. By taking vertex corrections into account - the value of $\lambda$ can be decreased by $1.3\%$.

\begin{figure}
\includegraphics[width=0.48\columnwidth]{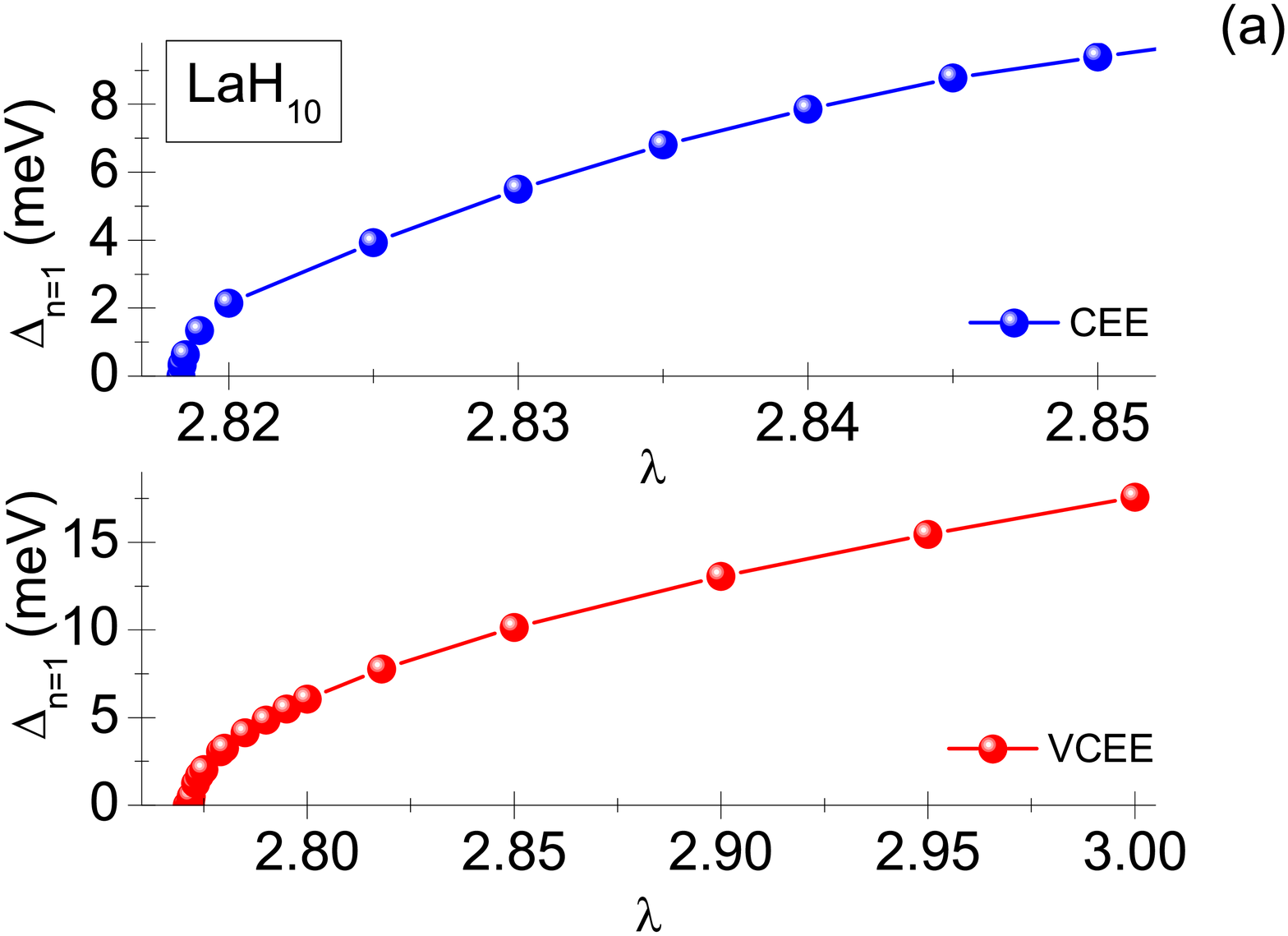}
\includegraphics[width=0.48\columnwidth]{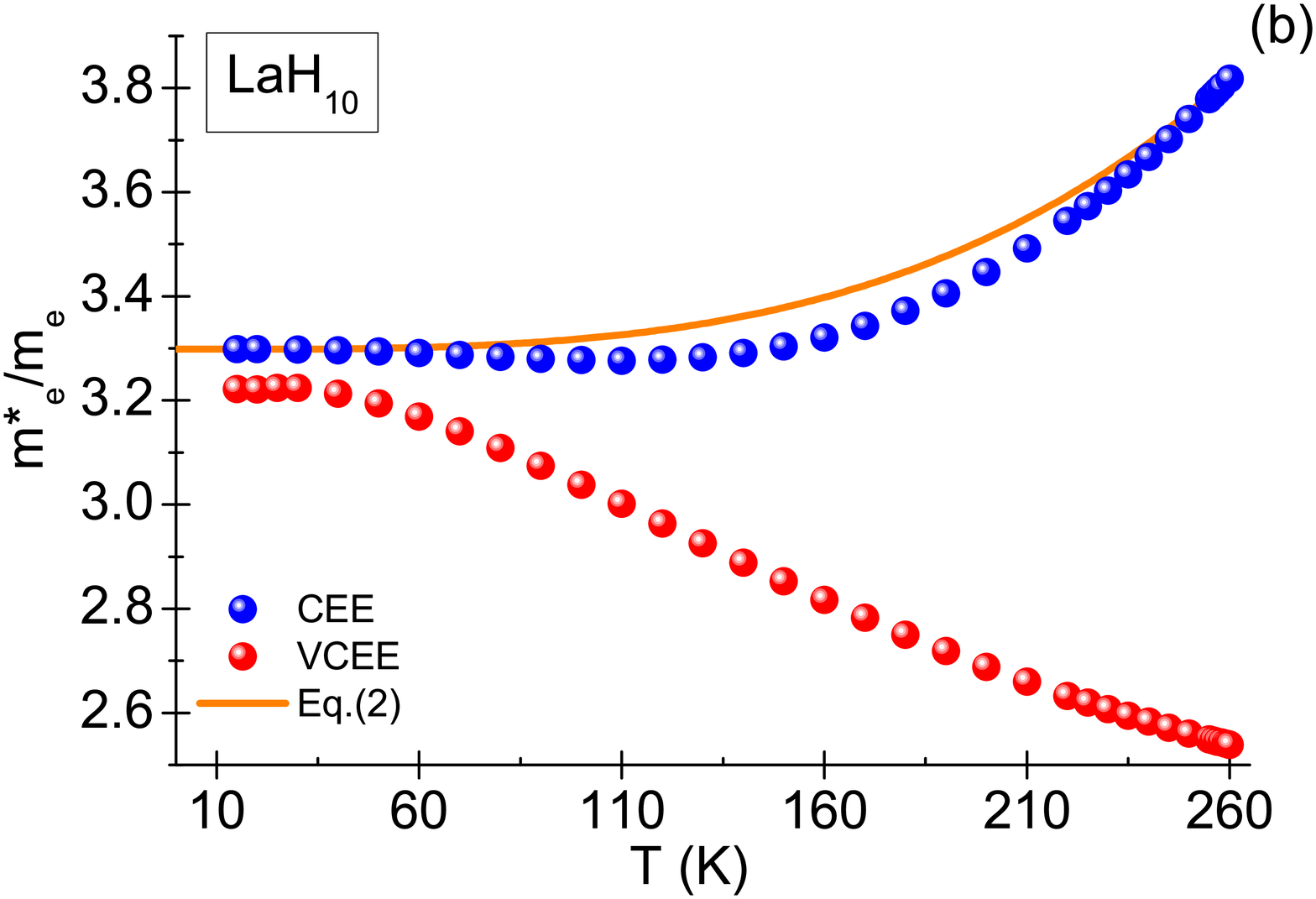}
\caption
{
(a) The dependence of maximum value of order parameter on the electron-phonon coupling constant ($T=T_{C}$).
(b) The influence of temperature on the effective electron mass. Spheres represent the numerical results.  
    The orange line was obtained using the analytical formula \eq{r02-IIb}. 
}
\label{Fig02-E}
\end{figure}

Fitting the value of electron-phonon coupling constant to experimental results is presented in \fig{Fig02-E}~(a). In particular, we assumed: 
$\mu^{\star}=0.1$, $\omega_{0}=100$~meV, and $\omega_{C}=1$~eV. We have obtained the stable solutions of Eliashberg equations for $T\geq T_{0}=15$~K.

The full dependence of order parameter on temperature we plotted in \fig{Fig02-IIIa}~(b). In turn the influence of temperature on the value of effective electron mass to band electron mass ratio is presented in \fig{Fig02-E}~(b). It can be seen that, analogically to C-S-H system (see \fig{Fig03-IIIa}~(a)), the vertex corrections to electron-phonon interaction noticeably lower the value of effective mass of electron. 

The dimensionless ratio $R_{\Delta}$ for the ${\rm LaH_{10}}$ superconductor is equal $5.25$ and $5.33$, respectively for schema CEE and VCEE. 
These results are due to significant retardation and strong-coupling effects. The other thermodynamic parameters of superconductor ${\rm LaH_{10}}$ have already been analyzed by us in CEE scheme and described in detail in the publication \cite{Kostrzewa2020A}.
%

\section{\label{Dod0F} The thermodynamic parameters of superconducting state for ${\rm YH_{6}}$ compound}

The superconducting properties of ${\rm YH_{6}}$ compound attracted the attention of researchers several years ago. 
Calculations carried out in 2014 by Li {\it et al.} suggested the high transition teperature value of $251$~K-$264$~K at $120$~GPa 
\textcolor{blue}{\cite{Li2015A}}. Additionally, in 2019 it was suggested that the value of $\rm{T_C}$ at $300$~GPa could be as high as $290$~K  \textcolor{blue}{\cite{Heil2019A}}. In contrast, experiments conducted in 2020 showed that the critical temperature value is equal to $224$~K at $166$~GPa ($Im\bar{3}m$ structure) \textcolor{blue}{\cite{Troyan2020A}}, so is lower than theoretically predicted value of ${T_C}$.
\begin{table}
\caption{\label{Tab01-F} The input parameters to Eliashberg equations for ${\rm YH_{6}}$ compound.}
\begin{tabular}{|c||c|} 
\hline
                                                 &             \\
$p$                                              &  $166$~GPa  \\
                                                 &             \\
$T_{C}$  \textcolor{blue}{\cite{Troyan2020A}}    &  $224$~K    \\
                                                 &             \\
\hline
                                                                                    &               \\
$\omega_{D}$ (anharmonic) from $\alpha^2F(\omega)$ for $Im\bar{3}m$ \textcolor{blue}{\cite{Troyan2020A}} & $210.24$~meV  \\
                                                                                    &               \\
$\lambda$ \textcolor{blue}{\cite{Troyan2020A}}                                      & $1.71$        \\
                                                                                    &               \\                                                                                  
\hline
                               &               \\
$\Omega_{C1}$                  & 3$\omega_{D}$ \\
                               &               \\
$\mu^{\star}_{1}$              & $0.277$       \\
                               &               \\
\hline
                               &               \\
$\Omega_{C2}$                  & 10$\omega_{D}$\\
                               &               \\
$\mu^{\star}_{2}$              & $0.363$       \\
                               &               \\
\hline
                                                         &               \\
The Migdal ratio: $\frac{\omega_D}{\varepsilon_{F}}$     & $0.013$       \\
                                                         &               \\
$\lambda\frac{\omega_{D}}{\varepsilon_{F}}$              & $0.023$       \\
                                                         &               \\
\hline
\end{tabular}
\end{table}

\begin{figure}
\includegraphics[width=0.45\textwidth]{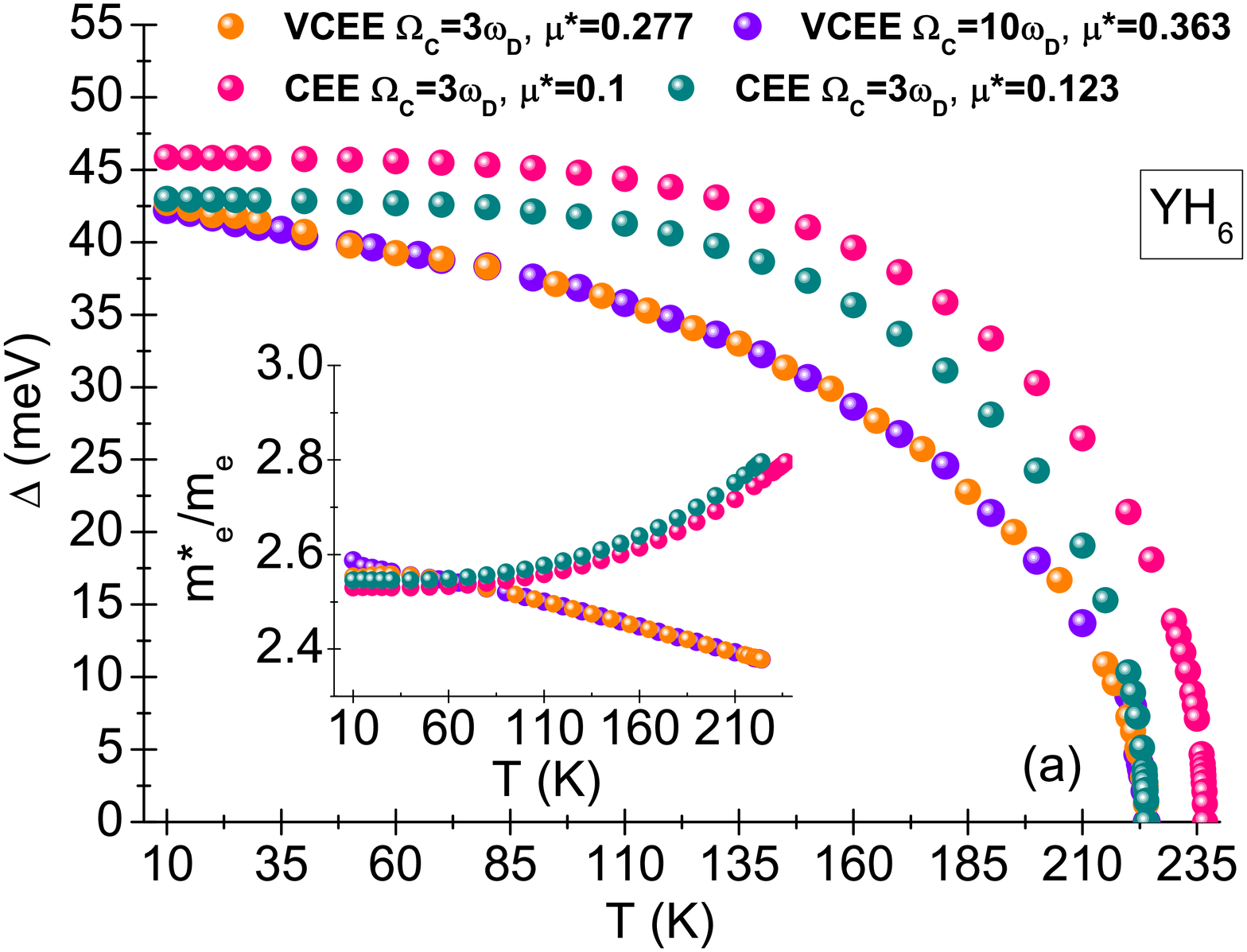}
\includegraphics[width=0.45\textwidth]{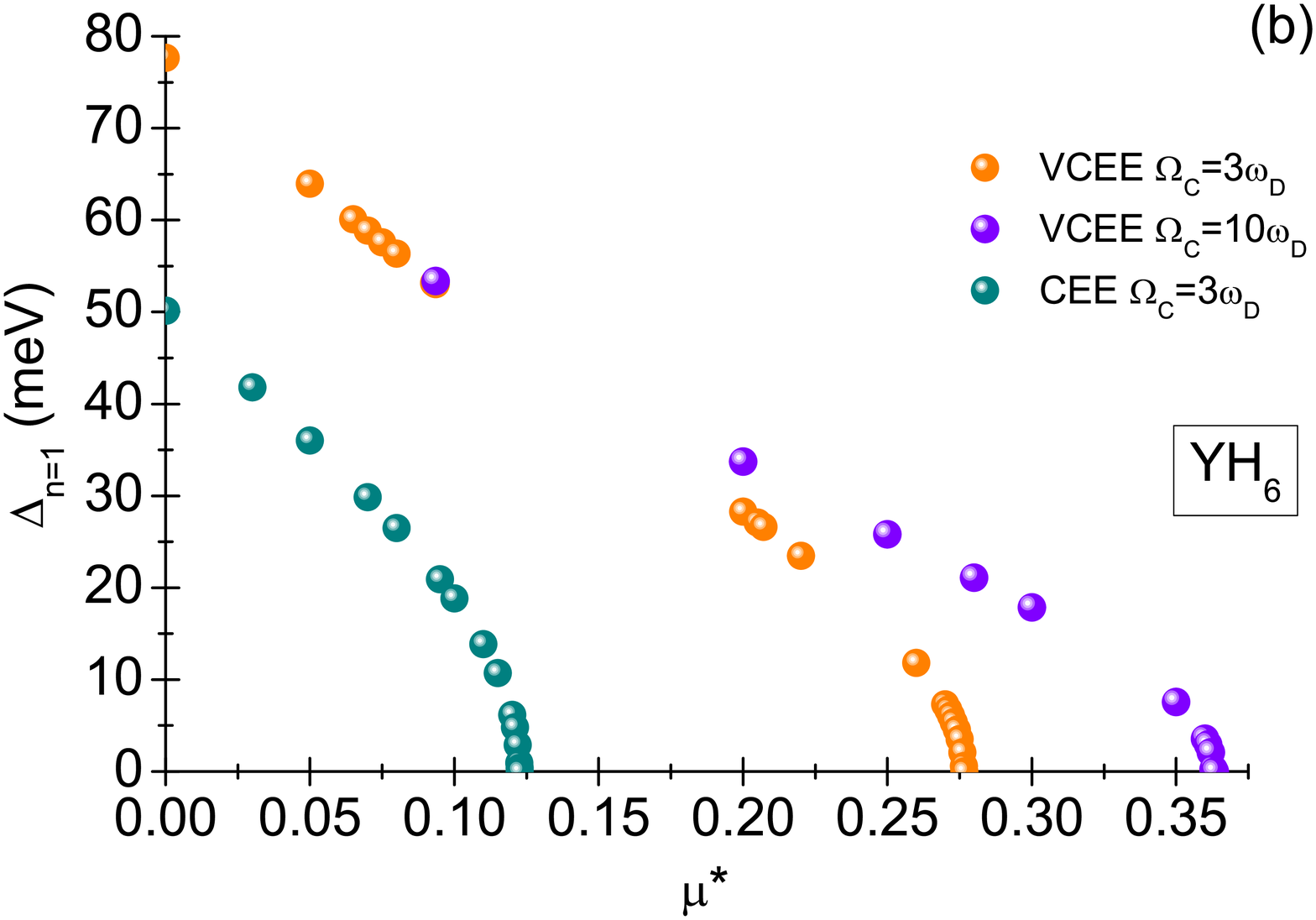}
\caption{(a) The dependence of order parameter on temperature for ${\rm YH_{6}}$ compound. 
             In the inset, we presented the values of electron effective mass.
         (b) The dependence of maximum value of the order parameter on Coulomb pseudopotential for $T=T_{C}$.  
        }
\label{Fig01-IV}
\end{figure}

In the first step, we performed the numerical calculations for ${\rm YH_{6}}$ within classical Eliashberg formalism (CEE). We used the anharmonic Eliashberg function determined in the paper \textcolor{blue}{\cite{Troyan2020A}} (see also \tab{Tab01-F}). For the standard value of Coulomb pseudopotential $\mu^{\star}=0.1$ (the pink spheres in \fig{Fig01-IV}~(a)), we obtained slightly higher critical temperature value 
($T_{C}=236.8$~K) than in the experiment ($T_{C}=224$~K). In the next step, we calculated the exact value of Coulomb pseudopotential 
for CEE scheme based on the equation $\left[\Delta_{n=1}\left(\mu^{\star}\right)\right]_{T=T_{C}}=0$, assuming that $\Omega_{C}=3\omega_{D}$ 
(the dark cyan spheres in \fig{Fig01-IV}~(b)). The obtained result is as follows: $\mu^{\star}=0.123$. 
Then, we determined the temperature dependence of order parameter (the dark cyan spheres in \fig{Fig01-IV}~(a)). It can be easily noticed that the obtained results allow to determine the width of superconducting gap, which in the analyzed case is: $2\Delta(T_0)=86.28$~meV. On this basis, we have obtained: $R_{\Delta}=4.45$. 

Note that in the publication \textcolor{blue}{\cite{Troyan2020A}}, the numerical calculations were also performed using the Eliashberg formalism 
(although the different form of Eliashberg equations was used). In the harmonic case, they received $T_{C}=251$~K for $\mu^{\star}=0.195$ and $T_{C}=261$~K-$272$~K for $\mu^{\star}=0.15-0.1$. In the anharmonic case, the results are: $T_{C}=226$~K for $\mu^{\star}=0.195$ and $T_{C}=236$~K-$247$~K for 
$\mu^{\star}=0.15-0.1$. As can easily be seen, the best outcomes (most close to experimental result) are given by the results obtained for anharmonic 
$\alpha^2F(\omega)$ of $Im\bar{3}m$-${\rm YH_{6}}$ ($165$~GPa). However, it can be seen that the relatively high value of Coulomb pseudopotential was used.

For ${\rm YH_{6}}$ we also performed the analysis within VCEE formalism. The input data for Eliashberg equations are collected in \tab{Tab01-F}. 
We obtained even higher value of Coulomb pseudopotential ($\mu^{\star}_{1}=0.277$), than Troyan {\it et al.} (we have chosen the standard cut-off frequency: $\Omega_{C1}=3\omega_{D}$). Note that the even higher cut-off frequency ($\Omega_{C2}=10\omega_{D}$) leads to even greater undesirable increase in the Coulomb pseudopotential: $\mu^{\star}_{2}=0.363$. The dependence courses of $\Delta_{n=1}$ on $\mu^{\star}$, characterizing the situation discussed by us, are collected in \fig{Fig01-IV}~(b). 
In \fig{Fig01-IV}~(a), we have plotted the temperature dependencies of order parameter. The results obtained for formalism including vertex corrections, for both calculated values of the Coulomb pseudopotential, converge the entire analyzed temperature range up to $T_C=224$~K. It can also be seen that the influence of vertex corrections on the values of order parameter is analogous to ${\rm LaH_{10}}$ superconductor and C-S-H system (strong-coupling case - see \fig{Fig02-IIIa}).   
  
\begin{figure}
\includegraphics[width=0.45\textwidth]{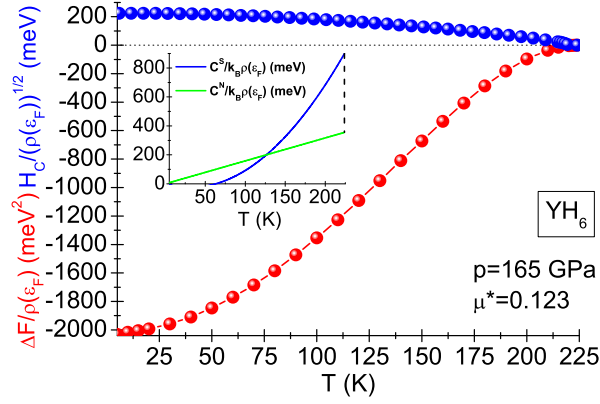}
\caption{(Lower panel) The free energy difference between superconducting and normal states as a function of temperature.
         (Upper panel) The thermodynamic critical field as a function of temperature.
          The inset shows the temperature dependence of specific heat for superconducting and normal state.   
        }
\label{Fig01-F}
\end{figure}

In area of very low temperatures and temperatures close to the critical temperature the values of order parameter obtained in CEE and VCEE schemes are physically indistinguishable. This result implies that the thermodynamics of superconducting state near temperature $T_{0}$ and close to critical temperature can be analyzed by using the classical Eliashberg approach. The \fig{Fig01-F} shows the difference of free energy $\Delta F$, the function of thermodynamic critical field $H_{C}$ and the specific heat in superconducting $C^{S}$ and normal $C^{N}$ states. Negative values of free energy difference indicate the thermodynamic stability of superconducting phase up to critical temperature. Additionally, we can observe a characteristic jump of specific heat occurring in $T_C=224$~K, its value equals $544.59$~meV. For ${\rm YH_{6}}$ superconductor, the values of characteristic dimensionless parameters $R_{H}$ and $R_{C}$ are $0.136$ and $1.53$, respectively. 

%

\section{\label{Dod0C} Influence of many-body corrections on the order parameter (classical Eliashberg equations)}

The classical Eliashberg equations were derived in self-consistent way, in the second order of electron-phonon coupling function ($g^{2}$). 
This procedure ignores the higher-order many-body corrections that affect the order parameter \textcolor{blue}{\cite{Carbotte1990A}}.
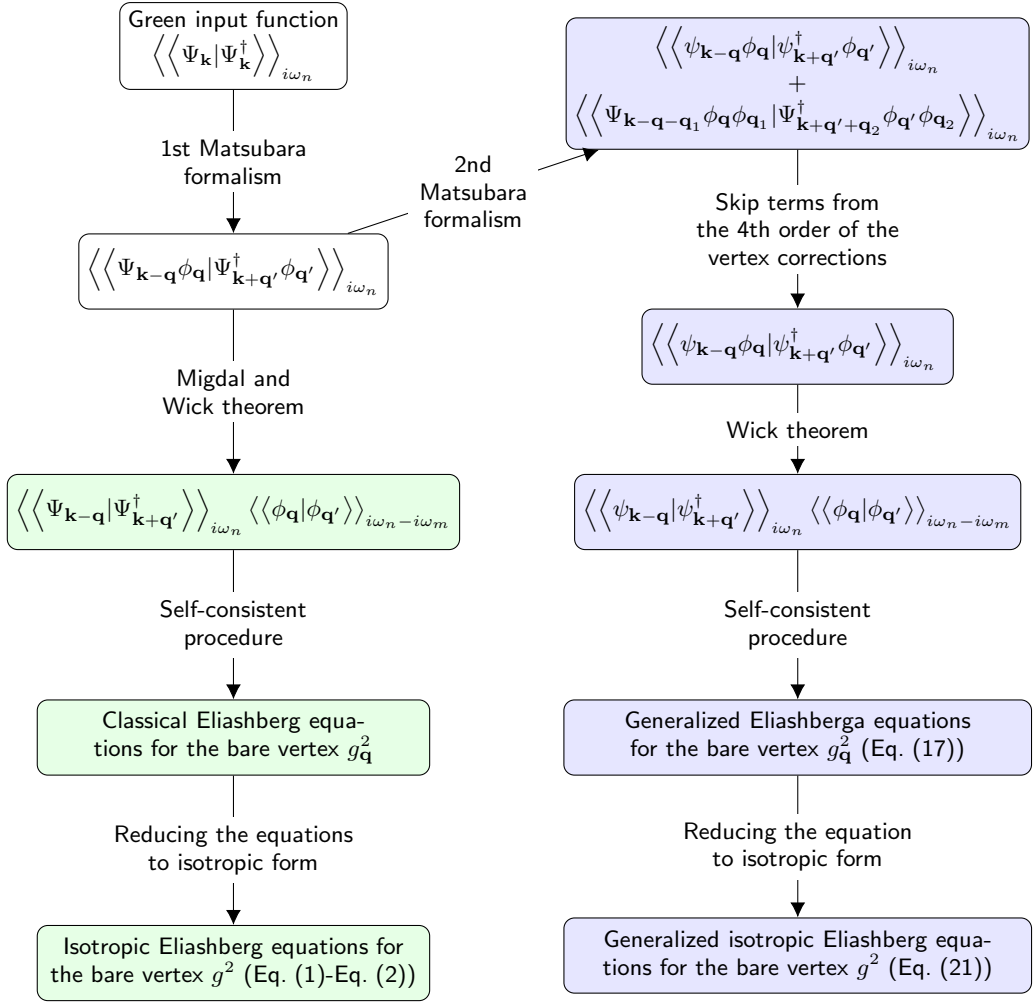
\begin{figure}
\begin{tikzpicture}[scale=0.6, node distance=1.5cm,
    every node/.style={fill=white, font=\sffamily}, align=center]
\node (start) [startstop] {Green input function
\\$\left<\left<\Psi_{\bf k}| \Psi^{\dagger}_{\bf k}\right>\right>_{i\omega_{n}}$};
 \node (start2)[startstop, below of=start, yshift=-1.5cm]{$\left<\left<\Psi_{\kvec-\qvec}\phi_{\qvec}|\Psi^{\dagger}_{\kvec+\qvec'}\phi_{\qvec'}\right>\right>_{i\omega_{n}}$};
 \node (start3)[activityStarts, right of=start, xshift=6cm,yshift=-0.5cm]{$\left<\left<\psi_{\kvec-\qvec}\phi_{\qvec}|\psi^{\dagger}_{\kvec+\qvec'}\phi_{\qvec'}\right>\right>_{i\omega_{n}}$ \\ + \\$\left<\left<\Psi_{\kvec-\qvec-\qvec_{1}}\phi_{\qvec}\phi_{\qvec_{1}}|\Psi^{\dagger}_{\kvec+\qvec'+\qvec_{2}}\phi_{\qvec'}\phi_{\qvec_{2}}\right>\right>_{i\omega_{n}}$};
\draw[->](start) -- node[text width=3.5cm] {1st Matsubara\\ formalism}(start2);
\draw[->](start2) -- node[text width=1.5cm] {2nd Matsubara formalism}(start3);
\node (start4)[activityRuns, below of=start2, yshift=-1.7cm]  {$\left<\left<\Psi_{\kvec-\qvec}|\Psi^{\dagger}_{\kvec+\qvec'}\right>\right>_{i\omega_{n}}\left<\left<\phi_{\qvec}|\phi_{\qvec'}\right>\right>_{i\omega_{n}-i\omega_{m}}$};
\draw[->] (start2) -- node[text width=3cm] {Migdal and Wick theorem} (start4);
\node (start5) [activityRuns, below of=start4, yshift=-1.5cm,text width=5cm] {Classical Eliashberg equations for the bare vertex $g^{2}_{\qvec}$};
\draw[->](start4) -- node[text width=3cm] {Self-consistent procedure}(start5);
\node (start6) [activityRuns, below of=start5, yshift=-1.5cm,text width=5cm] {Isotropic Eliashberg equations for the bare vertex $g^{2}$ (\eq{r01-A}-\eq{r02-A})};
\draw[->](start5) -- node[text width=3.5cm] {Reducing the equations to isotropic form}(start6);
\node (start7)[activityStarts, below of=start3, yshift=-2cm]{ $\left<\left<\psi_{\kvec-\qvec}\phi_{\qvec}|\psi^{\dagger}_{\kvec+\qvec'}\phi_{\qvec'}\right>\right>_{i\omega_{n}}$ };
\draw[->](start3) -- node[text width=3cm] {Skip terms from the 4th order of the vertex corrections }(start7);
\node (start8)[activityStarts, below of=start7, yshift=-0.7cm]  { $\left<\left<\psi_{\kvec-\qvec}|\psi^{\dagger}_{\kvec+\qvec'}\right>\right>_{i\omega_{n}}\left<\left<\phi_{\qvec}|\phi_{\qvec'}\right>\right>_{i\omega_{n}-i\omega_{m}}$};
\node (start9) [activityStarts, below of=start8, yshift=-1.5cm,text width=6cm] {Generalized Eliashberga equations for the bare vertex $g^{2}_{\qvec}$ (\eq{r04-C})};
\draw[->] (start7) -- node[text width=3cm] {Wick theorem} (start8);
\draw[->] (start8) -- node[text width=3cm] {Self-consistent procedure} (start9);
\node (start10) [activityStarts, below of=start9, yshift=-1.4cm,text width=6cm] {Generalized isotropic Eliashberg equations for the bare vertex $g^{2}$ (\eq{r08-C})};
\draw[->](start9) -- node[text width=3.5cm] {Reducing the equation to isotropic form}(start10);
  \end{tikzpicture}
\caption{The classic (green) and modified (blue) scheme for deriving the Eliashberg equations. The white blocks are common to both methods.}
\label{Fig01-C}
\end{figure}

We present the derivation of Eliashberg equations for the higher-order many-body terms in the channel of order parameter. 
The self-consistent scheme has been characterized by \fig{Fig01-C}. In particular, the Matsubara Green function is given by formula 
\textcolor{blue}{\cite{Fetter1971A, Elk1979A}}: 
\begin{equation}
\label{r01-C}
\left<\left<\Psi_{\kvec-\qvec}|\Psi^{\dagger}_{\kvec+\qvec'}\right>\right>_{i\omega_{n}}
=\left( \begin{array}{cc}
\left<\left<c_{\kvec-\qvec \uparrow}|c^{\dagger}_{\kvec+\qvec'\uparrow}\right>\right>_{i\omega_{n}}&
\left<\left<c_{\kvec-\qvec \uparrow}|c^{}_{-\kvec-\qvec'\downarrow}\right>\right>_{i\omega_{n}}\\
\left<\left<c^{\dagger}_{-\kvec+\qvec \downarrow}|c^{\dagger}_{\kvec+\qvec'\uparrow}\right>\right>_{i\omega_{n}}&
\left<\left<c^{\dagger}_{-\kvec+\qvec \downarrow}|c_{-\kvec-\qvec'\downarrow}\right>\right>_{i\omega_{n}}
\end{array}\right)
\equiv
\left( \begin{array}{cc}
g^{A}_{\kvec-\qvec}(i\omega_{n})&
g^{B}_{\kvec-\qvec}(i\omega_{n})\\
g^{C}_{\kvec-\qvec}(i\omega_{n})&
g^{D}_{\kvec-\qvec}(i\omega_{n})
\end{array}\right),
\end{equation}
where the symbol $c^{\dag}_{\kvec\sigma}$ ($c_{\kvec\sigma}$) represents the electron creation (anihilation) operator with 
momentum ${\bf k}$ and spin $\sigma\in\{\uparrow, \downarrow\}$. Including the many-body contributions of fourth order of 
$g$, one can obtain:
\begin{equation}
\label{r02-C}
\left<\left<\psi_{\kvec-\qvec}|\psi^{\dagger}_{\kvec+\qvec'}\right>\right>_{i\omega_{n}}=
\left( \begin{array}{cc}
f^{(1)}_{\kvec\qvec}(i\omega_{n})+\omega^{2}_{\qvec} \cdot g^{A}_{\kvec-\qvec}(i\omega_{n})&
-(\omega_{n}-\omega_{m})^{2} \cdot g^{B}_{\kvec-\qvec}(i\omega_{n})\\
-(\omega_{n}-\omega_{m})^{2} \cdot g^{C}_{\kvec-\qvec}(i\omega_{n})&
f^{(2)}_{\kvec\qvec}(i\omega_{n})+\omega^{2}_{\qvec}\cdot g^{D}_{\kvec-\qvec}(i\omega_{n})
\end{array}\right).
\end{equation}
The functions appearing in above matrix possesses the following form:
\begin{eqnarray}
\label{r03-C}
f^{(1)}_{\kvec\qvec}(i\omega_{n})=\left[g^{-}_{0\kvec-\qvec}(i\omega_{n})\right]^{-1}\left(1+n^{ph}_{\qvec}+n^{ph}_{-\qvec}\right)+ \omega_{\qvec}\left(1+n^{ph}_{\qvec}-n^{ph}_{-\qvec}\right)-2\omega_{\qvec}n^{e}_{\kvec-\qvec \uparrow},\\ \nonumber
f^{(2)}_{\kvec\qvec}(i\omega_{n})=\left[g^{+}_{0\kvec-\qvec}(i\omega_{n})\right]^{-1}\left(1+n^{ph}_{\qvec}+n^{ph}_{-\qvec}\right)+ \omega_{\qvec}\left(1+n^{ph}_{\qvec}-n^{ph}_{-\qvec}\right)-2\omega_{\qvec}(1-n^{e}_{-\kvec+\qvec \downarrow}),
\end{eqnarray}
where: $g^{\pm}_{0\kvec}(i\omega_{n})=\left[\varepsilon_{\kvec}\pm  i\omega_{n}\right]^{-1}$, 
are bare electron Green functions that ignore the electron-phonon interaction. 
Additionally, $\varepsilon_{\kvec}$ is the electron band energy, and $\omega_{\qvec}$ represents the phonon dispersion relation. 
The symbols $n^{e}_{\kvec \sigma}$ and $n^{ph}_{\qvec}$ refer to the Fermi-Dirac and the Bose-Einstein functions, respectively.
The self-consistent procedure allows to obtain the modified equation for order parameter:
\begin{equation}
\label{r04-C}
\varphi_{\bf k}\left(i\omega_{n}\right)=\frac{k_{B}T}{N}\sum_{m{\bf q}}
K^{\left(a\right)}_{\bf kq}\left(\omega_{n}-\omega_{m}\right)K_{\bf q}\left(\omega_{n}-\omega_{m}\right) 
\frac{\varphi_{{\bf k}-{\bf q}}\left(i\omega_{m}\right)}{D_{{\bf k}-{\bf q}}\left(i\omega_{m}\right)}.
\end{equation}
The additional many-body term can be written as:
\begin{equation}
\label{r05-C}
K^{\left(a\right)}_{\kvec\qvec}\left(\omega_{n}-\omega_{m}\right)=\frac{\left(\omega_{n}-\omega_{m}\right)^{2}}
{\omega^{2}_{n}+\varepsilon^{2}_{\kvec-\qvec}}.
\end{equation}
Assuming $K^{\left(a\right)}_{\kvec\qvec}\left(\omega_{n}-\omega_{m}\right)\rightarrow 1$, we get classic equation for 
the order parameter. We introduced also the designation: $D_{\kvec-\qvec}\left(i\omega_{m}\right)
=\left(\omega_{m}Z_{\kvec-\qvec}\left(i\omega_{m}\right)\right)^{2}+\varepsilon^{2}_{\kvec-\qvec}+\varphi^{2}_{\kvec-\qvec}
\left(i\omega_{m}\right)$. In the isotropic limit: $\varphi_{\bf k}\left(i\omega_{n}\right)\rightarrow\varphi_{n}$, the pairing kernel 
of electron-phonon interaction is transformed as follows:
\begin{equation}
\label{r06-C}
K_{\bf q}\left(\omega_{n}-\omega_{m}\right)=2g^{2}_{\qvec}\frac{\omega_{\qvec}}{\left(\omega_{n}-\omega_{m}\right)^{2}+\omega^{2}_{\qvec}}
\rightarrow K\left(\omega_{n}-\omega_{m}\right)=\int^{+\infty}_{-\infty}d\omega
\frac{\left<g_{\qvec}^2F_{\qvec}\left(\omega\right)\right>2\omega}{\left(\omega_{n}-\omega_{m}\right)^{2}+\omega^{2}}=\frac{1}{\rho\left(0\right)}\lambda_{nm},
\end{equation}
where symbol $\left<...\right>$ means averaging over the Fermi surface, and $F_{\qvec}\left(\omega\right)=\delta\left(\omega-\omega_{\qvec}\right)$ 
is the phonon density of states. Then the sum of wave vectors should be replaced by the energy integral: 
$\frac{1}{N}\sum_{\qvec}\rightarrow\rho\left(0\right)\int^{+\infty}_{-\infty}d\varepsilon$. Hence, the equation for order parameter takes the form:
\begin{equation}
\label{r07-C}
\varphi_{n}=k_{B}T\sum_{m}\lambda_{nm}
\int^{+\infty}_{-\infty}d\varepsilon \frac{\left(\omega_{n}-\omega_{m}\right)^{2}}{\omega^{2}_{n}+\varepsilon^{2}}
\frac{\varphi_{m}}{\left(\omega_{m}Z_{m}\right)^{2}+\varepsilon^{2}+\varphi_{m}^{2}}.
\end{equation}
The integration in \eq{r07-C} can be done analytically ($\int^{+\infty}_{-\infty}d\varepsilon \frac{1}{A+\varepsilon^{2}}\frac{1}{B+\varepsilon^{2}}=\frac{\pi}{A\sqrt{B}+\sqrt{A}B}$). As a result, we get the algebraic form of order parameter equation in the isotropic approximation:
\begin{equation}
\label{r08-C}
\varphi_{n}=\pi k_{B}T\sum_{m}\left(\omega_{n}-\omega_{m}\right)^{2}\lambda_{nm} 
\frac{\varphi_{m}}{\omega^{2}_{n}\sqrt{\left(\omega_{m}Z_{m}\right)^{2}+\varphi_{m}^{2}}
+|\omega_{n}|\left(\left(\omega_{m}Z_{m}\right)^{2}+\varphi_{m}^{2}\right)}.
\end{equation}
The simple transformations allow the separation of higher-order many-body contributions $p_{1}\left(\omega_{n},\omega_{m}\right)$ 
and $p_{2}\left(\omega_{n},\omega_{m}\right)$ in \eq{r08-C}:
\begin{equation}
\label{r09-C}
\varphi_{n}=\pi k_{B}T\sum_{m}\left[1+p_{1}\left(\omega_{n},\omega_{m}\right)\right]\lambda_{nm} 
\frac{\varphi_{m}}{\sqrt{\left(\omega_{m}Z_{m}\right)^{2}+\varphi_{m}^{2}}
+p_{2}\left(\omega_{n},\omega_{m}\right)},
\end{equation}
where:
\begin{eqnarray}
\label{r10-C}
p_{1}\left(\omega_{n},\omega_{m}\right)&=&\left(\frac{\omega_{m}}{\omega_{n}}\right)^{2}-2\frac{\omega_{m}}{\omega_{n}},\\ \nonumber
p_{2}\left(\omega_{n},\omega_{m}\right)&=&\frac{1}{|\omega_{n}|}\left(\left(\omega_{m}Z_{m}\right)^{2}+\varphi_{m}^{2}\right).
\end{eqnarray}

\begin{figure}
\includegraphics[width=0.45\textwidth]{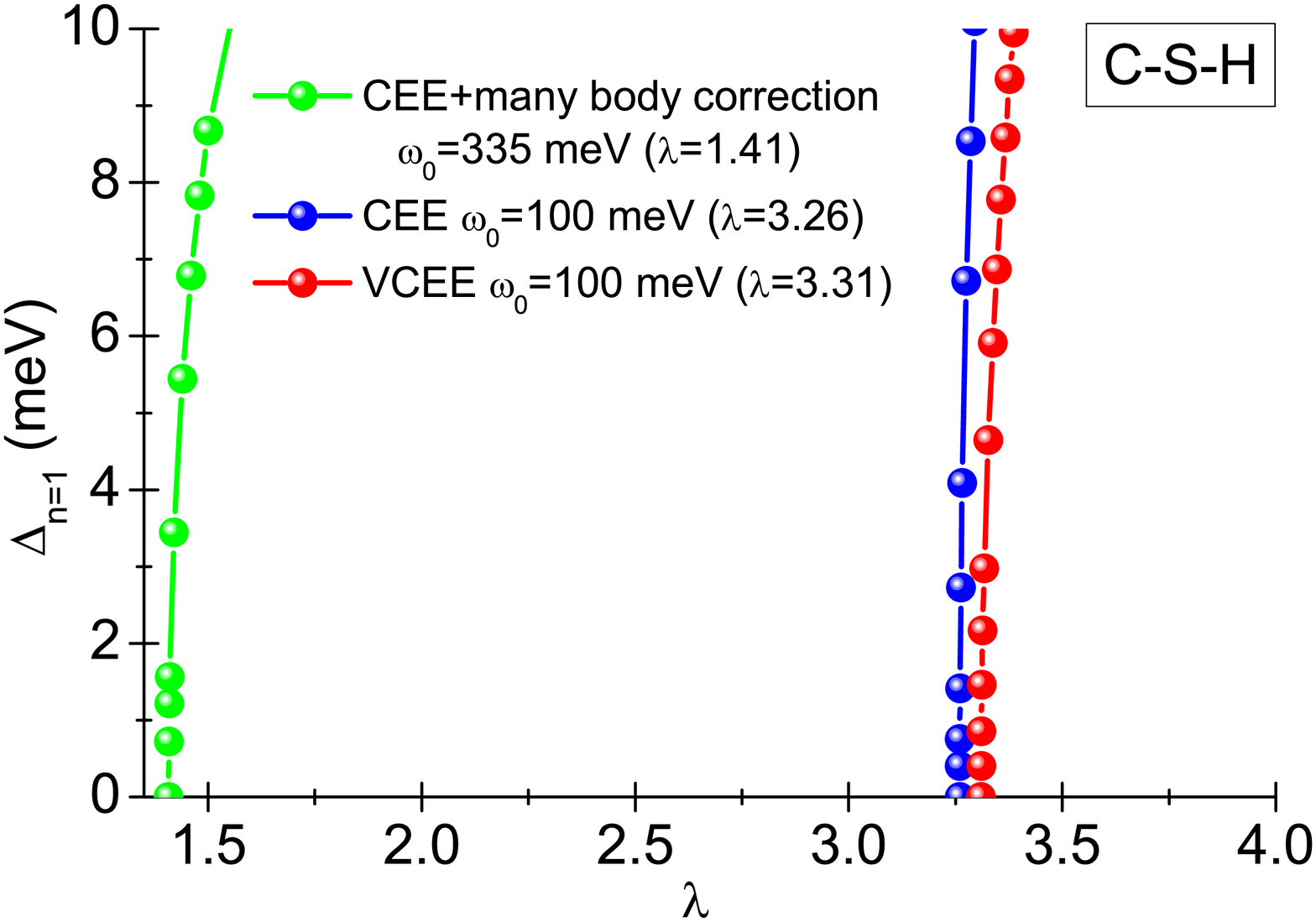}
\includegraphics[width=0.45\textwidth]{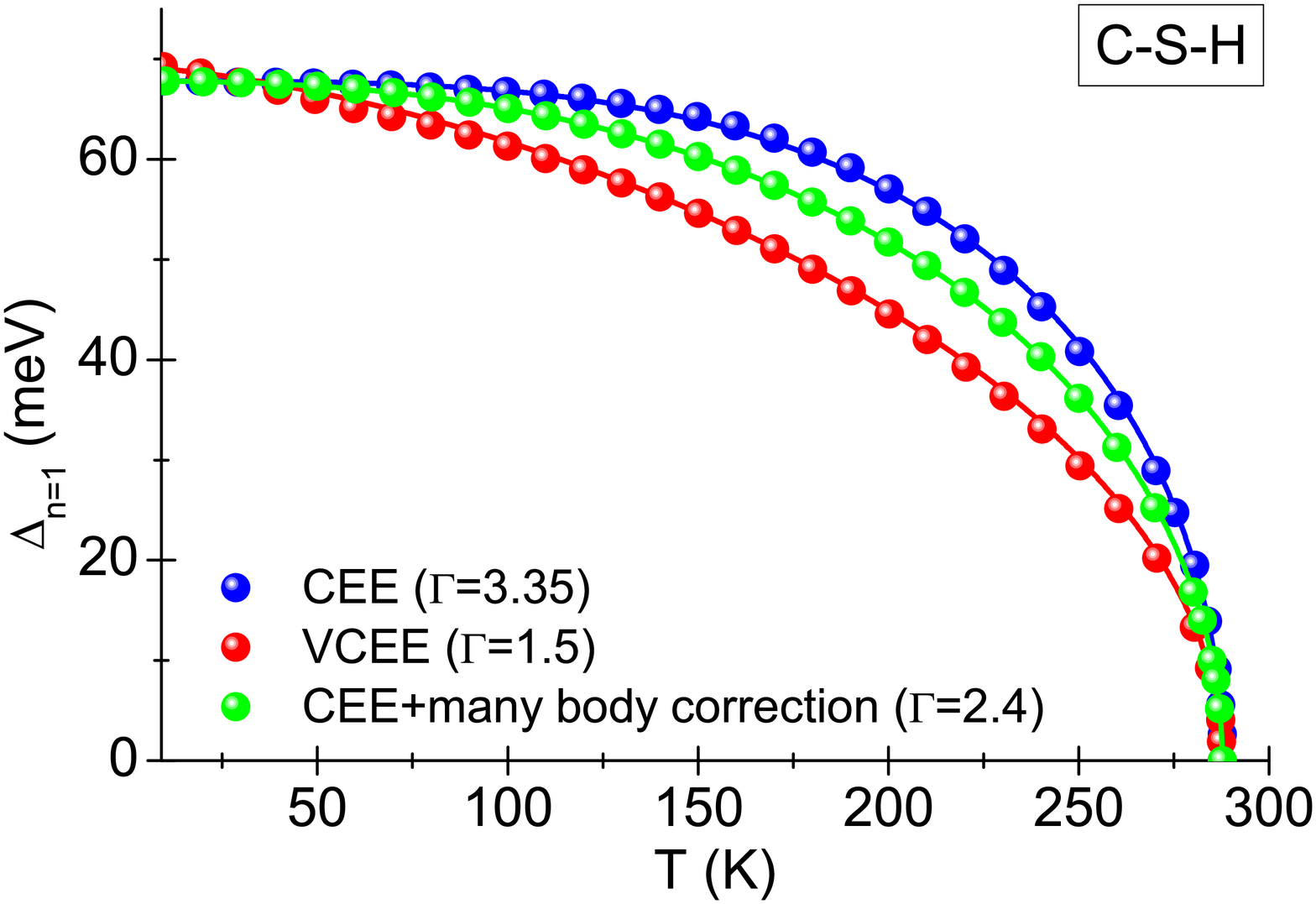}
\caption{
        (a) The dependence of maximum value of the order parameter on the electron-phonon coupling constant for C-S-H system
            ($T_{C}=267$~K and $\mu^{\star}=0.1$).
        (b) The dependence of order parameter on the temperature. The results has been obtained in the framework of CEE model, VCEE formalism 
            and equations taking into account the additional many-body terms.}  
\label{Fig02-IV}
\end{figure}

We will show below that the inclusion of additional many-body terms in the Eliashberg formalism (\eq{r09-C}) leads to the significant reduction 
of electron-phonon coupling constant (C-S-H system). In this case, the modified formalism suggests ($\omega_{0}\sim 350$~meV), 
since for lower values of $\omega_{0}$  ($100$~meV or $200$~meV), there is $\Delta_{n}=0$ for $T<T_{C}$. In particular, the data collected in 
\fig{Fig02-IV}~(a) proves that $\lambda=1.41$, with the function $\Delta\left(T\right)$ is very slightly different from the curve obtained under 
the classical Eliashberg formalism (\fig{Fig02-IV}~(b)). This means that the CEE model correctly defines the measurable thermodynamic parameters 
of superconducting phase.      

\end{document}